\documentclass{article}
\usepackage{amsmath,amssymb,amscd}
\newtheorem{Th}{Theorem}

\newtheorem{Le}{Lemma}

\newtheorem{Co}{Corollary}

\newtheorem{Rem}{Remark}

\newcommand{\ol}{\overline}

\begin{document}
\renewcommand{\theequation}{\arabic{section}.\arabic{equation}}

\title{$L_p$ estimates of solutions to mixed boundary value problems for the Stokes system in polyhedral domains.}

\date{}
\author{by V. G. Maz'ya and J. Rossmann}
\maketitle

{\small{\bf Key words}: Stokes system, nonsmooth domains\\

}

\begin{abstract}
A mixed boundary value problem for the Stokes system in a polyhedral domain is
considered. Here different boundary conditions (in particular, Dirichlet, Neumann,
free surface conditions) are prescribed on the sides of the polyhedron. The authors
prove the existence of solutions in (weighted and non-weighted) $L_p$ Sobolev
spaces and obtain regularity assertions for weak solutions. The results are
based on point estimates of Green's matrix.
\end{abstract}

\maketitle
\setcounter{section}{-1}
\section{Introduction}
Steady-state flows of incompressible viscous Newtonian fluids are modelled by the
Navier-Stokes equations
\begin{equation}
-\nu \, \Delta u + (u\cdot \nabla)\, u + \nabla p = f, \qquad \nabla\cdot u = 0
\end{equation}
for the velocity $u$ and the pressure $p$. To this system, one may add
a variety of boundary conditions on different parts of the boundary (see e.g. \cite{Gunzburger}).
For example, there is the Dirichlet condition $u=0$ on solid walls. On other parts
of the boundary (an artificial boundary such as the exit of a canal, or a free surface)
a no-friction condition $2\nu \varepsilon(u)\, n -pn =0$ may be useful. Here $\varepsilon(u)$
denotes the matrix with the components $\frac 12 (\partial_{x_i}u_j+\partial_{x_j}u_i)$,
and $n$ is the outward normal.
It is also of interest to consider boundary conditions containing components of the velocity
and of the friction. Frequently used combinations are the normal component of the velocity
and the tangential component of the friction (slip condition for uncovered fluid surfaces) or the
tangential component of the velocity and the normal component of the friction (condition for
in/out-stream surfaces).

In the present paper, we consider a mixed boundary value problem for the linear Stokes system
\begin{equation} \label{intro1}
-\Delta u + \nabla p = f, \qquad -\nabla\cdot u = g
\end{equation}
in a three-dimensional domain of polyhedral type, where components of the
velocity and/or the friction are given on the boundary. To be more precise, we
have one of the following boundary conditions on each side $\Gamma_j$:
\begin{itemize}
\item[(i)] $u=h$,
\item[(ii)] $u_\tau=h,\quad -p+2\varepsilon_{n,n}(u) = \phi$,
\item[(iii)] $u_n  = h, \quad \varepsilon_{n,\tau}(u)=\phi$,
\item[(iv)] $-p n + 2\varepsilon_n (u) = \phi$,
\end{itemize}
where $u_n=u\cdot n$ denotes the normal and $u_\tau=u-u_n n$ the tangential component of $u$,
$\varepsilon_n(u)$ is the vector $\varepsilon(u)\, n$, $\varepsilon_{n,n}(u)$ is the normal component
and $\varepsilon_{n,\tau}(u)$ the tangential component of $\varepsilon_n(u)$.

In the last decades a  number of mathematical papers appeared which treat elliptic boundary value problems
in piecewise smooth domains. For a historical account of this development we
refer to the books of Grisvard \cite{Grisvard-85}, Dauge \cite{Dauge-88}, Nazarov and Plamenevski\u{\i}
\cite{Nazarov/Plam}, Kozlov, Maz'ya and Rossmann  \cite{kmr1}.
Our main goal is to prove regularity assertions for weak solutions of the mixed problem to the
Stokes system.
For the Dirichlet problem such results were obtained in papers by Maz'ya and Plamenevski\u{\i} \cite{mp83}
and Dauge \cite{Dauge-89}. Fabes, Kenig and Verchota \cite{Fabes} studied the Dirichlet problem
for the Stokes system in Lipschitz domains. Spectral properties of operator pencils generated by the mixed
boundary value problem in a cone were investigated by Kozlov, Maz'ya and Rossmann \cite[Ch.6]{kmr2}.

It is well-known that the singularities of solutions of elliptic problems near edges and corners
have power (or power-logarithmic) form. For this reason, it is natural to use weighted Sobolev
spaces, where the weights are powers of the distances to the edges and corners. Special boundary
value problems (e.g., the Dirichlet problem) can be studied in weighted Sobolev spaces with
``homogeneous" norms (see e.g. \cite{mp83,mps}). However, the more general problem with boundary conditions (i)--(iv)
requires the use of weighted spaces with ``nonhomogenous" norms. This makes the consideration of
the boundary value problem more difficult. On the other hand, in some cases (e.g. the Dirichlet
problem in convex polyhedral domains), the results can be improved when considering solutions in
weighted spaces with nonhomogeneous norms. We also note that the class of weighted Sobolev spaces
with nonhomogeneous norms contains the nonweighted Sobolev spaces.

The largest part of the paper (Sections 3 and 4) concerns the boundary value problem for the Stokes
system in a polyhedral cone ${\cal K}$ with sides $\Gamma_1,\ldots,\Gamma_n$ and edges $M_1,\ldots,M_n$.
Section 3 deals with the existence  of solutions $(u,p) \in W_{\beta,\delta}^{2,s}({\cal K})^3\times
W_{\beta,\delta}^{1,s}({\cal K})$ of the boundary value problem if $f \in W_{\beta,\delta}^{0,s}({\cal K})^3$,
$g\in W_{\beta,\delta}^{1,s}({\cal K})$, and the boundary data $h$, $\phi$ are from the corresponding trace
spaces. Here, for integer $l\ge 0$, $\beta\in {\Bbb R}$, $\delta=(\delta_1,\ldots,\delta_n)\in {\Bbb R}^n$, and
$1<s<\infty$, the space $W_{\beta,\delta}^{l,s}({\cal K})$ is defined as the set all functions $u$ such that
\[
\rho^{\beta-l+|\alpha|} \prod_{k=1}^n \Big( r_k/\rho \Big)^{\delta_k}\, \partial_x^\alpha u \in
  L_s({\cal K})   \ \mbox{ for } |\alpha|\le l,
\]
$\rho$ is the distance to the vertex of the cone, and $r_k$ denotes the distance to the edge $M_kj$.
The estimates of the solutions in these spaces are essentially based on point estimates for
Green's matrix obtained in our previous paper \cite{mr-03a,mr-03}.
It is shown that there is a uniquely determined solution if $g$ and the boundary data satisfy certain
compatibility conditions on the edges, the line $\mbox{Re}\, \lambda = 2-\beta-3/s$ is free of eigenvalues
of a certain operator pencil ${\mathfrak A}(\lambda)$, and $\max(2-\mu_k,0)<\delta_k+2/s<2$ for $k=1,\ldots,n$,
where $\mu_k$ are certain positive numbers depending on the angle $\theta_k$ at the edge $M_k$.
For example, in the case of the Dirichlet problem, we have $\mu_k=\pi/\theta_k$ if $\theta_k<\pi$, while
$\mu_k$ is the smallest positive solution of the equation $\sin(\mu\theta_k)+\mu\sin\theta_k=0$ if
$\theta_k>\pi$. Estimates for the eigenvalues of the pencil ${\mathfrak A}(\lambda)$ can be found e.g.
in \cite{Dauge-89,kmr2,kms,mp83}.

In Section 4 we consider weak solutions of the boundary value problem, i.e. vector functions
$(u,p) \in W_{\beta,\delta}^{1,s}({\cal K})^3\times W_{\beta,\delta}^{0,s}({\cal K})$ satisfying
\begin{eqnarray*}
&& 2\int_{\cal K} \sum_{i,j=1}^3 \varepsilon_{i,j}(u)\, \varepsilon_{i,j}(v)\, dx
  - \int_{\cal K} p\, \nabla\cdot v\, dx = F(v) \quad\mbox{for all }v\in
  W_{-\beta,-\delta}^{1,s'}({\cal K})^3,\ S_j v=0\ \mbox{on }\Gamma_j, \\
&& -\nabla\cdot u = g \ \mbox{ in }{\cal K}, \quad S_ju=h_j\ \mbox{ on }\Gamma_j, \ j=1,\ldots,n.
\end{eqnarray*}
Here $S_ju =u$ in the case of the Dirichlet condition on $\Gamma_j$, $S_j u=u_\tau$ in the case of
condition (ii), and $S_j u=u_n$ in the case of condition (iii).
We prove that a unique weak solution exists if the boundary data $h_j$ satisfy certain compatibility
conditions on the edges, the line $\mbox{Re}\, \lambda = 1-\beta-3/s$ is free of eigenvalues
of the pencil ${\mathfrak A}(\lambda)$, and $\max(1-\mu_k,0)<\delta_k+2/s<1$
for $k=1,\ldots,n$. In the case $s=2$, the last condition can be replaced by $-\min(\mu_k,1)
<\delta_k\le 0$.

Moreover, we obtain regularity assertions for the weak solution.
For example, let $(u,p)\in W_{0,0}^{1,2}({\cal K})^3 \times L_2({\cal K})$ be the weak solution
of the boundary value problem, where
\[
F\in (W_{0,0}^{1,2}({\cal K})^*)^3\cap (W_{-\beta,-\delta}^{1,s'}({\cal K})^*)^3,\ \
  g \in L_2({\cal K}) \cap W_{\beta,\delta}^{0,s}({\cal K}), \ \ h_j \in W_{0,0}^{1/2,2}(\Gamma_j)
  \cap W_{\beta,\delta}^{1-1/s,s}(\Gamma_j),
\]
$s'=s/(s-1)$. If $\max(1-\mu_k,0) < \delta_k+2/s < 1$ and there are no eigenvalues of the pencil
${\mathfrak A}(\lambda)$ in the strip $-1/2 < \mbox{Re}\, \lambda \le 1-\beta-3/s$, then
\[
(u,p) \in W_{\beta,\delta}^{1,s}({\cal K})^3\times W_{\beta,\delta}^{0,s}({\cal K}).
\]
Suppose the functional $F\in (W_{0,0}^{1,2}({\cal K})^*)^3$ has the form
\[
F(v) = \int_{\cal G} f\cdot v\, dx + \sum \int_{\Gamma_j} \phi_j\cdot v\, dx
\]
where $f\in W_{\beta,\delta}^{l-2,s}({\cal K})^3$, $\phi_j\in W_{\beta,\delta}^{l-1-1/s,s}(\Gamma_j)$,
$l\ge 2$. If, moreover, $g\in W_{\beta,\delta}^{l-1,s}{\cal K})$, $h_j \in
W_{\beta,\delta}^{l-1/s,s}(\Gamma_j)$, the data $g$, $h_j$ and $\phi_j$
satisfy certain compatibility conditions on the edges of the cone, the components $\delta_k$ of $\delta$
satisfy the inequalities $\max(l-\mu_k,0)<\delta_k+2/s<l$, and the strip
$-1/2 < \mbox{Re}\, \lambda \le l-\beta-3/s$ is free of eigenvalues of the pencil ${\mathfrak A}(\lambda)$,
then
\[
(u,p) \in W_{\beta,\delta}^{l,s}({\cal K})^3\times W_{\beta,\delta}^{l-1,s}({\cal K}).
\]
In Section 5 we consider the boundary value problem for the Stokes system (\ref{intro1}) in a bounded
domain ${\cal G}$ of polyhedral type. Under certain compatibility conditions, there exists a
weak solution $(u,p)\in W^{1,2}({\cal G})^3\times L_2({\cal G})$ which is unique up to a certain subspace
of linear vector functions. Using the results of Section 4, we obtain regularity assertions for this solution.

As an example, we consider the weak solution of the Dirichlet problem to the Stokes
system in a polyhedron with boundary data $h_j=0$. From our results and from estimates for the
eigenvalues of the pencil ${\mathfrak A}(\lambda)$ (see \cite{Dauge-89,kmr2,kms,mp83}) it follows that
\[
(u,p) \in W^{1,s}({\cal G})^3 \times L_s({\cal G}), \quad 2 < s \le 3,
\]
if $f\in W^{-1,s}({\cal G})^3$, $g\in L_s({\cal G})$. If the polyhedron ${\cal G}$ is convex, then this
result is true for all $s>2$.
Furthermore, the following $W^{2,s}$-regularity result holds for the weak solution
$(u,p) \in W^{1,2}({\cal G})^3\times L_2({\cal G})$ of the Dirichlet problem:
\[
(u,p) \in W^{2,s}({\cal K})^3\times W^{1,s}({\cal K}), \ \ 1<s\le 4/3,
\]
if $f\in W^{-1,2}({\cal G})^3\cap L_s({\cal G})^3$, $g\in L_2({\cal G}) \cap W^{1,s}({\cal G})$.
If the edge angles are less than $3\, \mbox{arccos}\frac 14 \approx 1.2587\pi$, then this result is
true for $1<s\le 3/2$. In the case of a convex polyhedron, this result is true for $1<s\le 2$.
If, moreover, the edge angles are less then $\frac 34 \pi$, then the result holds even for $1<s<3$.
However, in the case $s>2$ the trace of the function $g$ on the edges must be equal to zero, while in
the case $s=2$ the function $g$ must be such that
\begin{equation} \label{condg}
\int_{\cal G} \rho_j^{-1} \prod_{k\in N_j} (r_k/\rho_j)^{-1} \, |g(x)|^2\, dx < \infty,
\end{equation}
for every $j$, where $\rho_j$ denotes the distance to the vertex $O_j$, $r_k$ denotes the distance
to the edge $M_k$, and $N_j$ is the set of all $k$ such that $\bar{M}_k\ni O_j$.
In the case $s=2$ the $W^{2,s}$-regularity result for convex polyhedrons was also proved by Dauge \cite{Dauge-89}.

Similar $W^{1,s}$ and $W^{2,s}$ regularity results can be obtained for Neumann and mixed problems.
Let us consider, for example, the mixed boundary value problem with boundary conditions (i)--(iii). We assume
that for every edge, the Dirichlet condition is given on at least one of the adjoining sides. Then the
following $W^{1,s}$ regularity result holds. The weak solution $(u,p) \in W^{1,2}({\cal G})^3\times L_2({\cal G})$
belongs to $W^{1,s}({\cal G})^3\times L_s({\cal G})$ if $F\in (W^{1,s'}({\cal G})^*)^3$, $g\in L_s({\cal G})$,
$h_j=0$, $2\le s\le 8/3$, $s'=s/(s-1)$. If at every edge with boundary condition (ii) or (iii)  on one of
the adjoining sides, the angle is less than $\frac 32\pi$, then this result is even true for $2\le s \le 3$.

Lastly, we present a $W^{2,s}$ regularity result for the mixed problem with Dirichlet and Neumann
boundary conditions. Suppose that $(u,p)\in W^{1,2}({\cal G})^3\times L_2({\cal G})$ is a weak solution
of this problem with data $f\in L_s({\cal G})^3$, $g\in W^{1,s}({\cal G})$,
$h_j \in W^{2-1/s,s}(\Gamma_j)^3$, and $\phi_j\in W^{1-1/s,s}(\Gamma_j)^3$, $1< s\le 8/7$.
Then $(u,p) \in W^{2,s}({\cal G})^3 \times W^{1,s}({\cal G})$.

Other examples are given at the end of Section 5.
In a forthcoming paper, we extend the results to mixed problems for the nonlinear Navier-Stokes system.

\setcounter{equation}{0}
\setcounter{Th}{0}
\setcounter{Le}{0}
\setcounter{Co}{0}
\section{Weighted Sobolev spaces}

\subsection{Weighted Sobolev spaces in a dihedron}
Let ${\cal D}$ be the dihedron
\begin{equation} \label{dih}
{\cal D} = \{ x=(x',x_3):\ x'\in K,\ x_3 \in {\Bbb R}\},
\end{equation}
where $K$ is an infinite angle which has the form $\{ x'=(x_1,x_2)\in {\Bbb R}^2:\
0<r<\infty,\ -\theta/2<\varphi<\theta/2\}$ in polar coordinates $r,\varphi$. The boundary of
${\cal D}$ consists of the half-planes $\Gamma^\pm:\, \varphi=\pm \theta/2$ and the edge $M$.
We denote by $V_\delta^{l,s}({\cal D})$ and $W_\delta^{l,s}({\cal D})$, $1<s<\infty$, the
weighted Sobolev spaces with the norms
\[
\| u\|_{V_\delta^{l,s}({\cal D})} = \Big( \int\limits_{\cal D} \sum_{|\alpha|\le l}
  |x'|^{s(\delta-l+|\alpha|)}\, \big| \partial_x^\alpha u\big|^s\, dx\Big)^{1/s},
\quad
\| u\|_{W_\delta^{l,s}({\cal D})} = \Big( \int\limits_{\cal D} \sum_{|\alpha|\le l}
  |x'|^{s\delta}\, \big| \partial_x^\alpha u\big|^s\, dx\Big)^{1/s}.
\]
Analogously, the spaces $V_\delta^{l,s}(K)$ and $W_\delta^{l,s}(K)$ are defined (here
in the above norms ${\cal D}$ has to be replaced by $K$ and $dx$ by $dx'$).
By Hardy's inequality, every function $u\in C_0^\infty(\ol{\cal D})$ satisfies the
inequality
\[
\int_{\cal D} r^{s(\delta-1)} |u|^2\, dx \le c\, \int_{\cal D} r^{s\delta}\,
   |\nabla u|^s\, dx
\]
for $\delta>1-2/s$ with a constant $c$ depending only on $s$ and $\delta$. Consequently,
the space $W_\delta^{l,s}({\cal D})$ is continuously imbedded into
$W_{\delta-1}^{l-1,s}({\cal D})$ if $\delta>1-2/s$. If $\delta >l-2/s$, then
$W_\delta^{l,s}({\cal D})\subset V_\delta^{l,s}({\cal D})$.

Let $V_\delta^{l-1/s,s}(\Gamma^\pm)$ and $W_\delta^{l-1/s,s}(\Gamma^\pm)$ be the trace
spaces corresponding to $V_\delta^{l,s}({\cal D})$ and $W_\delta^{l,s}({\cal D})$,
respectively. The trace spaces for  $V_\delta^{l,s}(K)$ and $W_\delta^{l,s}(K)$
on the sides $\gamma^\pm$ of $K$ are denoted by $V_\delta^{l-1/s,s}(\gamma^\pm)$ and
$W_\delta^{l-1/s,s}(\gamma^\pm)$, respectively.

Note that the trace of a function $u\in W_\delta^{l,s}({\cal D})$ or $u\in W_\delta^{l-1/s,s}(\Gamma^\pm)$
on the edge $M$ exists if $-2/s<\delta<l-2/s$. It belongs to the Sobolev-Slobodetski\u{\i} space
$W^{l-\delta-2/s,s}(M)$ if $l-\delta-2/s$ is not integer. There is the following relation between
the spaces $V_\delta^{l,s}$ and $W_\delta^{l,s}$ (see \cite{mr-88,r92}).

\begin{Le} \label{al1}
{\em 1)} Let $u\in W_\delta^{l,s}({\cal D})$, $-2/s<\delta\le l-2/s$, If $\delta+2/s$ is not integer, then
\begin{equation} \label{1al1}
u \in V_\delta^{l,s}({\cal D}) \Leftrightarrow \partial_{x'}^\alpha u(x) = 0 \ \mbox{ on $M$ for }
  |\alpha|<l-\delta-2/s.
\end{equation}
If $\delta+2/s$ is integer, then for the inclusion $u \in V_\delta^{l,s}({\cal D})$ it is necessary and sufficient
that the conditions {\em (\ref{1al1})} and
\[
\int_{\cal D} r^{-2} \big| \partial_{x'}^\alpha u(x)\big|^s\, dx <\infty \ \mbox{ for }|\alpha|=l-\delta-2/s
\]
are satisfied.

{\em 2)} Let $u\in W_\delta^{l-1/s,s}(\Gamma^+)$, $-2/s<\delta\le l-2/s$. If $\delta+2/s$ is not integer, then
\begin{equation} \label{2al1}
u \in V_\delta^{l,s}(\Gamma^+) \Leftrightarrow \partial_r^j u(r,x_3) = 0 \ \mbox{ on $M$ for }j<l-\delta-2/s.
\end{equation}
If $\delta+2/s$ is integer, then for the inclusion $u \in V_\delta^{l,s}({\cal D})$ is it necessary and sufficient
that the conditions {\em (\ref{2al1})} and
\[
\int_{\Bbb R}\int_0^\infty r^{-1} \big| \partial_r^{l-\delta-2/s} u(r,x_3)\big|^s\, dr\, dx_3 <\infty
\]
are satisfied.
\end{Le}

We introduce the following extension operator $E$ mapping $W^{l-\delta-2/s,s}(M)$ into
$W_\delta^{l,s}({\cal D})$ or $W_\delta^{l-1/p,p}(\Gamma^\pm)$.
\begin{equation} \label{a1}
(Ef)(x) = \chi(r)\, \int_{\Bbb R} f(x_3+tr)\, \psi(t)\, dt,
\end{equation}
where $r=|x'|$, $\chi$ is a smooth function on $(0,\infty)$ with support in $[0,1]$ equal to 1 in
$(0,\frac 12)$, and $\psi$ is a smooth function on ${\Bbb R}$ with support in $[-1,+1]$ satisfying
the condition
\[
\int_{\Bbb R} \psi(t)\, dt = 1, \quad \int_{\Bbb R} t^j \, \psi(t)\, dt =0 \ \mbox{ for }j=1,2,\ldots,l.
\]
Since the function $Ef$ depends only on $r$ and $x_3$, it can be also considered as a function
on the half-planes $\Gamma^+$ and $\Gamma^-$. For the following lemma we refer to \cite{mr-88}.

\begin{Le} \label{al2}
Let $-2/s<\delta<l-2/s$ and $\delta+2/s$ be not integer. Then
\begin{equation} \label{1al2}
(\partial_{x_3}^j Ef)|_M = \partial_{x_3}^j f \quad \mbox{for } j<l-\delta-2/s .
\end{equation}
Moreover, if $Ef$ is considered as a function on ${\cal D}$, then $\partial_{x'}^\alpha Ef \in
V_{\delta}^{l-|\alpha|,s}({\cal D})$ for $1\le  |\alpha|\le l$. In particular, the trace of
$\partial_{x'}^\alpha Ef$ on $M$ vanishes for $1\le |\alpha|<l-\delta-2/s$.

If $Ef$ is considered as a function on $\Gamma^\pm$, then $\partial_r^j Ef \in
V_\delta^{l-j-1/s,s}(\Gamma^\pm)$ for $j=1,\ldots,l-1$.
\end{Le}

If $f\in W_{\delta}^{2-1/s,s}(\Gamma^+)$, $\delta< 1-2/2$, then the traces of $f$ and $\partial_{x_3}f$
on $M$ exist. Obviously, $(\partial_{x_3}f)|_M = \partial_{x_3}(f|_M)$. The following result for
the limit case $\delta=1-2/s$ follows from \cite[Le.7, Rem.4]{r92}.

\begin{Le} \label{al3}
If $f \in W_{1-2/s}^{2-1/s,s}(\Gamma^+)$ and $f|_M=0$, then
\[
\int_{\Bbb R} \int_0^\varepsilon r^{-1} \, \big|\partial_{x_3}f(r,x_3)\big|^s\, dr\, dx_3
  \le c\, \| f\|^s_{W_{1-2/s}^{2-1/s,s}(\Gamma^+)}
\]
for arbitrary positive $\varepsilon$.
\end{Le}

For the following lemma we refer to \cite[Le.2.1]{mr-02}.

\begin{Le} \label{al4}
If $\partial_{x_3}^j u \in V_\delta^{2,s}({\cal D})$, $1<s< 2$, for $j=0,1,2$, then
$u \in V_{\delta-3+2/s}^{0,2}({\cal D})$.
\end{Le}

\begin{Co} \label{ac1}
If $\partial_{x_3}^j u \in W_\delta^{3,s}({\cal D})$ for $j=0,1,2$, where $1<s<2$ and
$\delta >2-2/s$, then $u\in W_{\delta-3+2/s}^{1,2}({\cal D})$.
\end{Co}

{\it Proof:}
By Lemma \ref{al4}, the inclusion $\partial_{x_3}^j u\in W_\delta^{2,s}({\cal D})
=V_\delta^{2,p}({\cal D})$ for $j\le 2$ implies $u\in V_{\delta-3+2/s}^{0,2}({\cal D})$.
Furthermore, by our assumptions, $\partial_{x_3}^j \nabla u \in V_{\delta}^{2,s}({\cal D})^3$
for $j=0,1,2$ and, therefore, $\nabla u \in V_{\delta-3+2/s}^{0,2}({\cal D})^3$. The result follows.
\rule{1ex}{1ex} \\

\subsection{Weighted Sobolev spaces in a cone}

Let ${\cal K}$ be the cone
\begin{equation} \label{cone}
{\cal K} = \{ x\in {\Bbb R}^3:\ x/|x| \in \Omega\}
\end{equation}
where $\Omega$ is a domain on  the unit sphere of polygonal type ...

We denote by ${\cal S}$ the set $M_1\cup\cdots\cup M_n\cup\{ 0\}$
of all singular boundary points. Furthermore, for an arbitrary point $x\in {\cal K}$
we denote by $\rho(x)=|x|$ the distance to the vertex of the cone, by $r_j(x)$
the distance to the edge $M_j$, and by $r(x)$ the regularized distance to ${\cal S}$,
i.e., an infinitely differentiable function in ${\cal K}$ which satisfies the estimates
\[
c_1\, \mbox{dist}(x,{\cal S}) < r(x) < c_2\, \mbox{dist}(x,{\cal S})\quad\mbox{and}\quad
  |\partial_x^\alpha r(x)|\le c_\alpha \, \mbox{dist}(x,{\cal S})^{1-|\alpha|}
\]
for all $x\in {\cal K}$ and all multi-indices $\alpha$. Here $c_1,c_2,c_\alpha$ are positive
constants independent of $x$.

Let $l$ be a nonnegative integer, $\beta\in {\Bbb R}$, $\delta=(\delta_1,\ldots,\delta_n)\in {\Bbb R}^n$,
$\delta_j>-2/s$ for $j=1,\ldots,n$, and $1<s<\infty$. We define $V_{\beta,\delta}^{l,s}({\cal K})$ and
$W_{\beta,\delta}^{l,s}({\cal K})$ as the weighted Sobolev spaces with the norms
\begin{eqnarray*}
&& \| u\|_{V_{\beta,\delta}^{l,s}({\cal K})}= \Big( \int_{\cal K} \sum_{|\alpha|\le l}
  |x|^{s(\beta-l+|\alpha|)} \ |\partial_x^\alpha u|^s\, \prod_{j=1}^n \big(\frac{r_j(x)}{|x|}
  \Big)^{s(\delta_j-l+|\alpha|)}\, dx\Big)^{1/s}, \\
&& \| u\|_{W_{\beta,\delta}^{l,s}({\cal K})} = \Big( \int_{\cal K}
  \sum_{|\alpha|\le l} |x|^{s(\beta-l+|\alpha|)} \ |\partial_x^\alpha u|^s
  \prod_{j=1}^n \big(\frac{r_j}{\rho}\big)^{s\delta_j}\, dx\Big)^{1/s},
\end{eqnarray*}
respectively. Furthermore, we introduce the following notation. If $d$ is real number,
then $V_{\beta,d}^{l,s}({\cal K})$ and $W_{\beta,d}^{l,s}({\cal K})$ denote the above
introduced spaces with $\delta=(d,\ldots,d)$. If
$\delta=(\delta_1,\ldots,\delta_n)$ and $d$ is a real number, then we define
$W_{\beta,\delta+d}^{l,s}({\cal K})=W_{\beta,\delta'}^{l,s}({\cal K})$, where
$\delta'=(\delta_1+d,\ldots,\delta_n+d)$.

Passing to spherical coordinates $\rho=|x|,\omega=x/|x|$, one obtains the following equivalent
norm in $W_{\beta,\delta}^{l,s}({\cal K})$:
\[
\| u\| = \Big( \int_0^\infty \rho^{s(\beta-l)+2} \sum_{k=0}^l \|
  (\rho\partial_\rho)^k u(\rho,\cdot)\|^s_{W_{\delta}^{l-k,s}(\Omega)}
  \, d\rho  \Big)^{1/s} ,
\]
where the norm in $W_{\delta}^{l,s}(\Omega)$ is given by
\[
\| v\|_{W_{\delta}^{l,s}(\Omega)} = \Big(  \int\limits_{\substack{{\cal K}\\ 1<|x|<2}} \sum_{|\alpha|\le l}
  \big| \partial_x^\alpha v(x)\big|^s\ \prod_{j=1}^n r_j^{s\delta_j}\, dx\Big)^{1/s}
\]
(here the function $v$ on $\Omega$ is extended by $v(x)=v\big( x/|x|\big)$ to the
cone ${\cal K}$).

By Hardy's inequality, the space $W_{\beta+1,\delta'}^{l+1,s}(\Omega)$
is continuously imbedded into $W_{\beta,\delta}^{l,s}(\Omega)$
if $\delta=(\delta_1,\ldots,\delta_n)$, $\delta'=(\delta'_1,\ldots,\delta'_n)$
are such that $\delta_j,\delta'_j>-2/s$ and $\delta'_j-\delta_j \le 1$ for $j=1,\ldots,n$.
This implies that, under the above assumptions on $\delta$ and $\delta'$, there is the imbedding
$W_{\beta+1,\delta'}^{l+1,s}({\cal K}) \subset W_{\beta,\delta}^{l,s}({\cal K})$.
In particular, we have  $V_{\beta,\delta}^{l,s}({\cal K}) = W_{\beta,\delta}^{l,s}({\cal K})$
if $\delta_j>l-2/s$ for $j=1,\ldots,n$.

Let $\zeta_k$ be smooth functions depending only on $\rho=|x|$ such that
\begin{equation} \label{zetak}
\mbox{supp}\, \zeta_k \subset (2^{k-1},2^{k+1}),\quad \sum_{k=-\infty}^{+\infty}
  \zeta_k=1,\quad |(\rho\partial_\rho)^j \zeta_k(\rho)|\le c_j
\end{equation}
with constants $c_j$ independent of $k$ and $\rho$.
It can be easily shown (cf. \cite[Le.6.1.1]{kmr1}) that the norms in $V_{\beta,\delta}^{l,s}({\cal K})$
and $W_{\beta,\delta}^{l,s}({\cal K})$ are equivalent to
\begin{equation} \label{equivalent}
\| u \| = \Big( \sum_{k=-\infty}^{+\infty} \| \zeta_k u  \|^s_{V_{\beta,\delta}^{l,s}({\cal K})}\Big)^{1/s}
  \quad\mbox{and\quad} \| u \| = \Big( \sum_{k=-\infty}^{+\infty} \| \zeta_k u  \|^s_{W_{\beta,\delta}^{l,s}({\cal K})}
  \Big)^{1/s},
\end{equation}
respectively. We denote the trace spaces for $V_{\beta,\delta}^{l,s}({\cal K})$ and
$W_{\beta,\delta}^{l,s}({\cal K})$, $l\ge 1$, on $\Gamma_j$ by $V_{\beta,\delta}^{l-1/s,s}(\Gamma_j)$,
and $W_{\beta,\delta}^{l-1/s,s}(\Gamma_j)$, respectively. The norms in these spaces are also equivalent to
\begin{equation} \label{equivalent1}
\| u \| = \Big( \sum_{k=-\infty}^{+\infty} \| \zeta_k u  \|^s_{V_{\beta,\delta}^{l-1/s,s}(\Gamma_j)}\Big)^{1/s}
  \quad\mbox{and\quad} \| u \| = \Big( \sum_{k=-\infty}^{+\infty} \| \zeta_k u
  \|^s_{W_{\beta,\delta}^{l-1/s,s}(\Gamma_j)}
  \Big)^{1/s},
\end{equation}
respectively. The trace of a function $u \in W_{\beta,\delta}^{l,s}({\cal K})$ (or $u\in
W_{\beta,\delta}^{l-1/s,s}(\Gamma_j)$) on the edge $M_k$ exists if $\delta_k < l-2/s$.
Using Lemma \ref{al3}, we obtain the following result.

\begin{Le} \label{al5}
Let $\Gamma$ be a side of the cone ${\cal K}$ adjacent to the edge $M_k$ and let
$f\in W_{\beta,\delta}^{2-1/s,s}(\Gamma)$, where $\delta_k=1-2/s$.
If $f|_{M_k}=0$, then
\[
\int_0^\infty \int_0^{\varepsilon t} t^{s(\beta-1)+2}\, r^{-1} \, \big| \partial_t f(r,t)\big|^s\,
  dr\, dt < \infty
\]
for sufficiently small $\varepsilon>0$. Here $r=\mbox{\em dist}\,(x,M_k)$ and $t$ is the coordinate on $M_k$.
\end{Le}

{\it Proof}:
Let $\eta_k=\zeta_{k-1}+\zeta_k+\zeta_{k+1}$, $\tilde{\eta}_k(x)=\eta_k(2^k x)$, and $\tilde{f}(r,t)=
f(2^k r,2^k t)$. By our assumptions on the functions  $\zeta_k$, we have $\tilde{\eta}_k(x)=1$ for
$1/2 < |x|<2$. From Lemma \ref{al3} it follows that
\[
\int_{1/2}^2 \int_0^{\varepsilon t} r^{-1} \, \big|\partial_t \tilde{f}(r,t)\big|^s\, dr\, dt
  \le c\, \| \tilde{\eta}_k \tilde{f}\|^s_{W_{\beta,\delta}^{2-1/s,s}(\Gamma)}
\]
Here,
\begin{eqnarray*}
&& \int_{1/2}^2 \int_0^{\varepsilon t} r^{-1} \, \big|\partial_t \tilde{f}(r,t)\big|^s\, dr\, dt
  = 2^{k(s-1)} \int_{2^{k-1}}^{2^{k+1}} \int_0^{\varepsilon t} r^{-1}\, \big| \partial_t f(r,t)\big|^s\, dr\, dt
  \quad\mbox{and} \\
&& \| \tilde{\eta}_k \tilde{f}\|^s_{W_{\beta,\delta}^{2-1/s,s}(\Gamma)} \le c\, 2^{-ks(\beta-2)-3k}\,
  \| \eta_k f \|^s_{W_{\beta,\delta}^{2-1/s,s}(\Gamma)}.
\end{eqnarray*}
This implies
\[
\int_{2^{k-1}}^{2^{k+1}} \int_0^{\varepsilon t} t^{s(\beta-1)+2}\, r^{-1}\, \big| \partial_t f(r,t)
  \big|^s\, dr\, dt \le c\, \| \eta_k f \|^s_{W_{\beta,\delta}^{2-1/s,s}(\Gamma)}.
\]
Summing up over all integer $k$ and using the equivalence of the norm in $W_{\beta,\delta}^{l-1/s,s}(\Gamma_j)$
with the the second norm in (\ref{equivalent1}), we obtain the desired inequality. \rule{1ex}{1ex}

\setcounter{equation}{0}
\setcounter{Th}{0}
\setcounter{Le}{0}
\setcounter{Co}{0}
\section{The boundary value problem in a dihedron}

We consider a boundary value problem for the Stokes system, where on each of the sides $\Gamma^\pm$ one
of the boundary conditions (i)--(iv) is given. Let $n^\pm=(n_1^\pm,n_2^\pm,0)$
be the exterior normal to $\Gamma^\pm$, $\varepsilon_n^\pm(u)=\varepsilon(u)\, n^\pm$ and
$\varepsilon_{nn}^\pm(u)=\varepsilon_n^\pm(u)\cdot n^\pm$. Furthermore, let $d^\pm\in \{ 0,1,2,3\}$
be integer numbers characterizing the boundary conditions on $\Gamma^+$ and $\Gamma^-$, respectively. We put
\begin{itemize}
\item $S^\pm u = u \quad \mbox{for }d^\pm=0$,
\item $S^\pm u = u-(u\cdot n^\pm)n^\pm, \quad N^\pm(u,p)= -p+2 \varepsilon_{nn}^\pm(u)
  \quad\mbox{for }d^\pm=1,$
\item $S^\pm u = u\cdot n^\pm,\quad N^\pm (u,p)= \varepsilon_n^\pm(u)
  -\varepsilon_{nn}^\pm(u)\, n^\pm \quad \mbox{for } d^\pm=2$
\item $N^\pm(u,p)=-pn^\pm +2\varepsilon_n^\pm(u)  \quad \mbox{for }d^\pm=3$
\end{itemize}
and consider the boundary value problem
\begin{eqnarray} \label{b1}
&& -\Delta u + \nabla p = f,\quad -\nabla\cdot u = g\quad\mbox{in }{\cal D}, \\
&& S^\pm u = h^\pm,\quad N^\pm (u,p)=\phi^\pm \quad \mbox{on }\Gamma^\pm. \label{b2}
\end{eqnarray}
Here the condition $N^\pm(u,p)=\phi^\pm$ is absent in the case $d^\pm=0$, while the condition
$S^\pm u=h^\pm$ is absent in the case $d^\pm=3$.
The Dirichlet problem for Stokes system and the mixed problem with Dirichlet condition (i) on $\Gamma^+$
and condition (ii) on $\Gamma^-$ were studied by Maz'ya, Plamenevski\u{i} and Stupelis
\cite{mps}. In contrast to \cite{mps}, we will use here weighted Sobolev spaces $W_{\delta}^{l,s}({\cal D})$
with nonhomogeneous norms.

\subsection{Reduction to homogeneous boundary conditions}

For the following lemma we refer to \cite{mr-03a,mr-03}.

\begin{Le} \label{bl1}
Let $h^\pm \in V_\delta^{l-1/s,s}(\Gamma^\pm)^{3-d^\pm}$, $\phi^\pm \in V_{\delta}^{l-1-1/s,s}
(\Gamma^\pm)^{d^\pm}$, $l\ge 2$. Then there exists a vector function $u\in V_\delta^{l,s}({\cal D})^3$
such that $S^\pm u=h^\pm$ and $N^\pm(u,0)=\phi^\pm$ on $\Gamma^\pm$ satisfying the estimate
\[
\|u\|_{V_\delta^{l,s}({\cal D})^3} \le c\, \Big( \| h^\pm \|_{V_\delta^{l-1/s,s}(\Gamma^\pm)^{3-d^\pm}}
  + \| \phi^\pm \|_{V_{\delta}^{l-1-1/s,s}(\Gamma^\pm)^{d^\pm}}\Big)
\]
with a constant $c$ independent of $h^\pm$ and $\phi^\pm$. If $h^\pm$ and $\phi^\pm$ vanish for $|x'|>1$,
then also $u$ can be chosen such that $u(x)=0$ for $|x'|>1$.
\end{Le}

Now let $h^\pm\in W_\delta^{2-1/s,s}(\Gamma^\pm)^{3-d^\pm}$, $\phi^\pm \in W_{\delta}^{1-1/s,s}
(\Gamma^\pm)^{d^\pm}$ and $g \in W_\delta^{1,s}({\cal D})$, $\delta<2-2/s$, be given functions
vanishing for $|x'|>1$. We want to answer the question under which conditions there exist
functions $u \in W_\delta^{2,s}({\cal D})^3$ and $p\in W_\delta^{1,s}({\cal D})$ such that
\begin{equation} \label{b3}
S^\pm u = h^\pm,\quad N^\pm(u,p)=\phi^\pm\ \mbox{ on }\Gamma^\pm\quad \mbox{and }\ \nabla\cdot u + g
  \in V_\delta^{1,s}({\cal D})
\end{equation}
For $\delta>1-2/s$ the answer follows immediately from the following lemma and from Lemma \ref{bl1}.

\begin{Le} \label{bl2}
Let $h^\pm \in W_\delta^{l-1/s,s}(\Gamma^\pm)^{3-d^\pm}$, $l-1-2/s< \delta< l-2/s$, $l\ge 1$,
$h^\pm(x)=0$ for $|x'|>1$. Suppose that $h^+$ and $h^-$ satisfy the compatibility condition
\begin{equation} \label{1bl2}
\big( h^+|_M\, ,\, h^-|_M\big) \in R(T),
\end{equation}
where $R(T)$ denotes the range of the operator $T=(S^+,S^-)$ (here $S^\pm$ are considered as operators
on $W^{l-\delta-2/s,s}(M)^3$). Then there exists a vector function $u\in W_\delta^{l,s}({\cal D})^3$
such that $u(x)=0$ for $|x'|>1$ and $S^\pm u = h^\pm$.
\end{Le}

{\it Proof:}
By (\ref{1bl2}), there exists a vector function $\psi\in W^{l-\delta-2/s,s}(M)^3$ such that
$S^\pm \psi=h^\pm|_M$. Let $v \in W_\delta^{l,s}({\cal D})^3$ be an extension of $\psi$ vanishing
for $|x'|>1$. Then the traces of $S^\pm v|_{\Gamma^\pm}-h^\pm$ are zero on $M$ and, consequently,
$S^\pm v|_{\Gamma^\pm}-h^\pm \in V_\delta^{l-1/s,s}(\Gamma^\pm)^{3-d^\pm}$ (see Lemma \ref{al1}).
Applying Lemma \ref{bl1} (in the case $l=1$ see \cite[Le.3.1]{mp78a}), we obtain the assertion
of the lemma. \rule{1ex}{1ex}\\

Note that condition (\ref{1bl2}) can be also written in the form
\begin{equation} \label{b4}
Ah^+|_M = Bh^-|_M \, ,
\end{equation}
where $A$ and $B$ are certain matrices. For example, $A=B=I$ in the case of the Dirichlet problem ($d^+=d^-=0$),
$A=(n^-)^t=(n_1^-,n_2^-,0)$, $B=1$ if $d^+=0$ and $d^-=2$.

If $h^\pm\in W_\delta^{2-1/s,s}(\Gamma^\pm)^{3-d^\pm}$, $\phi^\pm \in W_{\delta}^{l-1-1/s,s}
(\Gamma^\pm)^{d^\pm}$ and $g \in W_\delta^{1,s}({\cal D})$, $-2/s<\delta<1-2/s$, then the traces of $g$, $h^\pm$,
$\partial_r h^\pm$ and $\phi^\pm$ on $M$ exist. Suppose that $(u,p) \in W_\delta^{2,s}({\cal D})^3  \times
W_\delta^{1,s}({\cal D})$ satisfies (\ref{b3}). We put
\[
b=u|_M\, ,\quad c=(\partial_{x_1}u)|_M\, , \quad d=(\partial_{x_2}u)|_M\, \quad\mbox{and}\quad q=p|_M\, .
\]
Then from the equations $S^\pm u=h^\pm$ on $\Gamma^\pm$ it follows that $S^\pm \partial_r u = \partial_r h^\pm$
on $\Gamma^\pm$, and therefore,
\begin{eqnarray} \label{b5a}
&& S^\pm b = h^\pm|_M\, , \\ \label{b5b}
&& S^\pm \big( c\cos\mbox{$\frac\theta 2$}\pm d\sin\mbox{$\frac\theta 2$}\big) = (\partial_r h^\pm)|_M\, .
\end{eqnarray}
Moreover $\nabla\cdot u + g \in V_\delta^{1,s}({\cal D})$ if and only if the trace of $\nabla\cdot u+g$
on $M$ vanishes, i.e.,
\begin{equation} \label{b6}
c_1+d_2+\partial_{x_3}b = -g|_M\, .
\end{equation}
Obviously, the trace of $N^\pm(u,p)$ on $M$ can be written as a linear form $M^\pm(c,d,\partial_{x_3}b,q)$.
Thus, from $N^\pm(u,p)=\phi^\pm$ on $\Gamma^\pm$ it follows that
\begin{equation} \label{b7}
M^\pm(c,d,\partial_{x_3}b,q)=\phi^\pm|_M\, .
\end{equation}

\begin{Le} \label{bl3}
Suppose that there doesn't exist a pair $(u,p)\not=(0,0)$ of a linear vector function
$u=cx_1+dx_2$ and a constant $p$ satisfying
\begin{equation} \label{1bl3}
-\nabla\cdot u = 0\ \mbox{ in }{\cal D},\quad S^\pm u =0,\ \ N^\pm(u,p)=0\ \mbox{ on }\Gamma^\pm.
\end{equation}
Then the linear system {\em (\ref{b5b})--(\ref{b7})} has a unique solution $(c,d,q)$ for arbitrary
$h^\pm,\phi^\pm,g$, and $b$.
\end{Le}

{\it Proof}:
Inserting $u=cx_1+dx_2$ and $p=q=const.$ into (\ref{1bl3}), we obtain
\begin{equation} \label{2bl3}
c_1 +d_2=0,\quad S^\pm\big( c\cos\mbox{$\frac\theta 2$}\pm d\sin\mbox{$\frac\theta 2$}\big)=0,\quad
  \mbox{and}\quad M^\pm(c,d,0,q)=0.
\end{equation}
By the assumption of the lemma, the homogeneous system (\ref{2bl3}) of 7 linear equations with
7 unknowns has only the trivial solution $c=d=0$, $q=0$.
Consequently, the inhomogeneous system (\ref{b5b})--(\ref{b7}) is uniquely solvable. \rule{1ex}{1ex}\\

The last lemma together with Lemmas \ref{al1} and \ref{al2} allows us to obtain the following result.

\begin{Le} \label{bl4}
Let $h^\pm\in W_\delta^{2-1/s,s}(\Gamma^\pm)^{3-d^\pm}$, $\phi^\pm \in W_{\delta}^{1-1/s,s}
(\Gamma^\pm)^{d^\pm}$ and $g \in W_\delta^{1,s}({\cal D})$, $-2/s<\delta<2-2/s$, be given functions
vanishing for $|x'|>1$, and let $h^+$ and $h^-$ satisfy the compatibility condition {\em (\ref{b4})}
on $M$. If $\delta\le 1-2/s$ we assume additionally that the assumption of Lemma {\em \ref{bl3}}
is satisfied. Then there exist a vector function $u \in W_\delta^{2,s}({\cal D})^3$ and a function $p\in
W_\delta^{1,s}({\cal D})$ vanishing for $|x'|>1$ and satisfying {\em (\ref{b3})} and the estimate
\[
\| u \|_{W_\delta^{2,s}({\cal D})^3} + \| p\|_{W_\delta^{1,s}({\cal D})} \le c\, \Big(
  \sum_\pm\| h^\pm\|_{W_\delta^{2-1/s,s}(\Gamma^\pm)^{3-d^\pm}}+ \sum_\pm
  \|\phi^\pm \|_{W_{\delta}^{l-1-1/s,s}(\Gamma^\pm)^{d^\pm}} + \| g \|_{W_\delta^{1,s}({\cal D})}\Big).
\]
\end{Le}

{\it Proof}:
For $\delta>1-2/s$ the assertion of the lemma follows immediately from Lemmas \ref{bl1}
and \ref{bl2}. Let $\delta<1-2/s$. Then there exist $b\in W^{2-\delta-2/s,s}(M)^3$, $c,d\in
W^{1-\delta-2/s,s}(M)^3$ and $q\in W^{1-\delta-2/s,s}(M)$ satisfying (\ref{b5a})--(\ref{b7}). We put
\[
v = Eb + x_1\, Ec + x_2\, Ed,\quad p=Eq,
\]
where $E$ is the extension operator (\ref{a1}). Then, by Lemma \ref{al2},
\[
S^\pm v|_M = h^\pm|_{M}\, ,\quad (\partial_r S^\pm w)|_M= (\partial_r h^\pm)|_M\, ,\quad
  -(\nabla\cdot w)|_M = g|_M\, .
\]
and
\[
N^{\pm}(v,p)|_M = M^\pm(c,d,\partial_{x_3}b,q) = \phi^\pm|_M\, .
\]
Consequently, by Lemma \ref{al1}, we have
\[
S^\pm v - h^\pm \in V_\delta^{2-1/s,s}(\Gamma^\pm)^{3-d^\pm},\quad \nabla\cdot v + g\in V_\delta^{1,s}({\cal D}),
  \quad \mbox{and}\quad N^\pm (v,p) -\phi^\pm \in V_\delta^{1-1/s,s}(\Gamma^\pm)^{d^\pm} .
\]
By Lemma \ref{bl1}, there exist a vector function $w \in V_\delta^{2,s}({\cal D})^3$, $w(x)=0$ for $|x'|>1$
such that
\[
S^\pm w = h^\pm - S^\pm v, \quad N^\pm(w,0) = \phi^\pm -N^\pm (v,p)\quad \mbox{on } \Gamma^\pm.
\]
Then the pair $(u,p)=(v+w,p)$ has the desired properties. In the case $\delta=1-2/s$ the lemma can be proved
analogously using the relations between the spaces $V_\delta^{l,s}({\cal D})$ and $W_\delta^{l,s}({\cal D})$
given in \cite{r92}. \rule{1ex}{1ex}

\begin{Rem} \label{br2}
{\em The condition of Lemma \ref{bl3} is satisfied for $d^+ +d^-=3$, $\sin 2\theta\not=0$ and
for $d^+ + d^- \in \{1,5\}$, $\cos\theta\, \cos 2\theta\not=0$.
If $d^+ +d^-$ is an even number, then the condition of Lemma \ref{bl3} fails for
all $\theta$. If $d^+$ and $d^-$ are both even, then obviously $(u,p)=(0,1)$ satisfies (\ref{1bl3}), while
in the case of odd $d^+$ and $d^-$, the vector $(u,p)=(x_1,-x_2,0,0)$ satisfies (\ref{1bl3}). In these cases
the assertion of Lemma \ref{bl4} holds only under additional compatibility conditions on the
functions $h^\pm$, $\phi^\pm$ and $g$.}
\end{Rem}

We give here the corresponding result for the Dirichlet boundary condition.

\begin{Le} \label{bl5}
Let $h^\pm\in W_\delta^{2-1/s,s}(\Gamma^\pm)^3$ and $g \in W_\delta^{1,s}({\cal D})$, $-2/s<\delta<2-2/s$,
be given functions vanishing for $|x'|>1$ such that $h^+|_M=h^-|_M$. If $\delta<1-2/s$, we assume additionally that
$\theta\not=\pi$, $\theta\not=2\pi$, and
\begin{equation} \label{1bl5}
n^-\cdot \partial_r h^+|_M + n^+\cdot\partial_r h^-|_M = (g|_M +\partial_{x_3}h_3^+|_M)\, \sin\theta,
\end{equation}
while for $\delta=1-2/s$ the "generalized trace condition"
\[
\int_0^\infty \int_{\Bbb R} r^{-1}\, \Big| n^-\cdot \partial_r h^+(r,x_3) + n^+\cdot\partial_r h^-(r,x_3) -
  \big( \stackrel{\circ}{g}\!(r,x_3) - \partial_{x_3}h_3^+(r,x_3)\big)\, \sin\theta\Big|^s\, dx_3\, dr <\infty
\]
is assumed to be valid. Here
\[
\stackrel{\circ}{g}\!(r,x_3) = \frac 1\theta \int_{-\theta/2}^{\theta/2} g(r\cos\varphi,r\sin\varphi,x_3)\,
  d\varphi
\]
denotes the average of $g$ with respect to the variable $\varphi$. Then there exists a vector function
$u\in W_\delta^{2,s}({\cal D})^3$ vanishing for $|x'|>1$ such that $u=h^\pm$ on $\Gamma^\pm$ and
$\nabla\cdot u +g \in V_\delta^{2,s}({\cal D})$.
\end{Le}

{\it Proof}: If $\delta<1-2/s$, then the traces of $h^\pm$, $\partial_r h^\pm$, and $g$ on $M$ exist
and there are vector functions $c,d \in W^{1-\delta-2/s,s}(M)^3$ satisfying
\[
c\cos\mbox{$\frac\alpha 2$}\pm d\sin\mbox{$\frac\alpha 2$} = (\partial_r h^\pm)|_M\quad\mbox{and}\quad
  c_1+d_2 = -\partial_{x_3}h^+|_M - g|_M\, .
\]
Analogously to the proof of Lemma \ref{bl4}, it can be shown that $v=Eh^+|_M+x_1\, Ec+x_2\, Ed$ satisfies
the conditions $v|_{\Gamma^\pm}-h^\pm \in V_\delta^{2,s}(\Gamma^\pm)^3$ and $\nabla\cdot v + g
\in V_\delta^{1,s}({\cal D})$. Applying Lemma \ref{bl1}, we obtain the assertion of the lemma
for $\delta<1-2/s$. Analogously, it can be proved for $\delta=1-2/s$.
\rule{1ex}{1ex}

\begin{Rem} \label{br1}
{\em Analogous results are valid for even $d^+ +d^-\not=0$. Then of course the conditions $h^+|_M=h^-|_M$
and (\ref{1bl5}) have to be replaced by another compatibility conditions on $M$. If $\delta<1-2/s$, then
the traces of $h^\pm,\partial_r h^\pm,\phi^\pm$ and $g$ on $M$ must be such that the
system (\ref{b5a})--(\ref{b7}) with the unknowns $b,c,d,q$ is solvable. For example, in the case
of the Neumann problem ($d^+=d^-=3$), $\theta\not=\pi$, $\theta\not=2\pi$, the boundary
data $\phi^+$ and $\phi^-$ must satisfy the condition
\[
\phi^+\cdot n^- = \phi^-\cdot n^+ \ \mbox{ on }M.
\]
In the case $d^-=0$, $d^+=2$, the data $h^+$, $h^-$, $\phi^+$ and $g$ must satisfy the compatibility conditions
$h^-\cdot n^+ = h^+$ and
\[
\partial_r h^+\, \cos 2\theta - (2n^+\cos\theta\, + n^-)\, \partial_r h^- + 2\sin^2\theta\, (\phi_1^+\cos
\theta/2 + \phi_2^+\sin\theta/2) + \frac 12 (g+\partial_{x_3}h_3^-)\, \sin 2\theta=0
\]
on the edge $M$.}

\end{Rem}

\subsection{Regularity results}

The following two lemmas are proved in \cite[Le.3.1,Le.3.4]{mr-02} for boundary value problems to
elliptic systems of the form
\[
-\sum_{i,j=1}^3 A_{i,j}\, \partial_{x_i}\, \partial_{x_j} u = f
\]
The proof for the Stokes system is essentially the same.

\begin{Le} \label{bl6}
Let $(u,p)\in W_{loc}^{l,s}(\bar{\cal D}\backslash M)^3\times W_{loc}^{l-1,s}(\bar{\cal D}\backslash M)$
be a solution of problem {\em (\ref{b1}), (\ref{b2})}. Furthermore, let
$\zeta$, $\eta$ be infinitely differentiable functions with compact supports on
$\bar{\cal D}$ such that $\eta=1$ in a neighborhood of $\mbox{\em supp}\, \zeta$.

{\em 1)} If $\eta u \in V_{\delta-l}^{0,s}({\cal D})^3$, $\eta p \in V_{\delta-l+1}^{0,s}({\cal D})$,
$\eta f \in V_{\delta}^{l-2,s}({\cal D})^3$, $\eta g \in V_{\delta}^{l-1,s}({\cal D})$,
$\eta h^\pm \in V_{\delta}^{l-1/s,s}(\Gamma^\pm)^{3-d^\pm}$, and $\eta \phi^\pm \in V_{\delta}^{l-1-1/s,s}
(\Gamma^\pm)^{d^\pm}$, $l\ge 2$, then $\zeta u\in V_{\delta}^{l,s}({\cal D})^3$, $\zeta p\in
V_{\delta}^{l-1,s}({\cal D})$ and
\begin{eqnarray} \label{1bl6}
\|\zeta u\|_{V_\delta^{l,s}({\cal D})^3} + \|\zeta p\|_{V_{\delta}^{l-1,s}({\cal D})} & \le & c\, \Big(
  \|\eta u \|_{V_{\delta-l}^{0,s}({\cal D})^3} + \|\eta p \|_{V_{\delta-l+1}^{0,s}({\cal D})}
  + \|\eta f \|_{V_{\delta}^{l-2,s}({\cal D})^3} + \|\eta g \|_{V_{\delta}^{l-1,s}({\cal D})} \nonumber\\
&& \quad + \sum_\pm \| \eta h^\pm \|_{V_{\delta}^{l-1/s,s}(\Gamma^\pm)^{3-d^\pm}} + \sum_\pm
  \| \eta \phi^\pm \|_{V_{\delta}^{l-1-1/s,s}(\Gamma^\pm)^{d^\pm}}\Big).
\end{eqnarray}

{\em 2)} If $\eta u \in W_{\delta-l+k}^{k,s}({\cal D})^3$, $\eta p \in W_{\delta-l+k}^{k-1,s}
({\cal D})$, $\eta f \in W_{\delta}^{l-2,s}({\cal D})^3$, $\eta g \in W_{\delta}^{l-1,s}({\cal D})$,
$\eta h^\pm \in W_{\delta}^{l-1/s,s}(\Gamma^\pm)^{3-d^\pm}$, and $\eta \phi^\pm \in W_{\delta}^{l-1-1/s,s}
(\Gamma^\pm)^{d^\pm}$, $l\ge k+1\ge 2$, $\delta>l-k-2/s$, then $\zeta u\in W_{\delta}^{l,s}({\cal D})^3$,
$\zeta p\in W_{\delta}^{l-1,s}({\cal D})$ and an estimate analogous to {\em (\ref{1bl6})} holds.
\end{Le}

We define the operator $A(\lambda)$ as follows
\[
A(\lambda)\, \big( U(\varphi) , P(\varphi)\big) = \big(
  r^{2-\lambda}(-\Delta u+ \nabla p)\, ,\, -r^{1-\lambda}\nabla\cdot u \, , \,
  r^{-\lambda} S^\pm u|_{\varphi=\pm\theta/2}\, , \, r^{1-\lambda}N^\pm(u,p)|_{\varphi=\pm\theta/2}\big),
\]
where $u=r^\lambda U(\varphi)$, $p=r^{\lambda-1}P(\varphi)$, $\lambda\in {\Bbb C}$, $r,\varphi$
are the polar coordinates of the point $x'=(x_1,x_2)$.
The operator $A(\lambda)$ depends quadratically on the parameter $\lambda$ and realizes
a continuous mapping
\[
W^{2,s}((\mbox{$-\frac\theta 2 ,+\frac\theta 2$}))^3 \times W^{1,s}((\mbox{$-\frac\theta 2 ,+\frac\theta 2$}))
  \to W^{1,s}((\mbox{$-\frac\theta 2 ,+\frac\theta 2$}))^3 \times L^{s}((\mbox{$-\frac\theta 2 ,\frac\theta 2$}))
  \times {\Bbb C}^3\times {\Bbb C}^3
\]
for every $\lambda \in {\Bbb C}$.
In \cite{mr-03a,mr-03} a description of the spectrum of the pencil $A(\lambda)$ is given for
different $d^-$ and $d^+$. For example, in the cases of the Dirichlet problem ($d^+=d^-=0$) and Neumann
problem ($d^+=d^-=3$), the spectrum of $A(\lambda)$ consists of the solutions of the equation
\[
\sin(\lambda\theta)\, \big( \lambda^2\sin^2\theta-\sin^2(\lambda\theta)\big)=0,
\]
$\lambda\not=0$ for $d^+=d^-=0$. In the case $d^-=0,\ d^+=1$, the eigenvalues of $A(\lambda)$
are the nonzero solutions of the equation
\[
\sin(\lambda\theta)\, \big( \lambda\sin(2\theta)+\sin(2\lambda\theta)\big)=0.
\]
If $d^-=0,\ d^+=2$, then the eigenvalues are the nonzero solutions of the equation
\[
\sin(2\lambda\theta)\, \big( \lambda\sin(2\theta)-\sin(2\lambda\theta)\big)=0,
\]
while the nonzero solutions of the equation
\[
\sin(2\lambda\theta)\, \big( \lambda^2\sin^2\theta-\cos^2(\lambda\theta)\big)=0
\]
are eigenvalues of $A(\lambda)$ if $d^-=0$ and $d^+=3$.

\begin{Le} \label{bl7}
Let $\zeta$, $\eta$ be smooth functions on $\bar{\cal D}$ with compact supports such that
$\eta=1$ in a neighborhood of $\mbox{\em supp}\, \zeta$, and let $(u,p)$ be a solution of
problem {\em (\ref{b1}), (\ref{b2})} such that
\[
\eta u \in W_\delta^{l,s}({\cal D})^3, \ \ \eta p \in W_\delta^{l-1,s}({\cal D}), \ \
  \eta \partial_{x_3} u \in W_{\delta'}^{l,s}({\cal D})^3, \ \ \eta \partial_{x_3} p
  \in W_{\delta'}^{l-1,s}({\cal D})
\]
where $l\ge 2$, $-2/s < \delta\le \delta'\le \delta+1$. Furthermore, we assume that
\[
\eta f \in W_{\delta'}^{l-1,s}({\cal D})^3, \ \ \eta g \in W_{\delta'}^{l,s}({\cal D}), \ \
\eta h^\pm \in W_{\delta'}^{l+1-1/s,s}(\Gamma^\pm)^{3-d^\pm}, \ \
\eta \phi^\pm \in W_{\delta'}^{l-1/s,s}(\Gamma^\pm)^{d^\pm}.
\]
If there are no eigenvalues of the pencil $A(\lambda)$ in the strip
$l-\delta-2/s \le \mbox{\em Re}\, \lambda \le l+1-\delta'-2/s$, then
$\zeta u \in W_{\delta'}^{l+1,s}({\cal D})^3$, $\zeta p \in W_{\delta'}^{l,s}({\cal D})$.
\end{Le}

\setcounter{equation}{0}
\setcounter{Th}{0}
\setcounter{Le}{0}
\setcounter{Co}{0}
\section{Solvability of the boundary value problem in a polyhedral cone}

Let ${\cal K}$ be the cone (\ref{cone}) introduced in Section 1.2. For every $j=1,\ldots,n$ let
$d_j$ be one of numbers $0,1,2,3$. We consider the boundary value problem
\begin{eqnarray} \label{d1}
&& -\Delta u + \nabla p = f,\quad -\nabla\cdot u = g\ \ \mbox{in }{\cal K},\\
&& S_j u = h_j\, ,\quad N_j(u,p)= \phi_j \quad \mbox{on }\Gamma_j,\ j=1,\ldots,n. \label{d2}
\end{eqnarray}
Here $S_j$ is defined as
\[
S_j u=u \ \mbox{if }d_j=0,\quad S_j u=u_n=u\cdot n \ \mbox{if } d_j=2,\quad S_j u=u_\tau=u-u_n n \ \mbox{if }d_j=1,
\]
while the operators $N_j$ are defined as
\[
N_j(u,p)=-p+2\varepsilon_{n,n}(u) \ \mbox{if }d_j=1,\ \ N_j(u,p)=\varepsilon_{n,\tau}(u) \
  \mbox{if } d_j=2, \ \ N_j(u,p)=-pn+2\varepsilon_n(u) \ \mbox{if }d_j=3.
\]
In the case $d_j=0$ the condition $N_j(u,p)=\phi_j$ does not appear in (\ref{d2}), whereas the
condition $S_ju=h_j$ does not appear if $d_j=3$.

\subsection{Reduction to homogeneous boundary conditions}

\begin{Le} \label{dl1}
Let $h_j \in V_{\beta,\delta}^{l-1/s,s}(\Gamma_j)^{3-d_j}$, $\phi_j \in
V_{\beta,\delta}^{l-1-1/s,s}(\Gamma_j)^{d_j}$, $l\ge 2$.
Then there exists a vector function $u\in V_{\beta,\delta}^{l,s}({\cal K})^3$
such that $S_j u=h_j$ and $N_j(u,0)=\phi_j$ on $\Gamma_j$ and
\begin{equation} \label{1dl1}
\| u \|_{V_{\beta,\delta}^{l,s}({\cal K})^3} \le c\,
  \sum_{j=1}^n  \Big( \| h_j\|_{V_{\beta,\delta}^{l-1/s,s}(\Gamma_j)^{3-d_j}} +
  \| \phi_j\|_{V_{\beta,\delta}^{l-1-1/s,s}(\Gamma_j)^{d_j}}\Big)
\end{equation}
with a constant $c$ independent of $h_j$ and $\phi_j$.
\end{Le}

{\it Proof:}
Let $\zeta_k$ be smooth functions depending only on $\rho=|x|$ such that
\begin{equation} \label{2dl1}
\mbox{supp}\, \zeta_k \subset (2^{k-1},2^{k+1}),\quad \sum_{k=-\infty}^{+\infty}
  \zeta_k=1,\quad |(\rho\partial_\rho)^j \zeta_k(\rho)|\le c_j
\end{equation}
with constants $c_j$ independent of $k$ and $\rho$. We set $h_{k,j}(x) = \zeta_k(2^k x)\,
h_j(2^k x)$,  $\phi_{k,j}(x) = 2^k \, \zeta_k(2^k x)\, \phi_j(2^k x)$. These functions
vanish for $|x|<\frac 12$ and $|x|>2$. Consequently, by Lemma \ref{bl1}, there exist
vector functions $v_k \in V_{\beta,\delta}^{l,s}({\cal K})^3$ such that
$S_j v_k=h_{k,j}$ and $N_j(v_k,0)=\phi_{k,j}$ on $\Gamma_j$ for $j=1,\ldots,n$,
\begin{equation} \label{3dl1}
\| v_k \|_{V_{\beta,\delta}^{l,s}({\cal K})^3} \le c\, \sum_{j=1}^n \Big(
  \| h_{k,j}\|_{V_{\beta,\delta}^{l-1/s,s}(\Gamma_j)^{3-d_j}} +
  \| \phi_{k,j}\|_{V_{\beta,\delta}^{l-1-1/s,s}(\Gamma_j)^{d_j}}\Big),
\end{equation}
and $v_k(x)=0$ for $|x|<\frac 14$ and $|x|>4$. Hence for the functions
$u_k(x)=v_k(2^{-k}x)$ we obtain $S_j u_k= \zeta_k h_j$ and
$N_j(u_k,0)= \zeta_k \phi_j$ on $\Gamma_j$, $u_k(x)=0$ for $|x|<2^{k-2}$ and
$|x|>2^{k+2}$. Furthermore, $u_k$ satisfies (\ref{3dl1}) with $\zeta_k h_j$ and
$\zeta_k\phi_j$ instead of $h_{k,j}$ and $\phi_{k,j}$. Here the constant $c$ is
independent of $k$, $h_j$ and $\phi_j$.
Consequently, for $u=\sum u_k$ we have $S_ju=h_j$ on and $N_j(u,0)=\phi_j$ on $\Gamma_j$
for $j=1,\ldots,n$. Inequality (\ref{1dl1}) follows from the equivalence of the norms in
$V_{\beta,\delta}^{l,s}({\cal K})$ and $V_{\beta,\delta}^{l-1/s,s}(\Gamma_j)$ with the norms
\begin{equation} \label{zetanorm}
\| u\| = \Big( \sum_{k=-\infty}^{+\infty} \| \zeta_k u\|^2_{V_{\beta,\delta}^{l,s}({\cal K})} \Big)^{1/s}
  \quad\mbox{and}\quad \| h \| = \Big( \sum_{k=-\infty}^{+\infty} \| \zeta_k
  h \|^2_{V_{\beta,\delta}^{l-1/s,s}(\Gamma_j)} \Big)^{1/s},
\end{equation}
respectively (cf. \cite[Sect.6.1]{kmr1}). \rule{1ex}{1ex}\\

An analogous result in $W_{\beta,\delta}^{2,s}({\cal K})$ is only valid under additional
compatibility conditions on the boundary data. Denote by $\Gamma_{k_+}$ and $\Gamma_{k_-}$ the
sides of the cone ${\cal K}$ adjacent to the edge $M_k$ and by $\theta_k$ the inner angle
at $M_k$. If $u\in W_{\beta,\delta}^{2,s}({\cal K})$ and $\delta_k<2-2/s$, then the trace of $u$
on $M_k$ exists and from the equations $S_j u=h_j$ on $\Gamma_j$ it follows that the pair
$\big( h_{k_+}|_{M_j}, h_{k_-}|_{M_j}\big)$ belongs to the range of the matrix operator
$(S_{k_+},S_{k_-})$. This condition can be also written in the form
\begin{equation} \label{compat1}
A_k h_{k_+}|_{M_k} = B_k h_{k_-}|_{M_k} \, ,
\end{equation}
where $A_k,B_k$ are certain constant matrices (see Section 2.1).

Using Lemma \ref{bl4} (see also Remark \ref{br2}), one can prove the following result analogously
to Lemma \ref{dl1}.

\begin{Le} \label{dl2}
Let $h_j\in W_{\beta,\delta}^{2-1/s,s}(\Gamma_j)^{3-d_j}$, $\phi_j \in W_{\beta,\delta}^{l-1-1/s,s}
(\Gamma_j)^{d_j}$ and $g \in W_{\beta,\delta}^{1,s}({\cal K})$, $-2/s<\delta_k<2-2/s$ for $k=1,\ldots,n$,
be given functions such that the compatibility condition {\em (\ref{compat1})} is satisfied for
$k=1,\ldots,n$. In the case $\delta_k\le 1-2/s$ we assume additionally that $d_{k_+} + d_{k_-}$ is odd and
\[
\sin2\theta_k \not=0 \ \mbox{ if }d_{k_+} + d_{k_-}=3, \quad
\cos\theta_k\, \cos 2\theta_k\not=0 \ \mbox{ if } d_{k_+} + d_{k_-}\in\{1,5\}.
\]
Then there exist a vector function $u \in W_{\beta,\delta}^{2,s}({\cal K})^3$ and a function $p\in
W_{\beta,\delta}^{1,s}({\cal K})$ satisfying
\begin{equation} \label{1dl2}
S_j u = h_j,\quad N_j(u,p)=\phi_j \mbox{ on }\Gamma_j,\ j=1,\ldots,n,\quad  \nabla\cdot u + g
  \in V_{\beta,\delta}^{1,s}({\cal K})
\end{equation}
and the estimate
\[
\| u \|_{W_{\beta,\delta}^{2,s}({\cal K})^3} + \| p\|_{W_{\beta,\delta}^{1,s}({\cal K})}
  \le c\, \Big( \sum_\pm\| h_j\|_{W_{\beta,\delta}^{2-1/s,s}(\Gamma^\pm)^{3-d_j}}+ \sum_\pm
  \|\phi_j \|_{W_{\beta,\delta}^{l-1-1/s,s}(\Gamma_j)^{d_j}} + \| g \|_{W_{\beta,\delta}^{1,s}({\cal K})}\Big).
\]
\end{Le}

If $\delta_k\le 1-2/s$ and $d_{k_+} + d_{k_-}$ is even for at least one $k$, then the assertion
of Lemma \ref{dl2} holds only under an additional compatibility condition on the edge $M_k$
(cf. Lemma \ref{bl5}, Remark \ref{br1}).
We give here the corresponding result for the Dirichlet problem. An analogous result is valid in
the general case.

\begin{Le} \label{dl3}
Let $h_j\in W_{\beta,\delta}^{2-1/s,s}(\Gamma_j)^3$ and $g \in W_{\beta,\delta}^{1,s}({\cal K})$,
where $-2/s<\delta_k<2-2/s$ for $k=1,\ldots,n$, be given functions such that
\[
h_{k_+}|_{M_k} = h_{k_-}|_{M_k} \quad\mbox{for }k=1,\ldots,n.
\]
In the case $\delta_k< 1-2/s$ we assume additionally that $\theta_k\not=\pi$, $\theta_k\not=2\pi$ and
\begin{equation} \label{1dl3}
n_{k_-}\cdot (\partial_r h_{k_+})|_{M_k} + n_{k_+}\cdot (\partial_r h_{k_-})|_{M_k} =
  \big( g|_{M_k} + \partial_t(h_{k_+}\cdot e_k)|_{M_k}\big)\, \sin\theta_k,
\end{equation}
where $e_k$ is the unit vector on $M_k$, $r=\mbox{\em dist}(x,M_k)$, and $t$ denotes the coordinate on $M_k$.
For $\delta_k=1-2/s$ instead of {\em (\ref{1dl3})} the generalized trace condition
\[
\int_0^\infty \int_0^{\varepsilon t} t^{s(\beta-1)+2}\, r^{-1}\, \Big| n_{k_-}\cdot \partial_r h_{k_+}(r,t)
  + n_{k_+}\cdot\partial_r h_{k_-}(r,t) - \big( \stackrel{\circ}{g}\!(r,t)
  - \partial_t (h_{k_+}(r,t)\cdot e_k)\big)\, \sin\theta_k\Big|^s\, dr\, dt
  <\infty
\]
is assumed to be valid. Here $\varepsilon$ is a small positive number, the functions $h_{k_\pm}$
are considered near $M_k$ as functions in  the variables $r,t$, and $\stackrel{\circ}{g}\!\!(r,t)$ denotes
the average of $g$ with respect to the angle $\varphi$ in the plane perpendicular to $M_k$ (cf. Lemma {\em \ref{bl5}}).
Then there exists a vector function $u \in W_{\beta,\delta}^{2,s}({\cal K})^3$ such that
$u=h_j$ on $\Gamma_j$ for $j=1,\ldots,n$ and $\nabla\cdot u+g\in V_{\beta,\delta}^{1,s}({\cal K})$.
\end{Le}

\subsection{Operator pencils generated by the boundary value problem}

We introduce the following operator pencils ${\mathfrak A}$ and $A_j$. \\

1) Let $\Gamma_{k_\pm}$ be the sides of ${\cal K}$ adjacent to the edge $M_k$, and let $\theta_k$ be the
angle at the edge $M_k$. We consider the Stokes system in the dihedron ${\cal D}_k$ bounded by the
half-planes $\Gamma_{k_\pm}^\circ \supset \Gamma_{k_\pm}$ with the boundary conditions
\[
S_{k_\pm} u = h^\pm, \quad N_{k_\pm} (u,p) = \phi^\pm \ \mbox{ on } \Gamma_{k_\pm}^\circ.
\]
By $A_k(\lambda)$ we denote the operator pencil introduced before Lemma \ref{bl7} for this problem.
Furthermore, let $\lambda_1^{(k)}$ denote the eigenvalue with smallest
positive real part of this pencil, while $\lambda_2^{(k)}$ is the eigenvalue with smallest real part greater
than 1. Finally, we define
\begin{equation} \label{defmuk}
\mu_k = \left\{ \begin{array}{ll} \mbox{Re}\, \lambda_1^{(k)} & \mbox{if }d_{k_+} +d_{k_-} \mbox{ is odd }\
  \mbox{ or }  \ d_{k_+} + d_{k_-} \mbox{ is even and } \alpha_k \ge \pi/m_k, \\
  \mbox{Re}\, \lambda_2^{(k)} & \mbox{if } d_{k_+} + d_{k_-} \mbox{ is even and } \alpha_k < \pi/m_k, \end{array}\right.
\end{equation}
where $m_k=1$ if $d_{k_+} = d_{k_-}$, $m_k=2$ if $d_{k_+} \not= d_{k_-}$. \\

2) Let $\rho=|x|$, $\omega=x/|x|$, $V_\Omega = \{ u\in W^1(\Omega)^3:\, S_j u=0$ on $\gamma_j$ for
$j=1,\ldots,n\}$, and
\[
a\Big( \Big( \begin{array}{c} u \\ p\end{array}\Big),\Big( \begin{array}{c} v \\ q\end{array}\Big);\lambda\Big)
  = \frac{1}{\log 2}\, \int\limits_{\substack{{\cal K}\\ 1<|x|<2}} \Big( 2\sum_{i,j=1}^3 \varepsilon_{i,j}(U)
  \cdot \varepsilon_{i,j}(V) - P\nabla\cdot V - (\nabla\cdot U)\, Q\Big)\, dx,
\]
where $U=\rho^\lambda u(\omega)$, $V=\rho^{-1-\lambda} v(\omega)$, $P=\rho^{\lambda-1} p(\omega)$,
$Q=\rho^{-2-\lambda}q(\omega)$, $u,v \in V_\Omega$, $p,q\in L_2(\Omega)$, and $\lambda \in {\Bbb C}$.
The bilinear form $a(\cdot,\cdot;\lambda)$ generates the linear and continuous operator
\[
{\mathfrak A}(\lambda):\, V_\Omega \times L_2(\Omega) \to V_\Omega^*\times L_2(\Omega)
\]
by
\[
\int_\Omega {\mathfrak A}(\lambda)\Big( \begin{array}{c} u \\ p\end{array}\Big) \cdot
  \Big( \begin{array}{c} v \\ q\end{array}\Big)\, d\omega
  = a\Big( \Big( \begin{array}{c} u \\ p\end{array}\Big),\Big( \begin{array}{c} v \\ q\end{array}\Big)
  ;\lambda\Big), \quad   u,v\in V_\Omega ,\ p,q \in L_2(\Omega).
\]

\subsection{Regularity results for the problem in the cone}

The following results are based on Lemmas \ref{bl6}, \ref{bl7}

\begin{Le} \label{dl4}
Let $(u,p)\in W_{loc}^{l,s}(\bar{\cal K}\backslash {\cal S})^3\times W_{loc}^{l-1,s}(\bar{\cal K}
\backslash {\cal S})$ be a solution of problem {\em (\ref{d1}), (\ref{d2})}.

{\em 1)} If $u \in V_{\beta-l+1,\delta-l+1}^{1,s}({\cal K})^3$, $p \in
V_{\beta-l+1,\delta-l+1}^{0,s}({\cal K})$, $f \in V_{\beta,\delta}^{l-2,s}({\cal K})^3$,
$g\in V_{\beta,\delta}^{l-1,s}({\cal K})$, $h_j\in V_{\beta,\delta}^{l-1/s,s}(\Gamma_j)^{3-d_j}$,
and $\phi_j \in V_{\delta}^{l-1-1/s,s}(\Gamma_j)^{d_j}$, $l\ge 2$, then $u\in
V_{\beta,\delta}^{l,s}({\cal K})^3$, $p\in V_{\beta,\delta}^{l-1,s}({\cal K})$ and
\begin{eqnarray} \label{1dl4}
\| u\|_{V_{\beta,\delta}^{l,s}({\cal K})^3} + \| p\|_{V_{\beta,\delta}^{l-1,s}({\cal K})}
  & \le & c\, \Big( \| u \|_{V_{\beta-l+1,\delta-l+1}^{1,s}({\cal K})^3} + \| p
  \|_{V_{\beta-l+1,\delta-l+1}^{0,s}({\cal K})} + \| f \|_{V_{\beta,\delta}^{l-2,s}({\cal K})^3}
  + \| g \|_{V_{\beta,\delta}^{l-1,s}({\cal K})} \nonumber\\
&& \quad + \sum_{j=1}^n \| h_j \|_{V_{\beta,\delta}^{l-1/s,s}(\Gamma_j)^{3-d_j}} + \sum_{j=1}^n
  \| \phi_j \|_{V_{\beta,\delta}^{l-1-1/s,s}(\Gamma_j)^{d_j}}\Big).
\end{eqnarray}

{\em 2)} If $u \in W_{\beta-l+k,\delta-l+k}^{k,s}({\cal K})^3$, $p \in
W_{\beta-l+k+1,\delta-l+k}^{k-1,s}({\cal K})$, $f \in W_{\beta,\delta}^{l-2,s}({\cal K})^3$,
$g \in W_{\beta,\delta}^{l-1,s}({\cal K})$, $h_j \in W_{\beta,\delta}^{l-1/s,s}(\Gamma_j)^{3-d_j}$,
and $\phi_j \in W_{\beta,\delta}^{l-1-1/s,s}(\Gamma_j)^{d_j}$, $l\ge k+1\ge 2$, $\delta_j>l-k-2/s$,
then $u\in W_{\beta,\delta}^{l,s}({\cal K})^3$, $p\in W_{\beta,\delta}^{l-1,s}({\cal K})$
and an estimate analogous to {\em (\ref{1dl4})} holds.
\end{Le}

{\it Proof}:
1) Due to Lemma \ref{dl1}, we may assume, without loss of generality, that $h_j=0$ and $\phi_j=0$ for
$j=1,\ldots,n$. Let $\zeta_k$ be the same functions as in the proof of Lemma \ref{dl1}. Furthermore,
let $\eta_k=\zeta_{k-1}+\zeta_k+\zeta_{k+1}$, $\tilde{\zeta}_k(x)=\zeta_k(2^k x)$, $\tilde{\eta}_k(x)
=\eta_k(2^k x)$, $\tilde{u}(x)=u(2^k x)$, and $\tilde{p}(x)=2^k p(2^k x)$. By (\ref{2dl1}), the support
of $\tilde{\zeta}_k$ is contained in the set $\{x:\, 1/2<|x|<2\}$ and the derivatives of $\tilde{\zeta}_k$
are bounded by constants independent of $k$. Obviously,
\begin{eqnarray*}
&& -\Delta \tilde{u} + \nabla\tilde{p} = \tilde{f},\quad -\nabla \tilde{u} =\tilde{g}\ \mbox{ in }{\cal K}\\
&& S_j\tilde{u}=0,\ N_j(\tilde{u},\tilde{p})=0\ \mbox{ on }\Gamma_j,\ j=1,\ldots,n,
\end{eqnarray*}
where $\tilde{f}(x)=2^{2k}f(2^k x)$ and $\tilde{g}(x)= 2^k g(2^k x)$.
Consequently, from Lemma \ref{bl6} it follows that $\tilde{\zeta}_k\tilde{u}\in V_{\beta,\delta}^{l,s}({\cal K})^3$,
$\tilde{\zeta}_k\tilde{p}  \in V_{\beta,\delta}^{l-1,s}({\cal K})$, and
\begin{eqnarray*}
&& \| \tilde{\zeta}_k\tilde{u}\|_{V_{\beta,\delta}^{l,s}({\cal K})^3} + \| \tilde{\zeta}_k\tilde{p}
  \|_{V_{\beta,\delta}^{l-1,s}({\cal K})}  \\
&& \le  c\, \Big( \| \tilde{\eta}_k\tilde{u} \|_{V_{\beta-l,\delta-l}^{0,s}({\cal K})^3}
  + \| \tilde{\eta}_k\tilde{p} \|_{V_{\beta-l+1,\delta-l+1}^{0,s}({\cal K})}
  + \| \tilde{\eta}_k\tilde{f} \|_{V_{\beta,\delta}^{l-2,s}({\cal K})^3}
  + \| \tilde{\eta}_k\tilde{g} \|_{V_{\beta,\delta}^{l-1,s}({\cal K})} \Big).
\end{eqnarray*}
where $c$ is independent of $u$, $p$, and $k$. Using the coordinate change $2^k x=y$, we obtain
the same estimate with $\zeta_k,\eta_k,u,p,f,g$ instead of $\tilde{\zeta}_k,\tilde{\eta}_k,\tilde{u},
\tilde{p},\tilde{f}$ and $\tilde{g}$, respectively. Since the norm in $V_{\beta,\delta}^{l,s}({\cal K})$
is equivalent to the first norm in (\ref{equivalent}), this implies (\ref{1dl4}) for $\zeta=\eta=1$.

2) The second assertion can be proved analogously. \rule{1ex}{1ex}

\begin{Co} \label{dc1}
Let $(u,p) \in W_{loc}^{l,s}(\bar{\cal K}\backslash {\cal S})^3\times W_{loc}^{l-1,s}(\bar{\cal K}
\backslash {\cal S})$ be a solution of problem {\em (\ref{d1}), (\ref{d2})}, and let $\zeta$, $\eta$
be infinitely differentiable functions on $\bar{\cal K}$ with compact supports such that
$\eta=1$ in a neighborhood of $\mbox{\em supp}\, \zeta$. If $\eta u \in
V_{\beta-l+1,\delta-l+1}^{1,s}({\cal K})^3$, $\eta p \in V_{\beta-l+1,\delta-l+1}^{0,s}({\cal K})$,
$\eta f \in V_{\beta,\delta}^{l-2,s}({\cal K})^3$, $\eta g\in V_{\beta,\delta}^{l-1,s}({\cal K})$,
$\eta h_j\in V_{\beta,\delta}^{l-1/s,s}(\Gamma_j)^{3-d_j}$, and $\eta \phi_j \in V_{\delta}^{l-1-1/s,s}
(\Gamma_j)^{d_j}$, $l\ge 2$, then $\zeta u\in V_{\beta,\delta}^{l,s}({\cal K})^3$, $\zeta p
\in V_{\beta,\delta}^{l-1,s}({\cal K})$ and an estimate analogous to {\em (\ref{1dl4})} holds.
\end{Co}

{\it Proof}:
We apply Lemma \ref{dl4} to the vector function $(\zeta u,\zeta p)$. Obviously,
\[
-\Delta(\zeta u)+\nabla(\zeta p) =\zeta f -2\sum_{j=1}^3 (\partial_{x_j}\zeta) \,\partial_{x_j}u
  - u\, \Delta\zeta + p\,\nabla\zeta \quad\mbox{and}\quad -\nabla\cdot(\zeta u)=\zeta g - g\cdot \nabla\zeta,
\]
Moreover, $(\zeta u,\zeta p)$ satisfies the boundary conditions (\ref{d2}) with the data
$H_j=\zeta h_j$ and $\Phi_j = \zeta \phi_j+ N'_j(\nabla \zeta)\, u$, where $N'_j(\nabla \zeta)$ are certain
matrices depending on $\nabla\zeta$. Thus, in the case $l=2$ the assertion of the corolllary follows
immediately from Lemma \ref{dl4}. Using induction in $l$, we obtain the assertion for $l\ge 3$.
\rule{1ex}{1ex}

\begin{Le} \label{dl5}
Let $(u,p) \in W_{\beta,\delta}^{l,s}({\cal K})^3\times W_{\beta,\delta}^{l-1,s}({\cal K})$ be a solution of
problem {\em (\ref{d1}), (\ref{d2})} such that
\[
\partial_{\rho} u \in W_{\beta+1,\delta'}^{l,s}({\cal K})^3, \ \
  \partial_{\rho} p\in W_{\beta+1,\delta'}^{l-1,s}({\cal K}),
\]
where $l\ge 2$, $-2/s < \delta_k\le \delta'_k\le \delta_k+1$ for $k=1,\ldots,n$. Furthermore, we assume that
\[
f \in W_{\beta+1,\delta'}^{l-1,s}({\cal K})^3, \ \ g \in W_{\beta+1,\delta'}^{l,s}({\cal K}), \ \
  h_j \in W_{\beta+1,\delta'}^{l+1-1/s,s}(\Gamma^\pm)^{3-d_j}, \ \
  \phi_j \in W_{\beta+1,\delta'}^{l-1/s,s}(\Gamma^\pm)^{d_j}, \ \ j=1,\ldots,n.
\]
If there are no eigenvalues of the pencils $A_k(\lambda)$ in the strip
$l-\delta_k-2/s \le \mbox{\em Re}\, \lambda \le l+1-\delta'_k-2/s$, $k=1,\ldots,n$, then
$u \in W_{\beta+1,\delta'}^{l+1,s}({\cal K})^3$, $\zeta p \in W_{\beta+1,\delta'}^{l,s}({\cal K})$.
\end{Le}

\subsection{Representation of the solution by Green's matrix}

Let $f\in W_{\beta,\delta}^{0,s}({\cal K})^3$, $g\in W_{\beta,\delta}^{1,s}({\cal K})$,
$h_j \in W_{\beta,\delta}^{2-1/s,s}(\Gamma_j)^{3-d_j}$, $\phi_j \in W_{\beta,\delta}^{1-1/s,s}
(\Gamma_j)^{d_j}$. Our goal is to show that problem (\ref{d1}), (\ref{d2}) has a unique solution in
$W_{\beta,\delta}^{2,s}({\cal K})^3\times W_{\beta,\delta}^{1,s}({\cal K})$
if the line $\mbox{Re}\, \lambda = 2-\beta-3s$ does not contain eigenvalues of the
pencil ${\mathfrak A}(\lambda)$ and $\max(2-\mu_k,0) < \delta_k +2/s <2$ for $k=1,\ldots,n$.

Let $\kappa$ be a fixed real number such that the closed strip between the lines
$\mbox{Re}\, \lambda = -\kappa-1/2$ and $\mbox{Re}\, \lambda=2-\beta-3/s$ is free of eigenvalues of the
pencil ${\mathfrak A}$. Then, according to \cite[Th.4.5]{mr-03},
there exists a unique solution $G(x,\xi)=\big( G_{i,j}(x,\xi)\big)_{i,j=1}^4$ of the problem
\begin{eqnarray} \label{e1}
&& -\Delta_x \vec{G}_j(x,\xi)+ \nabla_x G_{4,j}(x,\xi)= \delta(x-\xi)\,
  (\delta_{1,j},\delta_{2,j},\delta_{3,j})^t \quad \mbox{for }x,\xi\in {\cal K}, \\ \label{e2}
&& -\nabla_x\cdot \vec{G}_j(x,\xi) = \delta_{4,j}\, \delta(x-\xi) \quad \mbox{for }x,\xi\in {\cal K},\\
&&  S_k \vec{G}_j(x,\xi)=0,\quad N_k(\partial_x)\, \big(\vec{G}_j(x,\xi),G_{4,j}(x,\xi)\big)=0
  \quad \mbox{for }x\in \Gamma_k,\ \xi\in {\cal K},\ \ k=1,\ldots,n,\label{e3}
\end{eqnarray}
(here $\vec{G}_j$ denotes the vector with the components $G_{1,j},G_{2,j},G_{3,j}$)
such that the function $x\to \zeta(|x-\xi|/r(\xi))\, G_{i,j}(x,\xi)$ belongs to $W_{\kappa,0}^1({\cal K})$
for $i=1,2,3$ and to $W_{\kappa,0}^0({\cal K})$ for $i=4$, where $\zeta$ is an arbitrary smooth function
on $(0,\infty)$ equal to one in $(1,\infty)$ and to zero in $(0,\frac 12)$.
We denote by $\Lambda_- < \mbox{Re}\, \lambda < \Lambda_+$ the widest strip in the complex plane containing
the line $\mbox{Re}\, \lambda=-\kappa-1/2$ which is free of eigenvalues of the pencil ${\mathfrak A}(\lambda)$.

If $h_j=0$ and $\phi_j=0$ for $j=1,\ldots,n$, then the solution of problem (\ref{d1}), (\ref{d2}) has the form
\begin{eqnarray} \label{e4}
u_i(x) & = & \sum_{j=1}^3 \int_{\cal K} \big( f_j(\xi)+\partial_{\xi_j}g(\xi)\big)\, G_{i,j}(x,\xi)\, d\xi
  + \int_{\cal K} g(\xi)\, G_{i,4}(x,\xi)\, d\xi,\quad i=1,2,3, \\ \label{e5}
p(x) & = & -g(x) + \sum_{j=1}^3 \int_{\cal K} \big( f_j(\xi)+\partial_{\xi_j}g(\xi)\big)\, G_{4,j}(x,\xi)\, d\xi
  + \int_{\cal K} g(\xi)\, G_{4,4}(x,\xi)\, d\xi
\end{eqnarray}
(see \cite[Th.4.5]{mr-03}). In the following, we will show that (\ref{e4}) and (\ref{e5}) define a continuous mapping
\[
W_{\beta,\delta}^{0,s}({\cal K})^3\times V_{\beta,\delta}^{1,s}({\cal K}) \ni (f,g) \to
  (u,p) \in W_{\beta,\delta}^{2,s}({\cal K})^3\times W_{\beta,\delta}^{1,s}({\cal K})
\]
if
\begin{equation} \label{e6}
\Lambda_- < 2-\beta-3/s < \Lambda_+ \quad \mbox{and} \quad \max(2-\mu_k,0) < \delta_k +2/s <2 \
  \mbox{ for }k=1,\ldots,n.
\end{equation}

\subsection{Estimates of Green's matrix}

The following estimates of Green's matrix are proved in \cite{mr-03a,mr-03}.

1) For $|x|>2|\xi|$ there is the estimate
\begin{eqnarray*}
\big| \partial_x^\alpha\partial_\xi^\gamma G_{i,j}(x,\xi)\big| & \le & c\,
  |x|^{\Lambda_- -\delta_{i,4} -|\alpha|+\varepsilon}\ |\xi|^{-\Lambda_- -1 -\delta_{j,4}
  -|\gamma|-\varepsilon} \\
&& \times \prod_{k=1}^n \Big( \frac{r_k(x)}{|x|}\Big)^{\min(0,\mu_k-|\alpha|-\delta_{i,4}-\varepsilon)}
  \prod_{k=1}^n \Big( \frac{r_k(\xi)}{|\xi|}\Big)^{\min(0,\mu_k-|\gamma|-\delta_{j,4}-\varepsilon)}\, ,
\end{eqnarray*}
where $\varepsilon$ is an arbitrarily small positive number. Analogously for $|\xi|> 2|x|$, there
is the inequality
\begin{eqnarray*}
\big| \partial_x^\alpha\partial_\xi^\gamma G_{i,j}(x,\xi)\big| & \le &
c\, |x|^{\Lambda_+ -\delta_{i,4} -|\alpha|-\varepsilon}\ |\xi|^{-\Lambda_+ -1 -\delta_{j,4}
  -|\gamma|+\varepsilon} \\
&& \times \prod_{k=1}^n \Big( \frac{r_k(x)}{|x|}\Big)^{\min(0,\mu_k-|\alpha|-\delta_{i,4}-\varepsilon)}
  \prod_{k=1}^n \Big( \frac{r_k(\xi)}{|\xi|}\Big)^{\min(0,\mu_k-|\gamma|-\delta_{j,4}-\varepsilon)}.
\end{eqnarray*}

2) For $|x|/2<|\xi|<2|x|$, $|x-\xi|>\min(r(x),r(\xi))$, we have
\[
\big| \partial_{x}^\alpha \partial_{\xi}^\gamma G_{i,j}(x,\xi) \big| \le c\, |x-\xi|^{-T-|\alpha|-|\gamma|}\,
  \Big( \frac{r(x)}{|x-\xi|}\Big)^{\min(0,\mu_x-|\alpha|-\delta_{i,4}-\varepsilon)} \
  \Big( \frac{r(\xi)}{|x-\xi|}\Big)^{\min(0,\mu_\xi-|\gamma|-\delta_{j,4}-\varepsilon)},
\]
where $T=1+\delta_{i,4}+\delta_{j,4}$.

3) Let $|x|/2<|\xi|<2|x|$ and $|x-\xi|<\min(r(x),r(\xi))$. Then
\[
\big| \partial_x^\alpha\partial_\xi^\gamma G_{i,j}(x,\xi)\big|\le c\, |x-\xi|^{-T-|\alpha|-|\gamma|}.
\]
Moreover for $i,j=1,\ldots,4$, there are the representations
\[
G_{4,j}(x,\xi) = - \nabla_x\cdot \vec{P}_j(x,\xi) + Q_j(x,\xi),\quad
G_{i,4}(x,\xi) = - \nabla_\xi\cdot \vec{\cal P}_i(x,\xi) + {\cal Q}_i(x,\xi),
\]
where $\vec{P}_j(x,\xi)\cdot n$ for $x\in \Gamma_k$, $k=1,\ldots,n$, $\xi\in {\cal D}$,
$\vec{\cal P}_i(x,\xi)\cdot n$ for $\xi\in \Gamma_k$, $x\in {\cal D}$, and
\begin{eqnarray*}
&& |\partial_x^\alpha\partial_\xi^\gamma \vec{P}_j(x,\xi)| \le c_{\alpha,\gamma} \,
  |x-\xi|^{-1-\delta_{j,4}-|\alpha|-|\gamma|}, \quad
  |\partial_x^\alpha\partial_\xi^\gamma Q_j(x,\xi)| \le c_{\alpha,\gamma} \,
  r(\xi)^{-2-\delta_{j,4}-|\alpha|-|\gamma|}, \\
&& |\partial_x^\alpha\partial_\xi^\gamma \vec{\cal P}_i(x,\xi)| \le c_{\alpha,\gamma} \,
  |x-\xi|^{-1-\delta_{i,4}-|\alpha|-|\gamma|}, \quad
  |\partial_x^\alpha\partial_\xi^\gamma {\cal Q}_i(x,\xi)| \le c_{\alpha,\gamma} \,
  r(\xi)^{-2-\delta_{i,4}-|\alpha|-|\gamma|}
\end{eqnarray*}
for $|x|/2<|\xi|<2|x|$, $|x-\xi|<\min(r(x),r(\xi))$.

\subsection{Auxiliary inequalities}

In this subsection we prove estimates for an integral operator with kernel $K(x,\xi)$ which
satisfies the same point estimates as the elements $G_{i,j}(x,\xi)$ of Green's matrix with
$\sigma=\delta_{i,4}$ and $\tau=\delta_{j,4}$.

\begin{Le} \label{el1}
 Let $\zeta_k$ be the same function as in the proof of Lemma {\em \ref{bl1}} and let
 \[
 v(x) = \zeta_m(x) \, \int_{\cal K} \zeta_l(\xi)\, f(\xi)\, K(x,\xi)\, d\xi.
 \]
Suppose that $m\ge l+3$, $f\in W_{\beta-\tau,\delta-\tau}^{0,s}({\cal K})$, and
\begin{equation} \label{1el1}
\big| \partial_x^\alpha K(x,\xi)\big|  \le  c\, \frac{|x|^{\Lambda_- -\sigma-|\alpha|+\varepsilon}}
  {|\xi|^{\Lambda_- +1+\tau+\varepsilon}}
  \prod_{k=1}^n \Big( \frac{r_k(x)}{|x|}\Big)^{\min(0,\mu_k-\sigma-|\alpha|-\varepsilon)}
  \prod_{k=1}^n \Big( \frac{r_k(\xi)}{|\xi|}\Big)^{\min(0,\mu_k-\tau-\varepsilon)}
\end{equation}
for $|x|>2|\xi|$, $|\alpha|\le 2-\sigma$, where $\sigma,\tau\in \{0,1\}$ and $\varepsilon$ is
a sufficiently small positive real number. If $\beta$ and $\delta$ satisfy condition
{\em (\ref{e6})}, then
\[
\| v\|_{W_{\beta,\delta}^{2-\sigma,s}({\cal K})} \le c\, 2^{-|m-l|\varsigma} \| \zeta_l f
  \|_{W_{\beta-\tau,\delta-\tau}^{0,s}({\cal K})}
\]
with positive constants $c$ and $\varsigma$ independent of $f$. The same estimates holds if $l\ge m+3$ and
\[
\big| \partial_x^\alpha K(x,\xi)\big|  \le  c\, \frac{|x|^{\Lambda_+ -\sigma-|\alpha|-\varepsilon}}
  {|\xi|^{\Lambda_+ +1+\tau-\varepsilon}}
  \prod_{k=1}^n \Big( \frac{r_k(x)}{|x|}\Big)^{\min(0,\mu_k-\sigma-|\alpha|-\varepsilon)}
  \prod_{k=1}^n \Big( \frac{r_k(\xi)}{|\xi|}\Big)^{\min(0,\mu_k-\tau-\varepsilon)}
\]
for $|\xi|>2|x|$, $|\alpha|\le 2-\sigma$.
\end{Le}

{\it Proof}: For $x\in \mbox{supp}\, \zeta_m$, $\xi\in \mbox{supp}\, \zeta_l$, we have $2^{m-1}<|x|<2^{m+1}$,
$2^{l-1}<|\xi|<2^{l+1}$. In particular, $|x|>2|\xi|$ if $m\ge l+3$. Therefore, by (\ref{1el1}) and by H\"older's
inequality,
\begin{eqnarray*}
&& \int_{\cal K} |x|^{s(\beta-2+\sigma+|\alpha|)} \prod_{k=1}^n \Big( \frac{r_k(x)}{|x|}\Big)^{s\delta_k}
  \big|\partial_x^\alpha v(x)\big|^s\, dx \\
&& \le c\, 2^{sm(\Lambda_- +\beta-2+\varepsilon)} \int\limits_{\substack{{\cal K}\\ 2^{m-1}<|x|<2^{m+1}}}
  \prod_{k=1}^n \Big( \frac{r_k(x)}{|x|} \Big)^{s(\delta_k+\min(0,\mu_k-\sigma-|\alpha|-\varepsilon))}\, dx \\
&& \qquad \times \Big( \int_{\cal K} |\xi|^{-\Lambda_- -1-\tau-\varepsilon} \prod_{k=1}^n  \Big(
  \frac{r_k(\xi)}{|\xi|}\Big)^{\min(0,\mu_k-\tau-\varepsilon)} \big| \zeta_l(\xi) f(\xi)\big|\, d\xi \Big)^s\\
&& \le c\,  2^{sm(\Lambda_- +\beta-2+\varepsilon)}\, \|\zeta_l f\|^s_{W_{\beta-\tau,\delta-\tau}^{0,s}({\cal K})}
  \int\limits_{\substack{{\cal K}\\ 2^{m-1}<|x|<2^{m+1}}} \prod_{k=1}^n \Big( \frac{r_k(x)}{|x|}
  \Big)^{s(\delta_k+\min(0,\mu_k-\sigma-|\alpha|-\varepsilon))}\, dx \\
&& \qquad\times \Big( \int\limits_{\substack{{\cal K}\\ 2^{l-1}<|\xi|<2^{l+1}}}
  |\xi|^{s'(-\Lambda_- -1-\beta-\varepsilon)} \prod_{k=1}^n
  \Big( \frac{r_k(\xi)}{|\xi|}\Big)^{s'(\min(0,\mu_k-\tau-\varepsilon)-\delta_k+\tau)}\, d\xi\Big)^{s/s'}
\end{eqnarray*}
for $|\alpha|\le 2-\sigma$, where $s'=s/(s-1)$. Since $s(\delta_k+\min(0,\mu_k-\sigma-|\alpha|)>-2$ and
$s'(\min(0,\mu_k-\tau)-\delta_k+\tau)>-2$, we obtain
\[
\int_{\cal K} |x|^{s(\beta-2+\sigma+|\alpha|)} \prod_{k=1}^n \Big( \frac{r_k(x)}{|x|}\Big)^{s\delta_k}
  \big|\partial_x^\alpha v(x)\big|^s\, dx
  \le c\, 2^{s(m-l)(\Lambda_-+\beta-2+\varepsilon+3/s)} \,
  \|\zeta_l f\|^s_{W_{\beta-\tau,\delta-\tau}^{0,s}({\cal K})}
\]
This proves the lemma for $m\ge l+3$. The proof for the case $l\ge m+3$ proceeds analogously. \rule{1ex}{1ex} \\

We will show an analogous result for the case $|l-m|\le 2$. For this we need the following lemma.

\begin{Le} \label{el2}
Let ${\cal D}$ be the dihedron {\em (\ref{dih})}, and let $r(x)$ denote the distance
of $x$ to the edge. If $\alpha+\beta> 3$ and $\beta<2$, then
\[
\int\limits_{\substack{{\cal D}\\ |\xi-x|>r(x)/3}} |\xi-x|^{-\alpha}\,
  r(\xi)^{-\beta}\, d\xi \le c\, r(x)^{3-\alpha-\beta}
\]
with a constant $c$ independent of $x$.
\end{Le}

{\it Proof:} The substitution $y=x/r(x)$, $\eta=\xi/r(x)$ yields
\begin{eqnarray*}
\int\limits_{\substack{{\cal D}\\ |\xi-x|>r(x)/3}} |\xi-x|^{-\alpha}\,
  r(\xi)^{-\beta}\, d\xi = r(x)^{3-\alpha-\beta}
  \int\limits_{\substack{{\cal D}\\ |\eta-y|>1/3}} |\eta-y|^{-\alpha}\,
  r(\eta)^{-\beta}\, d\eta.
\end{eqnarray*}
Since $r(y)=1$, the integral on the right is majorized by a finite constant $c$.
This proves the lemma. \rule{1ex}{1ex} \\

In the sequel, let $k(x)$ denote the smallest integer $k$ such that $r_{k(x)}=r(x)$.

\begin{Co} \label{ec1}
Let $c_1,c_2,\alpha,\beta_j,\gamma_j,\delta_j$ be real numbers such that
$\gamma_j+\delta_j<2$ and $3-\alpha+\beta_j-\delta_j<0$ for $j=1,\ldots,n$.
Furthermore, let ${\cal K}_x=\big\{ \xi\in {\cal K}:\, c_1 |x|<|\xi|< c_2 |x|,\
|\xi-x|>r(x)/3\big\}$. Then
\begin{equation} \label{1ec1}
\int\limits_{{\cal K}_x} |x-\xi|^{-\alpha} \Big( \frac{r(x)}{|x-\xi|}\Big)^{-\beta_{k(x)}}
  \Big( \frac{r(\xi)}{|x-\xi|}\Big)^{-\gamma_{k(\xi)}} \
  \prod_{j=1}^n \Big( \frac{r_j(\xi)}{|\xi|}\Big)^{-\delta_j} \
  \, d\xi \le c\, |x|^{3-\alpha}\, \prod_{j=1}^n \Big( \frac{r_j(x)}{|x|}\Big)^{3-\alpha-\delta_j}
\end{equation}
with $c$ independent of $x$.
\end{Co}

{\it Proof:}
Without loss of generality, we may assume that $r(x)=r_1(x)$, i.e. $k(x)=1$. Then
the left-hand side of (\ref{1ec1}) is equal to
\begin{equation} \label{2ec1}
|x|^{3-\alpha}\, r_1(y)^{-\beta_1} \int\limits_{{\cal K}_y} |y-\eta|^{-\alpha+\beta_1}
  \Big( \frac{r(\eta)}{|y-\eta|}\Big)^{-\gamma_{k(\eta)}} \
  \prod_{j=1}^n \Big( \frac{r_j(\eta)}{|\eta|}\Big)^{-\delta_j} \, d\eta,
\end{equation}
where $y=x/|x|$, $\eta=\xi/|x|$.

Suppose first that $r(y)=r_1(y)<r_i(y)$ for $i=2,\ldots,n$. We denote by ${\cal K}_y^{(1)}$ the
set of all $\eta\in {\cal K}_y$ such that $r(\eta)=r_1(\eta) < r_j(\eta)$ for $j=2,\ldots,n$.
Obviously, this set is contained in a dihedron ${\cal D}$ with edge $M\supset M_1$. Therefore, by Lemma \ref{el2},
\begin{eqnarray*}
\int\limits_{{\cal K}_y^{(1)}} |y-\eta|^{-\alpha+\beta_1} \Big( \frac{r(\eta)}{|y-\eta|}
  \Big)^{-\gamma_{k(\eta)}} \ \prod_{j=1}^n \Big( \frac{r_j(\eta)}{|\eta|}\Big)^{-\delta_j} \, d\eta
& \le & c\, \int\limits_{\cal D} |\eta-y|^{-\alpha+\beta_1+\gamma_1}\, r_1(\eta)^{-\gamma_1-\delta_1}\, d\eta\\
& \le & c\, r_1(y)^{3- \alpha+\beta_1-\delta_1}.
\end{eqnarray*}
Let ${\cal K}_y^{(i)}$, $i=2,\ldots,n$, be the set of all $\eta\in {\cal K}_y$ such that
$r(\eta)=r_i(\eta)$. Obviously, there exists a constant $c_0>0$ such that
\[
c_0 < |y-\eta| < c_2+1 \quad\mbox{if } r_1(y) < r_i(y),\ \eta \in {\cal K}_y^{(i)},\ i\ge 2.
\]
Consequently,
\[
\int\limits_{{\cal K}_y^{(i)}} |y-\eta|^{-\alpha+\beta_1} \Big( \frac{r(\eta)}{|y-\eta|}
  \Big)^{-\gamma_{k(\eta)}} \ \prod_{j=1}^n \Big( \frac{r_j(\eta)}{|\eta|}\Big)^{-\delta_j} \, d\eta
\le  c \int\limits_{{\cal K}_y^{(i)}}  r(\eta)^{-\gamma_{k(\eta)}-\delta_{k(\eta)}}\, d\eta
\le  c \le c\, r_1(y)^{3- \alpha+\beta_1-\delta_1}
\]
for $i\ge 2$. This together with (\ref{2ec1}) proves (\ref{1ec1}) if $r(y)=r_1(y)<r_j(y)$ for $j=2,\ldots,n$.

Suppose now that $r(y)=r_1(y)=r_i(y)$ for a certain $i\ge 2$. Then, there are the inequalities
\[
c_0 < r(y) \le |y|=1\quad\mbox{and}\quad c_0/3 < |y-\eta| < c_2+1 \ \mbox{ for }\eta\in {\cal K}_y
\]
with a positive constant $c_0$. Therefore,
\[
\int\limits_{{\cal K}_y} |y-\eta|^{-\alpha+\beta_1} \Big( \frac{r(\eta)}{|y-\eta|}
  \Big)^{-\gamma_{k(\eta)}} \ \prod_{j=1}^n \Big( \frac{r_j(\eta)}{|\eta|}\Big)^{-\gamma_j} \, d\eta
 \le  c\, \int\limits_{{\cal K}_y} r(\eta)^{-\beta_{k(\eta)}-\delta_{k(\eta)}}\, d\eta
 \le  c\, \le c\, r_1(y)^{3- \alpha+\beta_1-\delta_1}
\]
what implies (\ref{1ec1}). The proof is complete. \rule{1ex}{1ex}\\

We introduce the functions
\begin{equation} \label{e7}
\chi^+(x,\xi) = \chi\Big( \frac{|x-\xi|}{r(x)}\Big),\quad \chi^-(x,\xi) = 1-\chi^+(x,\xi),
\end{equation}
where $\chi$ is an arbitrary smooth cut-off function on $[0,\infty)$, $\chi(t)=1$ for $t<1/4$,
$\chi(t)=0$ for $t>1/2$. Furthermore, let $\mu_x=\mu_{k(x)}$, where $k(x)$ is the smallest integer $k$
such that $r(x)=r_k(x)$.

\begin{Le} \label{el3}
Let $\zeta_k$ be the same function as in the proof of Lemma {\em \ref{bl1}} and let
\[
v(x) = \zeta_m(x) \, \int_{\cal K} \zeta_l(\xi)\, f(\xi)\, \chi^-(x,\xi)\, K(x,\xi)\, d\xi,
\]
where $|l-m|\le 2$ and $f\in W_{\beta-\tau,\delta-\tau}^{0,s}({\cal K})$. Suppose that
\begin{equation} \label{1el3}
\big| \partial_x^\alpha K(x,\xi)\big|  \le  c\, |x-\xi|^{-1-\sigma-\tau-|\alpha|}
  \Big( \frac{r(x)}{|x-\xi|}\Big)^{\min(0,\mu_x-\sigma-|\alpha|-\varepsilon)}
  \Big( \frac{r(\xi)}{|x-\xi|}\Big)^{\min(0,\mu_\xi-\tau-\varepsilon)}
\end{equation}
for $|x|/32 < |\xi| < 32|x|$, $|x-\xi|>r(x)/4$, $|\alpha|\le 2-\sigma$, where $\sigma,\tau\in \{0,1\}$,
$\varepsilon$ is a sufficiently small positive real number.
If $\max(0,2-\mu_k)<\delta_k+2/s<2$ for $k=1,\ldots,n$, then
\[
\| v\|_{W_{\beta,\delta}^{2-\sigma,s}({\cal K})} \le c\,  \| \zeta_l f
  \|_{W_{\beta-\tau,\delta-\tau}^{0,s}({\cal K})}\, .
\]
\end{Le}

{\it Proof}:
Let $|\alpha|\le 2-\sigma$. Obviously, $\displaystyle \big|\partial_x^\alpha v(x)\big|
\le c\, \sum_{j+|\gamma|=|\alpha|} |x|^{-j} A_{\gamma}(x),$ where
\[
A_{\gamma}(x) =   \int\limits_{\substack{{\cal K}\\ |x-\xi|>r(x)/4}} |x-\xi|^{-1-\sigma-\tau-|\gamma|}
  \Big( \frac{r(x)}{|x-\xi|}\Big)^{\min(0,\mu_x-\sigma-|\gamma|-\varepsilon)}
  \Big( \frac{r(\xi)}{|x-\xi|}\Big)^{\min(0,\mu_\xi-\tau-\varepsilon)} |\zeta_l(\xi) f(\xi)|\, d\xi
\]
We have to prove that
\begin{equation} \label{3el3}
\int_{\cal K} |x|^{s(\beta-2+\sigma+|\gamma|)} \prod_{k=1}^n \Big( \frac{r_k(x)}{|x|}\Big)^{s\delta_k}
 \big| A_\gamma(x)\big|^s\, dx \le c\, \| \zeta_l f\|^s_{V_{\beta-\tau,\delta-\tau}^{0,s}({\cal K})}
\end{equation}
for $|\gamma|\le 2-\sigma.$ Let first $\sigma+\tau+|\gamma|\not=0$, and let $s'=s/(s-1)$.
Using H\"older's inequality and Corollary \ref{ec1}, we obtain
\begin{eqnarray} \label{2el3}
&& \hspace{-2em}| A_{\gamma}(x)|^s  \le c\int\limits_{\substack{{\cal K}\\ |x-\xi|>r(x)/4}} |x-\xi|^{-1-\sigma-\tau-|\gamma|}
  \Big( \frac{r(x)}{|x-\xi|}\Big)^{\min(0,\mu_x-\sigma-|\gamma|-\varepsilon)}
  \Big( \frac{r(\xi)}{|x-\xi|}\Big)^{\min(0,\mu_\xi-\tau-\varepsilon)} \nonumber\\
&& \hspace{5em} \times \prod_{k=1}^n \Big( \frac{r_k(\xi)}{|\xi|}\Big)^{ss_k} |\zeta_l f|^s\, d\xi
  \Big( \int\limits_{\substack{|x|/32<|\xi|<32|x| \\ |x-\xi|>r(x)/4}} |x-\xi|^{-1-\sigma-\tau-|\gamma|}
  \Big( \frac{r(x)}{|x-\xi|} \Big)^{\min(0,\mu_x-\sigma-|\gamma|-\varepsilon)}\nonumber\\
&&  \hspace{18em} \times\Big( \frac{r(\xi)}{|x-\xi|}\Big)^{\min(0,\mu_\xi-\tau-\varepsilon)}
  \prod_{k=1}^n \Big( \frac{r_k(\xi)}{|\xi|}\Big)^{-s's_k} \, d\xi\Big)^{s-1} \nonumber\\
& \le & |x|^{(s-1)(2-\sigma-\tau-|\gamma|)} \prod_{k=1}^n
  \Big( \frac{r_k(x)}{|x|} \Big)^{(s-1)(2-\sigma-\tau-|\gamma|)-ss_k}
  \int\limits_{\substack{{\cal K}\\ |x-\xi|>r(x)/4}} |x-\xi|^{-1-\sigma-\tau-|\gamma|} \nonumber\\
&& \quad\times\Big( \frac{r(x)}{|x-\xi|} \Big)^{\min(0,\mu_x-\sigma-|\gamma|-\varepsilon)}
  \Big( \frac{r(\xi)}{|x-\xi|}\Big)^{\min(0,\mu_\xi-\tau-\varepsilon)}
  \prod_{k=1}^n \Big( \frac{r_k(\xi)}{|\xi|}\Big)^{ss_k} \big|\zeta_l(\xi)\, f(\xi)\big|^s\, d\xi
\end{eqnarray}
provided $s_k$ satisfies the inequalities
\begin{equation} \label{4el3}
\max(2-\sigma-\tau-|\gamma|,2-\tau-\mu_k)< s's_k < \min(2,2-\tau+\mu_k).
\end{equation}
We put $t_k=-\tau$ if $\sigma+|\gamma|=2$, $t_k=\max(1-\tau,2-\tau-\mu_k)-\max(0,2-\mu_k)$ if $\sigma+|\gamma|=1$.
In the case $\sigma=|\gamma|=0$, $\tau=1$ let $t_k$ be arbitrary numbers in the intervals
$1-\delta_k-2/s< t_k < 1-\delta_k-2/s+\min(1,\mu_k)$. Obviously, $-\tau\le t_k\le 2-\sigma-|\gamma|-\tau$.
Due to condition (ii), the numbers $s_k$ can be chosen such that, additionally to condition (\ref{4el3}),
the inequalities
\[
\delta_k+t_k+\frac 1s \max(0,\tau-\mu_k) < s_k  < \delta_k+t_k + \frac 1s \min(\sigma+\tau+|\gamma|,\tau+\mu_k).
\]
are satisfied. Consequently, by Corollory \ref{ec1},
\begin{eqnarray*}
&& \hspace{-1em}\int_{\cal K} |x|^{s(\beta-2+\sigma+|\gamma|)} \prod_{k=1}^n \Big( \frac{r_k(x)}{|x|}\Big)^{s\delta_k}
 \big| A_\gamma(x)\big|^s\, dx \\
&& \hspace{-1em} \le c\!\!\int\limits_{\substack{{\cal K}\\ 2^{m-1}<|x|<2^{m+1}}}\!\!
  |x|^{s(\beta-\tau)-2+\sigma+\tau+|\gamma|}
  \prod_{k=1}^n \Big( \frac{r_k(x)}{|x|}\Big)^{s(\delta_k-s_k)+(s-1)(2-\sigma-\tau-|\gamma|) }\!\!\!
  \int\limits_{\substack{{\cal K}\\ |x-\xi|>r(x)/4}} \!\!\! |x-\xi|^{-1-\sigma-\tau-|\gamma|} \\
&&\qquad \times \Big( \frac{r(x)}{|x-\xi|}
  \Big)^{\min(0,\mu_x-\sigma-|\gamma|-\varepsilon)}
  \Big( \frac{r(\xi)}{|x-\xi|}\Big)^{\min(0,\mu_\xi-\tau-\varepsilon)}
  \prod_{k=1}^n \Big( \frac{r_k(\xi)}{|\xi|}\Big)^{ss_k} |\zeta_l f|^s\, d\xi\, dx \\
&&\hspace{-1em} \le c \int\limits_{\cal K} |\xi|^{s(\beta-\tau)-2+\sigma+\tau+|\gamma|} \prod_{k=1}^n \Big(
  \frac{r_k(\xi)}{|\xi|}\Big)^{ss_k}\, \big| \zeta_l(\xi) f(\xi)\big|^s \
  \int\limits_{\substack{2^{m-1}< |x|< 2^{m+1}\\ |x-\xi|>r(\xi)/5}} |x-\xi|^{-1-\sigma-\tau-|\gamma|} \\
&& \quad\times \Big( \frac{r(x)}{|x-\xi|} \Big)^{\min(0,\mu_x-\sigma-|\gamma|-\varepsilon)}
  \Big( \frac{r(\xi)}{|x-\xi|}\Big)^{\min(0,\mu_\xi-\tau-\varepsilon)}
  \prod_{k=1}^n \Big( \frac{r_k(x)}{|x|}\Big)^{s(\delta_k-s_k+t_k)-2+\sigma+\tau+|\gamma|}\, dx\, d\xi \\
&& \hspace{-1em} \le c\, \int_{\cal K} |\xi|^{s(\beta-\tau)}\ \prod_{k=1}^n \Big( \frac{r_k(x)}{|x|}
  \Big)^{s(\delta_k+t_k)}\, \big| \zeta_l(\xi)f(\xi)\big|^s\, d\xi.
\end{eqnarray*}
This proves (\ref{3el3}) for $\sigma+\tau+|\gamma|\not=0$. If $\sigma=\tau=|\gamma|=0$, then
H\"older's inequality yields
\begin{eqnarray*}
&& \big| A_\gamma(x)\big|^s \le \int\limits_{\cal K} |x-\xi|^{-1} \prod_{k=1}^n \Big(
  \frac{r_k(\xi)}{|\xi|}\Big)^{s\delta_k} \, \big| \zeta_l(\xi)f(\xi)\big|\, d\xi \
  \Big( \int\limits_{\substack{{\cal K}\\ 2^{l-1}<|\xi|<2^{l+1}}} |x-\xi|^{-1} \prod_{k=1}^n \Big(
  \frac{r_k(\xi)}{|\xi|}\Big)^{-s'\delta_k} \, d\xi\Big)^{s-1} \\
&& \le c\, 2^{2l(s-1)}\ \int\limits_{\cal K} |x-\xi|^{-1} \prod_{k=1}^n \Big(
  \frac{r_k(\xi)}{|\xi|}\Big)^{s\delta_k} \, \big| \zeta_l(\xi)f(\xi)\big|\, d\xi.
\end{eqnarray*}
Therefore,
\begin{eqnarray*}
&& \int_{\cal K} |x|^{s(\beta-2)} \prod_{k=1}^n \Big( \frac{r_k(x)}{|x|}\Big)^{s\delta_k}
 \big| A_\gamma(x)\big|^s\, dx \\
&& \le  c\, \int_{\cal K} |\xi|^{s\beta-2} \prod_{k=1}^n \Big( \frac{r_k(\xi)}{|\xi|}\Big)^{s\delta_k}
  \big|\zeta_l(\xi)f(\xi)\big|^s\, \Big( \int\limits_{\substack{{\cal K}\\ 2^{m-1}<|x|<2^{m+1}}}
  |x-\xi|^{-1} \prod_{k=1}^n  \Big( \frac{r_k(x)}{|x|}\Big)^{s\delta_k}  \, dx\Big)\, d\xi \\
&& \le  c\, \int_{\cal K} |\xi|^{s\beta} \prod_{k=1}^n \Big( \frac{r_k(\xi)}{|\xi|}\Big)^{s\delta_k}
  \big|\zeta_l(\xi)f(\xi)\big|^s\, d\xi.
\end{eqnarray*}
This proves the lemma. \rule{1ex}{1ex}

\begin{Le} \label{el4}
Let $\chi^+$ be defined by {\em (\ref{e7})}, $\sigma \in \{ 0,1\}$, and
\[
v(x) = \zeta_m(x) \, \int_{\cal K} \zeta_l(\xi)\, f(\xi)\, \chi^+(x,\xi)\, K(x,\xi)\, d\xi,\quad
  |l-m|\le 2.
\]
{\em 1)} If $f\in V_{\beta,\delta}^{0,s}({\cal K})$, and $K(x,\xi)$ satisfies the estimate
\begin{equation} \label{1el4}
| \partial_x^\alpha K(x,\xi)|  \le  c\, |x-\xi|^{-1-\sigma-|\alpha|}\quad\mbox{for } |x|/32 < |\xi| <32|x|,\
  |x-\xi|< r(x)/2, \ |\alpha|\le 1-\sigma,
\end{equation}
then
\[
\| v\|_{V_{\beta-1,\delta-1}^{1-\sigma,s}({\cal K})} \le c\,  \| \zeta_l f
  \|_{V_{\beta,\delta}^{0,s}({\cal K})}\, .
\]
{\em 2)} If $K(x,\xi)$ has the representation $K(x,\xi)= \nabla_\xi P(x,\xi) + Q(x,\xi)$, where
\begin{equation} \label{2el4}
| \partial_x^\alpha P(x,\xi) | \le c\, |x-\xi|^{-1-\sigma-|\alpha|},\quad
  |\partial_x^\alpha Q(x,\xi)|\le c\, r(\xi)^{-2-\sigma-|\alpha|}
\end{equation}
for $|x|/32 < |\xi| <32|x|$, $|x-\xi|< r(x)/2$, $|\alpha|\le 1-\sigma$, and $P(x,\xi)\cdot n^{(j)}=0$ for
$\xi \in \Gamma_j$, $j=1,\ldots,n$ (here $n^{(j)}$ denotes the exterior normal to $\Gamma_j$), then
\[
\| v\|_{V_{\beta-1,\delta-1}^{1-\sigma,s}({\cal K})} \le c\,  \| \zeta_l f
  \|_{V_{\beta,\delta}^{1,s}({\cal K})}\, .
\]
\end{Le}

{\it Proof}:
1) By our definition, the function $\chi^+(x,\xi)$ vanishes for $|x-\xi|> r(x)/2$. Note
that
\begin{equation} \label{3el4}
\frac 1{32}|x|< |\xi| < 32|x|, \quad \frac 12 r_k(x)\le r_k(\xi) \le \frac 32 \, r_k(x),\quad
 \mbox{and}\quad \frac 12 r(x)\le r(\xi) \le \frac 32 \, r(x),
\end{equation}
for $x\in \mbox{supp}\, \zeta_m$, $\xi\in \mbox{supp}\,\zeta_l$, and $|x-\xi|< r(x)/2$.
Let ${\cal K}_x=\{ \xi \in {\cal K}:\, |x|/32 < |\xi| <32|x|,\  |x-\xi|< r(x)/2\}$. Then
\[
\big|\partial_x^\alpha v(x)\big| \le c\int_{{\cal K}_x} |x-\xi|^{-1-\sigma-|\alpha|}\,\big|
  \zeta_l(\xi)\, f(\xi)\big|\, d\xi
\]
and, consequently,
\begin{eqnarray*}
\big|\partial_x^\alpha v(x)\big|^s & \le & c\int_{{\cal K}_x} |x-\xi|^{-1-\sigma-|\alpha|}\,\big|
  \zeta_l(\xi)\, f(\xi)\big|^s\, d\xi\ \Big( \int\limits_{|x-\xi|<r(x)/2} |x-\xi|^{-1-\sigma-|\alpha|}\, d\xi
  \Big)^{s-1}\\
& \le & c\, r(x)^{(s-1)(2-\sigma-|\alpha|)} \int_{{\cal K}_x} |x-\xi|^{-1-\sigma-|\alpha|}\,\big|
  \zeta_l(\xi)\, f(\xi)\big|^s\, d\xi
\end{eqnarray*}
for $|\alpha|\le 1-\sigma$. Using (\ref{3el4}), we obtain
\begin{eqnarray*}
&& \int_{\cal K} |x|^{s(\beta-2+\sigma+|\alpha|)} \prod_{k=1}^n \Big( \frac{r_k(x)}{|x|}
  \Big)^{s(\delta_k-2+\sigma+|\alpha|)}\, \big| \partial_x^\alpha v(x)\big|^s \, dx\\
&& \le c\, \int_{\cal K} |\xi|^{s(\beta-2+\sigma+|\alpha|)} \prod_{k=1}^n \Big( \frac{r_k(\xi)}{|\xi|}
  \Big)^{s(\delta_k-2+\sigma+|\alpha|)} \, r(\xi)^{(s-1)(2-\sigma-|\alpha|)} \, |\zeta_l f|^s\,
  \Big( \int\limits_{|x-\xi|< r(\xi)} |x-\xi|^{-1-\sigma-|\alpha|}\, dx\Big)\, d\xi \\
&& \le c\, \int_{\cal K} |\xi|^{s\beta} \prod_{k=1}^n \Big( \frac{r_k(\xi)}{|\xi|}
  \Big)^{s\delta_k} \, \big|\zeta_l(\xi) f(\xi)\big|^s\, d\xi.
\end{eqnarray*}
This proves the first part.

2) The second part can be proved analogously using the estimate
\[
\big|\partial_x^\alpha v(x)\big| \le c\int_{{\cal K}_x} \Big( |x-\xi|^{-1-\sigma-|\alpha|}
  \, \big| \nabla_\xi(\zeta_l f)\big| + r(\xi)^{-2-\sigma-|\alpha|}\, |\zeta_l f|\Big)\, d\xi
\]
which follows from our assumptions on $K(x,\xi)$. \rule{1ex}{1ex}

\subsection{Existence of solutions}

Let $f\in W_{\beta,\delta}^{0,s}({\cal K})^3$, $g\in W_{\beta,\delta}^{1,s}({\cal K})$,
$h_j \in W_{\beta,\delta}^{2-1/s,s}(\Gamma_j)^{3-d_j}$, $\phi_j \in W_{\beta,\delta}^{1-1/s,s}
(\Gamma_j)^{d_j}$. Our goal is to show that there exists a solution
$(u,p) \in W_{\beta,\delta}^{2,s}({\cal K})^3\times W_{\beta,\delta}^{1,s}({\cal K})$ of
problem (\ref{d1}), (\ref{d2}) if the following conditions are satisfied.
\begin{itemize}
\item[(i)] there are no eigenvalues of the pencil ${\mathfrak A}(\lambda)$ on the line
  $\mbox{Re}\, \lambda=2-\beta-3/s$,
\item[(ii)] $\max(2-\mu_k,0) < \delta_k +2/s <2$ for $k=1,\ldots,n$.
\item[(iii)] $h_j,\phi_j$ and $g$ are such that there exist $u\in W_{\beta,\delta}^{2,s}({\cal K})^3$
and $p\in W_{\beta,\delta}^{1,s}({\cal K})$ satisfying (\ref{1dl2}).
\end{itemize}
The last is a condition on the traces of $g$, $\phi_j$, $h_j$ and the derivatives of $h_j$ on the edges
of the cone ${\cal K}$ (see Section 3.1).

\begin{Le} \label{fl1}
Let ${\cal X}$, ${\cal Y}$ be Banach spaces of functions on ${\cal K}$ in each of them
the multiplication with a scalar function from $C_0^\infty(\ol{\cal K}\backslash
\{ 0\})$ is defined. We suppose that the inequalities
\[
\| f\|_{\cal X} \ge c\, \Big( \sum_{j=-\infty}^{+\infty} \| \zeta_j f
  \|^s_{\cal X}\Big)^{1/s} ,\qquad \| u\|_{\cal Y} \le c\, \Big(
  \sum_{j=-\infty}^{+\infty} \| \zeta_j u \|^s_{\cal Y}\Big)^{1/s}
\]
are satisfied for all $f\in {\cal X}$, $u\in {\cal Y}$.
Furthermore, let  ${\cal O}$ be a linear operator from ${\cal X}$ into ${\cal Y}$
defined on functions with compact support in $\ol{\cal K}\backslash\{ 0\}$ such that
\[
\| \zeta_m {\cal O}\zeta_l f\|_{\cal Y} \le c\, 2^{-\varsigma|l-m|}\,
  \| \zeta_l f\|_{\cal X}
\]
with positive constants $c$, $\varsigma$ independent of $l$, $m$ and $f$. Then
$\displaystyle \| {\cal O} f\|_{\cal Y} \le c\,  \| f\|_{\cal X}$
for all $f\in {\cal X}$ with compact support in $\ol{\cal K}\backslash\{ 0\}$.
\end{Le}

The proof of this lemma can be found in \cite{mp78c}.

\begin{Th} \label{ft1}
Let $f\in W_{\beta,\delta}^{0,s}({\cal K})^3$, $g\in W_{\beta,\delta}^{1,s}({\cal K})$,
$h_j \in W_{\beta,\delta}^{2-1/s,s}(\Gamma_j)^{3-d_j}$, and $\phi_j \in W_{\beta,\delta}^{1-1/s,s}
(\Gamma_j)^{d_j}$. Suppose that conditions {\em (i)--(iii)} are satisfied. Then there exists  a solution
$(u,p)\in W_{\beta,\delta}^{2,s}({\cal K})^3 \times W_{\beta,\delta}^{1,s}({\cal K})$ of problem
{\em (\ref{d1}), (\ref{d2})} satisfying the estimate
\[
\| u\|_{W_{\beta,\delta}^{2,s}({\cal K})^3}+\| p\|_{W_{\beta,\delta}^{1,s}({\cal K})}
\le c\, \Big( \|f\\|_{W_{\beta,\delta}^{0,s}({\cal K})^3} +\| g\|_{W_{\beta,\delta}^{1,s}({\cal K})}
+ \sum_{j=1}^n\Big( \| h_j \|_{W_{\beta,\delta}^{2-1/s,s}(\Gamma_j)^{3-d_j}}
+ \| \phi_j \|_{W_{\beta,\delta}^{1-1/s,s}(\Gamma_j)^{d_j}}\Big)\Big).
\]
\end{Th}

{\it Proof}:
Without loss of generality we may assume that $h_j=0$, $\phi_j=0$, and $g\in V_{\beta,\delta}^{1,s}
({\cal K})$. We consider the operator
\[
{\cal X}\stackrel{def}{=}W_{\beta,\delta}^{0,s}({\cal K})^3\times  V_{\beta,\delta}^{1,s}({\cal K}) \ni (f,g)
  \to {\cal O}(f,g)=(u,p),
\]
where $u$ and $p$ are defined by (\ref{e4}), (\ref{e5}) and $G_{i,j}$ are the elements of
Green's matrix introduced in Section 3.4. Then by Lemma \ref{el1},
\begin{equation} \label{1ft1}
\| \zeta_m {\cal O} \zeta_l(f,g) \|_{\cal Y} \le c\, 2^{-|l-m|\varsigma} \|\zeta_l(f,g)\|_{\cal X}
\end{equation}
for $|l-m|\ge 3$, where ${\cal Y}=W_{\beta,\delta}^{2,s}({\cal K})^3\times  W_{\beta,\delta}^{1,s}({\cal K})$
and $c,\varsigma$ are positive constants independent of $f,g,l,m$. In order to prove the same inequality for
$|l-m|\le 2$, we introduce the functions
\begin{eqnarray*}
&& u_i^\pm(x) = \sum_{j=1}^3 \int_{\cal K} \zeta_l(\xi)\, f_j(\xi)\, \chi^\pm(x,\xi)\, G_{i,j}(x,\xi)\, d\xi
  + \int_{\cal K} \zeta_l(\xi) g(\xi)\, \chi^\pm(x,\xi)\, G_{i,4}(x,\xi)\, d\xi,\quad i=1,2,3, \\
&& p^\pm(x) = -\zeta_l(x)\, g(x) + \sum_{j=1}^3 \int_{\cal K} \zeta_l(\xi)\, f_j(\xi)\, \chi^\pm(x,\xi)\,
  G_{4,j}(x,\xi)\, d\xi + \int_{\cal K} \zeta_l(\xi)\, g(\xi)\, \chi^\pm(x,\xi)G_{4,4}(x,\xi)\, d\xi,
\end{eqnarray*}
where $\chi^+$ and $\chi^-$ are defined by (\ref{e7}). Then
\[
-\Delta(u^+ +u^-)+\nabla(p^+ +p^-)=\zeta_l f, \quad -\nabla\cdot (u^+ +u^-)= \zeta_l g\ \mbox{ in }{\cal K}
\]
and $S_j (u^++u^-)=0$, $N_j(u^++u^-,p^++p^-)=0$ on $\Gamma_j$. Furthermore, by Lemmas \ref{el3}, \ref{el4},
there are the inequalities
\begin{equation} \label{2ft1}
\| \zeta_m (u^-,p^-)\|_{\cal Y} \le c\, \|\zeta_l(f,g)\|_{\cal X},\quad
  \| \zeta_m (u^+,p^+)\|_{V_{\beta-1,\delta-1}^{1,s}({\cal K})^3\times V_{\beta-1,\delta-1}^{0,s}({\cal K})}
  \le c\, \|\zeta_l(f,g)\|_{\cal X}
\end{equation}
if $|l-m|\le 3$, where $c$ is independent of $f,g,l,m$. Let $\eta_m=\zeta_{m-1}+\zeta_m+\zeta_{m+1}$.
Then, by Corollary \ref{dc1},
\begin{eqnarray*}
&&\| \zeta_m u^+\|_{V_{\beta,\delta}^{2,s}({\cal K})^3}+ \|\zeta_m p^+\|_{V_{\beta,\delta}^{1,s}({\cal K})} \\
&& \le c\, \Big( \| \eta_m u^+\|_{V_{\beta-1,\delta-1}^{1,s}({\cal K})^3} + \|\eta_m p^+
  \|_{V_{\beta-1,\delta-1}^{0,s}({\cal K})} + \| \eta_m \zeta_l(f,g)\|_{\cal X} + \| \eta_m (u^-,p^-)\|_{\cal Y}\Big) \\
\end{eqnarray*}
for $|l-m|\le 2$. Due to (\ref{2ft1}), the right hand side of the last inequality can be estimated by
the norm of $\zeta_l(f,g)$ in ${\cal X}$. Consequently,
\[
\|\zeta_m(u^+ +u^-,p^++p^-)\|_{\cal Y} \le c\, \| \zeta_l(f,g)\|_{\cal X}\quad\mbox{for }|l-m|\le 2.
\]
Thus, estimate (\ref{1ft1}) is valid for arbitrary $l$ and $m$. Now the assertion of the theorem
follows immediately from Lemma \ref{fl1}. \rule{1ex}{1ex}

\subsection{Uniqueness of the solution}

First we prove the uniqueness of the solution in Theorem \ref{ft1} in the case $s\ge 2$.

\begin{Le}  \label{fl3}
Let $s\ge 2$, and let the conditions {\em (i), (ii)} be satisfied. Then the homogeneous boundary
value problem {\em (\ref{d1}), (\ref{d2})} has only the trivial solution $(u,p)=(0,0)$ in
$W_{\beta,\delta}^{2,s}({\cal K})^3\times W_{\beta,\delta}^{1,s}({\cal K})$.
\end{Le}

{\it Proof:}
Let $(u,p) \in W_{\beta,\delta}^{2,s}({\cal K})^3\times W_{\beta,\delta}^{1,s}({\cal K})$ be a
solution of the homogeneous problem (\ref{d1}), (\ref{d2}).
By $\chi$ we denote a smooth cut-off function on $\ol{\cal K}$ equal to one for $|x|<1$ and to zero for
$|x|>2$. Furthermore, we set $\beta'=\beta-\frac 32 +\frac 3s$ and $\delta'_j=
\delta_j-1+\frac 2s$ for $j=1,\ldots,n$. Then, by H\"older's inequality,
\begin{eqnarray*}
& & \int_{\cal K} \rho^{2(\beta'+\varepsilon-2+|\alpha|)} \prod \Big( \frac{r_j}{\rho}
  \Big)^{2(\delta'_j+\varepsilon)} |\partial_x^\alpha (\chi u)|^2\, dx \\
& & \le \Big( \int_{\cal K} \rho^{s(\beta-2+|\alpha|)}
  \prod \Big( \frac{r_j}{\rho} \Big)^{s\delta_j} |\partial_x^\alpha (\chi u)|^s\,
  dx\Big)^{2/s}\ \Big(\int\limits_{\substack{{\cal K}\\ |x|\le 2}}\rho^{-3+s'\varepsilon}
  \prod \Big( \frac{r_j}{\rho}\Big)^{-2+q\varepsilon}\, dx\Big)^{2/s'} ,
\end{eqnarray*}
where $s'=2s/(s-2)$. The second integral on the right is finite if $\varepsilon>0$. Consequently,
$\chi u\in W_{\beta'+\varepsilon,\delta+\varepsilon}^{2,2}({\cal K})^3$ and $\chi p\in
W_{\beta'+\varepsilon,\delta+\varepsilon}^{1,2}({\cal K})$. Analogously, we obtain
$(1-\chi)u\in W_{\beta'-\varepsilon,\delta-\varepsilon}^{2,2}({\cal K})^3$ and
$(1-\chi)u\in W_{\beta'-\varepsilon,\delta-\varepsilon}^{2,2}({\cal K})$.
This implies
\[
-\Delta(\chi u) + \nabla(\chi p) = \Delta\big( (1-\phi)u\big) -\nabla\big((1-\chi)p\big) \in
  W_{\beta'-\varepsilon,\delta-\varepsilon}^{0,2}({\cal K})^3,
\]
and, analogously, $\nabla\cdot(\chi u)\in W_{\beta'-\varepsilon,\delta-\varepsilon}^{0,2}({\cal K})$,
$S_j(\chi u)\in W_{\beta'-\varepsilon,\delta-\varepsilon}^{3/2,2}(\Gamma_j)^{3-d_j}$ and
$N_j(\chi u,\chi p) \in W_{\beta'-\varepsilon,\delta-\varepsilon}^{3/2,2}(\Gamma_j)^{d_j}$.
From this and from \cite[Th.4.1]{mr-03} it follows that $\chi u\in
W_{\beta'-\varepsilon,\delta-\varepsilon}^{2,2}({\cal K})^3$ and $\chi p \in
W_{\beta'-\varepsilon,\delta-\varepsilon}^{1,2}({\cal K})$. The same is then obviously true for
$u$ and $p$. Hence, by Theorem \cite[Th.4.1]{mr-03}, we have $u=0$ and $p=0$. \rule{1ex}{1ex} \\

It remains to prove the uniqueness of the solution in Theorem \ref{ft1} in the case $s\le 2$.
In this case we pass to the coordinates $t,\omega$, where $t=\log \rho=\log|x|$ and $\omega=x/|x|$.
We denote by $W_{\delta}^{l,s}({\Bbb R}\times \Omega)$ the weighted Sobolev space with the norm
\[
\| u\|_{W_{\delta}^{l,s}({\Bbb R}\times \Omega)} = \Big( \int_{\Bbb R}\sum_{j=0}^l
\| \partial_t^j u(t,\cdot)\|^s_{W_{\delta}^{l-j,s}(\Omega)}\, dt\Big)^{1/s}.
\]
Note that $u\in W_{\beta,\delta}^{l,s}({\cal K})$ if and only if $\rho^{\beta-l+3/s}u$ (as function
of the variables $t$ and $\omega$) belongs to $W_\delta^{l,s}({\Bbb R}\times \Omega)$.

For an arbitrary function $v\in W_{\delta}^{l,s}({\Bbb R}\times \Omega)$ we define
by $v_\varepsilon$ the mollification with respect to the variable $t$ of $v$, i.e.,
\[
v_\varepsilon(t,\omega) = \int_{{\Bbb R}} v(\tau,\omega)\, h_\varepsilon(t-\tau)\,
d\tau,
\]
where $h_\varepsilon(t)= \varepsilon^{-1} h(t/\varepsilon)$ and $h$ is a smooth function
with compact support, $\int h(t)\, dt =1$. Since
\[
\partial_\omega^\alpha \partial_t^{j+k} v_\varepsilon(\omega,t) = \int_{\Bbb R}
  (\partial_\omega^\alpha\partial_t^k v)(\omega,\tau)\, h_\varepsilon^{(j)}(t-\tau)
  \, d\tau ,
\]
it follows that $\partial_t^j v_\varepsilon \in W_\delta^{l,s}({\Bbb R}\times \Omega)$
for $v \in W_{\delta}^{3,s}({\Bbb R}\times \Omega)$, $\varepsilon>0$, $j=0,1,\ldots$.

\begin{Th} \label{ft2}
Let $\beta\in {\Bbb R}$, $\delta\in {\Bbb R}^n$, $f\in W_{\beta,\delta}^{0,s}({\cal K})$, $g\in
W_{\beta,\delta}^{1,s}({\cal K})$, $h_j \in W_{\beta,\delta}^{2-1/s,s}(\Gamma_j)^{3-d_j}$, and
$\phi_j\in W_{\beta,\delta}^{1-1/s,s}(\Gamma_j)^{d_j}$ be such that conditions {\em (i)--(iii)} are satisfied.
Then problem {\em (\ref{d1}), (\ref{d2})} has a unique solution $(u,p) \in W_{\beta,\delta}^{2,s}({\cal K})^3
\times W_{\beta,\delta}^{1,s}({\cal K})$.
\end{Th}

{\it Proof:}
The existence existence of the solution and the uniqueness for $s\ge 2$ are already proved.
We show the uniqueness for the case $1<s<2$.
Let $(u,p) \in W_{\beta,\delta}^{2,s}({\cal K})^3 \times W_{\beta,\delta}^{2,s}({\cal K})$
be a solution of the homogeneous problem (\ref{d1}), (\ref{d2}). Since $W_{\beta,\delta}^{l,s}({\cal K})
\subset W_{\beta,\delta'}^{l,s}({\cal K})$ if $\delta_j\le \delta'_j$ for $j=1,\ldots,n$, we may assume,
without loss of generality, that $\max(2-\mu_j,1)< \delta_j+2/s <2$.

From Lemma \ref{dl4} it follows that $u \in W_{\beta+1,\delta+1}^{3,s}({\cal K})^\ell$ and
$p\in W_{\beta+1,\delta+1}^{2,s}({\cal K})$. We set $v=\rho^{\beta-2+3/s}u$ and $q=\rho^{\beta-1+3/s}p$.
Then, in the coordinates $t=\log |x|$ and $\omega=x/|x|$, we have $v \in W_{\delta+1}^{3,s}({\Bbb R}\times\Omega)^3$
and $q\in W_{\delta+1}^{2,s}({\Bbb R}\times\Omega)= V_{\delta+1}^{2,s}({\Bbb R}\times\Omega)$.
Consequently, $\partial_t^j v_\varepsilon \in W_{\delta+1}^{3,s}({\Bbb R}\times\Omega)^3$ for $j=0,1,2,\ldots$.
From Corollary \ref{ac1} we conclude that $v_\varepsilon \in W_{\delta-2+2/s}^{1,2}({\Bbb R}\times \Omega)^3
\subset W_0^{1,2}({\Bbb R}\times \Omega)^3$. Thus, the function $u_\varepsilon=\rho^{-\beta+2-3/s}v_\varepsilon$
(as function in $x$) belongs to the spaces $W_{\beta+3/s-5/2,0}^{1,2}({\cal K})^3$.
Analogously, using Lemma \ref{al4}, we obtain $p_\varepsilon=\rho^{-\beta+1-3/s}q_\varepsilon \in
W_{\beta+3/s-5/2,0}^{0,2}({\cal K})$. It can be easily seen that
$(u_\varepsilon,p_\varepsilon)$ is also a solution of the homogeneous problem (\ref{d1}), (\ref{d2}).
According to \cite[Th.4.2]{mr-03}, this problem has no nonzero
solutions in $W_{\beta+3/s-5/2,0}^{1,2}({\cal K})^3\times W_{\beta+3/s-5/2,0}^{0,2}({\cal K})$.
Therefore, $u_\varepsilon=0$, $p_\varepsilon=0$ what implies $u=0$, $v=0$.
The proof is complete. \rule{1ex}{1ex}

\begin{Th} \label{ft3}
Let $(u,p) \in W_{\beta',\delta'}^{2,\sigma}({\cal K})^3\times W_{\beta',\delta'}^{1,\sigma}({\cal K})$
be a solution of problem {\em (\ref{d1}), (\ref{d2})},
where
\begin{eqnarray*}
&& f\in W_{\beta,\delta}^{0,s}({\cal K})^3\cap W_{\beta',\delta'}^{0,\sigma}({\cal K})^3,\quad
g\in W_{\beta,\delta}^{1,s}({\cal K})\cap W_{\beta',\delta'}^{1,\sigma}({\cal K}), \\
&& h_j\in W_{\beta,\delta}^{2-1/s,s}(\Gamma_j)^{3-d_j} \cap W_{\beta',\delta'}^{2-1/\sigma,\sigma}
  (\Gamma_j)^{3-d_j},\quad \phi_j \in W_{\beta,\delta}^{1-1/s,s}(\Gamma_j)^{d_j}\cap
  W_{\beta',\delta'}^{1-1/\sigma,\sigma}(\Gamma_j)^{d_j}.
\end{eqnarray*}
Suppose that there are no eigenvalues of the pencil ${\mathfrak A}(\lambda)$ in the closed strip between the lines
$\mbox{\em Re}\, \lambda =2-\beta-3/s$ and $\mbox{\em Re}\, \lambda =2-\beta'-3/\sigma$, $\delta$ and $\delta'$
satisfy the inequalities $\max(0,2-\mu_k)>\delta_k+2/s<2$, $\max(0,2-\mu_k)>\delta'_k+2/\sigma<2$
and $g,h_j,\phi_j$ satisfy condition {\em (iii)} of Section {\em 3.6}.
Then $u\in W_{\beta,\delta}^{2,s}({\cal K})^3$ and $p\in W_{\beta,\delta}^{1,s}({\cal K})$.
\end{Th}

{\it Proof}:
By Theorem \ref{ft2}, there are unique solutions of problem (\ref{d1}), (\ref{d2}) in
$W_{\beta,\delta}^{2,s}({\cal K})^3\times W_{\beta,\delta}^{1,s}({\cal K})$ and
$W_{\beta',\delta'}^{2,\sigma}({\cal K})^3\times W_{\beta,\delta}^{1,\sigma}({\cal K})$.
These solutions coincide, since they are represented by the same Green's matrix.
\rule{1ex}{1ex}

\setcounter{equation}{0}
\setcounter{Th}{0}
\setcounter{Le}{0}
\setcounter{Co}{0}
\section{Weak solutions of the boundary value problem in a cone}

\subsection{Definition of weak solutions}

Obviously, the bilinear form
\begin{equation} \label{g1}
b(u,v) = 2\int_{\cal K} \sum_{i,j=1}^3 \varepsilon_{i,j}(u)\, \varepsilon_{i,j}(v)\, dx
\end{equation}
is continuous on $W_{\beta,\delta}^{1,s}({\cal K})^3\times W_{-\beta,-\delta}^{1,s'}({\cal K})^3$, where
$s'=s/(s-1)$. We suppose in this section that the line $\mbox{Re}\, \lambda=1-\beta-3/s$ is free of eigenvalues
of the pencil ${\mathfrak A}(\lambda)$ and that
\begin{equation} \label{g2}
\max(1-\mu_k,0)< \delta_k+2/s < 1 \ \mbox{ for } k=1,\ldots,n.
\end{equation}
Then $-\delta_k>1-2/s'$ and, therefore, $W_{-\beta,-\delta}^{1,s'}({\cal K})=V_{-\beta,-\delta}^{1,s'}({\cal K})$.
By $V_{\beta,\delta}^{-1,s}({\cal K})$ we denote the dual space of $V_{-\beta,-\delta}^{1,s'}({\cal K})^3$.
It can be shown analogously to \cite[Th.3.8]{adams} that every functional
$F \in V_{\beta,\delta}^{-1,s}({\cal K})^3$ has the form
\begin{equation} \label{g3}
F(v) = \int_{\cal K} f^{(0)}\cdot v\, dx + \sum_{k=1}^3 \int_{\cal K} f^{(k)}\, \partial_{x_k}v\cdot dx \
  \mbox{ for all }  v\in V_{-\beta,-\delta}^{1,s'}({\cal K})^3,
\end{equation}
where $f^{(0)}\in V_{\beta+1,\delta+1}^{0,s}({\cal K})^3$ and $f^{(k)}\in V_{\beta,\delta}^{0,s}({\cal K})^3$,
$k=1,2,3$.

Let $F \in V_{\beta,\delta}^{-1,s}({\cal K})^3$, $g\in W_{\beta,\delta}^{0,s}({\cal K})$ and
$h_j\in W_{\beta,\delta}^{1-1/s,s}(\Gamma_j)$, $j=1,\ldots,n$. By a weak solution of problem (\ref{d1}),
(\ref{d2}) we mean a pair $(u,p)\in W_{\beta,\delta}^{1,s}({\cal K})^3\times W_{\beta,\delta}^{1,s}({\cal K})^3$
satisfying
\begin{eqnarray} \label{g4}
&& b(u,v) - \int_{\cal K} p\, \nabla\cdot v = F(v)\ \mbox{ for all }v\in V_{-\beta,-\delta}^{1,s'}({\cal K})^3,
  \ S_j v=0 \mbox{ on }\Gamma_j,\ j=1,\ldots,n,\\ \label{g5}
&& -\nabla\cdot u = g \ \mbox{ in }{\cal K},\quad S_j u=h_j\ \mbox{ on }\Gamma_j,\ j=1,\ldots,n.
\end{eqnarray}
From Green's formula
\[
b(u,v)-\int_{\cal K} p\, \nabla\cdot v\, dx = \int_{\cal K} (-\Delta u - \nabla\nabla\cdot u
  + \nabla p)\cdot v\, dx + \sum_{j=1}^n \int_{\Gamma_j} (-pn^{(j)} + 2\varepsilon(u)n^{(j)})\cdot v\, dx
\]
it follows that every solution $(u,p)\in W_{\beta,\delta}^{1,s}({\cal K})^3\times W_{\beta,\delta}^{0,s}
({\cal K})^3$ of problem (\ref{g4}), (\ref{g5}) satisfies (\ref{d1}), (\ref{d2}) if
$g \in W_{\beta+1,\delta+1}^{1,s}({\cal K})$, $h_j \in W_{\beta+1,\delta+1}^{2-1/s}(\Gamma_j)$, and
$F$ has the form
\[
F(v) = \int_{\cal K} (f+\nabla g)\cdot v\, dx + \sum_{j=1}^n \int_{\Gamma_j} \phi_j\cdot v\, dx\quad
  \mbox{for all }v\in V_{-\beta,-\delta}^{1,s'}({\cal K})^3, \ S_j v=0 \mbox{ on }\Gamma_j,\ j=1,\ldots,n,
\]
where $f \in W_{\beta+1,\delta+1}^{0,s}({\cal K})^3$, $\phi_j \in W_{\beta+1,\delta+1}^{1-1/s}(\Gamma_j)$.

Let $\kappa$ be a fixed real number such that the closed strip between the lines
$\mbox{Re}\, \lambda = -\kappa-1/2$ and $\mbox{Re}\, \lambda=1-\beta-3/s$ is free of eigenvalues of the
pencil ${\mathfrak A}$. Then, according to \cite[Th.4.5]{mr-03},
there exists a unique solution $G(x,\xi)=\big( G_{i,j}(x,\xi)\big)_{i,j=1}^4$ of the problem
(\ref{e1})--(\ref{e3})
such that the function $x\to \zeta(|x-\xi|/r(\xi))\, G_{i,j}(x,\xi)$ belongs to $W_{\kappa,0}^{1,2}({\cal K})$
for $i=1,2,3$ and to $W_{\kappa,0}^{0,2}({\cal K})$ for $i=4$, where $\zeta$ is an arbitrary smooth function
on $(0,\infty)$ equal to one in $(1,\infty)$ and to zero in $(0,\frac 12)$.
We denote by $\Lambda_- < \mbox{Re}\, \lambda < \Lambda_+$ the widest strip in the complex plane containing
the line $\mbox{Re}\, \lambda=-\kappa-1/2$ which is free of eigenvalues of the pencil ${\mathfrak A}(\lambda)$.

Suppose that $h_j=0$ for $j=1,\ldots,n$ and $F\in V_{\beta,\delta}^{-1,s}({\cal K})^3$ is given in the form
(\ref{g3}). Then, analogously to (\ref{e4}), (\ref{e5}), the following representation for the solution of problem
(\ref{g4}), (\ref{g5}) holds.
\begin{eqnarray} \label{g6}
&& \hspace{-2em} u_i(x) = \sum_{j=1}^3 \int_{\cal K} \Big( f^{(0)}_j(\xi)\, G_{i,j}(x,\xi) +
  \sum_{k=1}^3 f^{(k)}_j(\xi)\partial_{\xi_k}G_{i,j}(x,\xi)\Big)\, d\xi
  + \int_{\cal K} g(\xi)\, G_{i,4}(x,\xi)\, d\xi, \\
&& \hspace{-2em} p(x) = -g(x) + \sum_{j=1}^3 \int_{\cal K} \Big( f^{(0)}_j(\xi)\, G_{4,j}(x,\xi) +
  \sum_{k=1}^3 f^{(k)}_j(\xi)\partial_{\xi_k}G_{4,j}(x,\xi)\Big)\, d\xi
  + \int_{\cal K} g(\xi)\, G_{4,4}(x,\xi)\, d\xi.  \label{g7}
\end{eqnarray}

\subsection{Auxiliary inequalities}

Our goal is to prove that (\ref{g6}), (\ref{g7}) define a continuous mapping
\[
V_{\beta+1,\delta+1}^{0,s}({\cal K})^3 \times V_{\beta,\delta}^{0,s}({\cal K})^9 \times
  V_{\beta,\delta}^{0,s}({\cal K}) \ni \big( f^{(0)},f^{(1)},f^{(2)},f^{(3)},g\big) \to
  (u,p) \in W_{\beta,\delta}^{1,s}({\cal K})^3\times W_{\beta,\delta}^{0,s}({\cal K})
\]
if $\delta$ satisfies condition (\ref{g2}) and $\beta$ satisfies the inequalities
\begin{equation} \label{g8}
\Lambda_- < 1-\beta-3/s < \Lambda_+ \, .
\end{equation}
The following lemmas allow us to estimate the integrals containing $f^{(k)}$, $k=1,2,3$,
and $g$ in (\ref{g6}) and (\ref{g7}).

\begin{Le} \label{gl1}
 Let $\zeta_k$ be the same function as in the proof of Lemma {\em \ref{bl1}} and let
 \[
 v(x) = \zeta_m(x) \, \int_{\cal K} \zeta_l(\xi)\, f(\xi)\, K(x,\xi)\, d\xi.
 \]
Suppose that $m\ge l+3$, $f\in W_{\beta,\delta}^{0,s}({\cal K})$, and
\begin{equation} \label{1gl1}
\big| \partial_x^\alpha K(x,\xi)\big|  \le  c\, \frac{|x|^{\Lambda_- -\sigma-|\alpha|+\varepsilon}}
  {|\xi|^{\Lambda_- +2+\varepsilon}}
  \prod_{k=1}^n \Big( \frac{r_k(x)}{|x|}\Big)^{\min(0,\mu_k-\sigma-|\alpha|-\varepsilon)}
  \prod_{k=1}^n \Big( \frac{r_k(\xi)}{|\xi|}\Big)^{\min(0,\mu_k-1-\varepsilon)}
\end{equation}
for $|x|>2|\xi|$, $|\alpha|\le 1-\sigma$, where $\sigma\in \{0,1\}$ and $\varepsilon$ is
a sufficiently small positive real number. If $\delta$ and $\beta$ satisfy conditions
{\em (\ref{g2})} and {\em(\ref{g8})}, then
\[
\| v\|_{W_{\beta,\delta}^{1-\sigma,s}({\cal K})} \le c\, 2^{-|m-l|\varsigma} \| \zeta_l f
  \|_{W_{\beta,\delta}^{0,s}({\cal K})}
\]
with positive constants $c$ and $\varsigma$ independent of $f$. The same estimates holds if $l\ge m+3$ and
\[
\big| \partial_x^\alpha K(x,\xi)\big|  \le  c\, \frac{|x|^{\Lambda_+ -\sigma-|\alpha|-\varepsilon}}
  {|\xi|^{\Lambda_+ +2-\varepsilon}}
  \prod_{k=1}^n \Big( \frac{r_k(x)}{|x|}\Big)^{\min(0,\mu_k-\sigma-|\alpha|-\varepsilon)}
  \prod_{k=1}^n \Big( \frac{r_k(\xi)}{|\xi|}\Big)^{\min(0,\mu_k-1-\varepsilon)}
\]
for $|\xi|>2|x|$, $|\alpha|\le 1-\sigma$.
\end{Le}

The proof of Lemma \ref{gl1} proceeds analogously to Lemma \ref{el1}.
Note that the elements $G_{i,4}(x,\xi)$ of Green's matrix and the derivatives $\partial_{\xi_k}G_{i,j}(x,\xi)$,
$j=1,2,3$, satisfy the assumptions on the kernel $K(x,\xi)$ (with $\sigma=\delta_{i,4}$) of Lemma \ref{gl1}
and of the following lemma.

\begin{Le} \label{gl2}
Let $\zeta_k$ be the same function as in the proof of Lemma {\em \ref{bl1}} and let
\[
v(x) = \zeta_m(x) \, \int_{\cal K} \zeta_l(\xi)\, f(\xi)\, \chi^-(x,\xi)\, K(x,\xi)\, d\xi,
\]
where $|l-m|\le 2$, $\chi^-$ is defined by {\em (\ref{e7})}, and $f\in W_{\beta-\tau,\delta-\tau}^{0,s}({\cal K})$.
Suppose that
\begin{equation} \label{1gl2}
\big| \partial_x^\alpha K(x,\xi)\big|  \le  c\, |x-\xi|^{-2-\sigma-|\alpha|}
  \Big( \frac{r(x)}{|x-\xi|}\Big)^{\min(0,\mu_x-\sigma-|\alpha|-\varepsilon)}
  \Big( \frac{r(\xi)}{|x-\xi|}\Big)^{\min(0,\mu_\xi-1-\varepsilon)}
\end{equation}
for $|x|/32 < |\xi| < 32|x|$, $|x-\xi|>r(x)/4$, $|\alpha|\le 1-\sigma$, where $\sigma\in \{0,1\}$,
$\varepsilon$ is a sufficiently small positive real number.
If $\delta$ satisfies condition {\em (\ref{g2})}, then
\[
\| v\|_{W_{\beta,\delta}^{1-\sigma,s}({\cal K})} \le c\,  \| \zeta_l f
  \|_{W_{\beta,\delta}^{0,s}({\cal K})}\, .
\]
\end{Le}

{\it Proof}:
Let $|\alpha|\le 1-\sigma$. Obviously, $\displaystyle \big|\partial_x^\alpha v(x)\big|
\le c\, \sum_{j+|\gamma|=|\alpha|} |x|^{-j} A_{\gamma}(x),$ where $A_\gamma$ satisfies the inequality
(\ref{2el3}) with $\tau=1$ provided
\begin{equation} \label{2gl2}
\max(1-\sigma-|\gamma|,1-\mu_k) < s' s_k < \min(2,1+\mu_k), \quad s'=s/(s-1).
\end{equation}
If additionally
\begin{equation} \label{3gl2}
\delta_k+t_k + \frac 1s \, \max(0,1-\mu_k) < s_k < \delta_k + t_k + \frac 1s \, \min(1+\sigma+|\gamma|,1+\mu_k),
\end{equation}
where
\[
t_k = \left\{ \begin{array}{ll} 1-\max(0,1-\mu_k) & \mbox{for }\sigma=|\gamma|=0, \\
    0 & \mbox{for } \sigma+|\gamma|=1,  \end{array} \right.
\]
then, using Corollory \ref{ec1}, we obtain
\begin{eqnarray*}
&& \hspace{-1em}\int_{\cal K} |x|^{s(\beta-1+\sigma+|\gamma|)} \prod_{k=1}^n \Big( \frac{r_k(x)}{|x|}\Big)^{s\delta_k}
 \big| A_\gamma(x)\big|^s\, dx \\
&& \hspace{-1em} \le c\int\limits_{\substack{{\cal K}\\ 2^{m-1}<|x|<2^{m+1}}}
 |x|^{s\beta-1+\sigma+|\gamma|} \prod_{k=1}^n \Big( \frac{r_k(x)}{|x|}
 \Big)^{s(\delta_k-s_k)+(s-1)(1-\sigma-|\gamma|)}
  \int\limits_{\substack{{\cal K}\\ |x-\xi|>r(x)/4}} |x-\xi|^{-2-\sigma-|\gamma|} \\
&&\qquad \times \Big( \frac{r(x)}{|x-\xi|} \Big)^{\min(0,\mu_x-\sigma-|\gamma|-\varepsilon)}
  \Big( \frac{r(\xi)}{|x-\xi|}\Big)^{\min(0,\mu_\xi-1-\varepsilon)}
  \prod_{k=1}^n \Big( \frac{r_k(\xi)}{|\xi|}\Big)^{ss_k} |\zeta_l f|^s\, d\xi\, dx \\
&&\hspace{-1em} \le c \int\limits_{\cal K} |\xi|^{s\beta-1+\sigma+|\gamma|} \prod_{k=1}^n \Big(
  \frac{r_k(\xi)}{|\xi|}\Big)^{ss_k}\, \big| \zeta_l(\xi) f(\xi)\big|^s \
  \int\limits_{\substack{2^{m-1}< |x|< 2^{m+1}\\ |x-\xi|>r(\xi)/5}} |x-\xi|^{-2-\sigma-|\gamma|} \\
&& \quad\times \Big( \frac{r(x)}{|x-\xi|} \Big)^{\min(0,\mu_x-\sigma-|\gamma|-\varepsilon)}
  \Big( \frac{r(\xi)}{|x-\xi|}\Big)^{\min(0,\mu_\xi-1-\varepsilon)}
  \prod_{k=1}^n \Big( \frac{r_k(x)}{|x|}\Big)^{s(\delta_k-s_k+t_k)-1+\sigma+|\gamma|}\, dx\, d\xi \\
&& \hspace{-1em} \le c\, \int_{\cal K} |\xi|^{s\beta}\ \prod_{k=1}^n \Big( \frac{r_k(x)}{|x|}
  \Big)^{s(\delta_k+t_k)}\, \big| \zeta_l(\xi)f(\xi)\big|^s\, d\xi.
\end{eqnarray*}
It can be easily verified that for $\sigma+|\gamma|\le 1$ there exist real numbers $s_k$ satisfying
(\ref{2gl2}), (\ref{3gl2}). This proves the lemma. \rule{1ex}{1ex} \\

The assumptions on $K(x,\xi)$ in the following lemma are satisfied by $\partial_{\xi_k} G_{4,j}$, $j=1,2,3$,
and $G_{4,4}$.

\begin{Le} \label{gl3}
Let
\[
v(x) = \zeta_m(x) \, \int_{\cal K} \zeta_l(\xi)\, f(\xi)\, \chi^+(x,\xi)\, K(x,\xi)\, d\xi,\quad
\]
where $|l-m|\le 2$, $f\in V_{\beta,\delta}^{0,s}({\cal K})$, and $\chi^+$ is defined by {\em (\ref{e7})}.
Suppose that $K(x,\xi)$ has the representation $K(x,\xi)= -\nabla_x \cdot P(x,\xi) + Q(x,\xi)$, where
\begin{equation} \label{1gl3}
| P(x,\xi) | \le c\, |x-\xi|^{-2},\quad  | Q(x,\xi)|\le c\, r(\xi)^{-3}
\end{equation}
for $|x|/32 < |\xi| <32|x|$, $|x-\xi|< r(x)/2$, $|\alpha|\le 1-\sigma$, and $P(x,\xi)\cdot n^{(j)}=0$ for
$x \in \Gamma_j$, $j=1,\ldots,n$ (here $n^{(j)}$ denotes the exterior normal to $\Gamma_j$), then
\begin{equation} \label{2gl3}
\| v\|_{V_{\beta-1,\delta-1}^{-1,s}({\cal K})} \le c\,  \| \zeta_l f
  \|_{V_{\beta,\delta}^{0,s}({\cal K})}\, .
\end{equation}
with a constant $c$ independent of $l,m$ and $f$.
\end{Le}

{\it Proof}:
Let $w \in V_{1-\beta,1-\delta}^{1,s'}({\cal K})$, $s'=s/(s-1)$. Then
\begin{equation} \label{3gl3}
\int_{\cal K} v(x)\, w(x)\, dx = \int_{\cal K} A(x)\, w(x) + B(x)\cdot \nabla w(x)\, dx,
\end{equation}
where
\begin{eqnarray*}
&& A(x) = \int_{\cal K} \Big( P(x,\xi)\, \nabla_x\big( \zeta_m(x)\, \chi^+(x,\xi)\big)
  + \zeta_m(x)\, \chi^+(x,\xi)\, Q(x,\xi)\Big)\, \zeta_l(\xi)\, f(\xi)\, d\xi,\\
&& B(x) = \int_{\cal K}  \zeta_m(x)\, \chi^+(x,\xi)\, P(x,\xi)\, \zeta_l(\xi)\, f(\xi)\, d\xi.
\end{eqnarray*}
We have to show that $A\in V_{\beta,\delta}^{0,s}({\cal K})$, $B\in V_{\beta-1,\delta-1}^{0,s}({\cal K})^3$, and
\begin{equation} \label{4gl3}
\| A\|_{V_{\beta,\delta}^{0,s}({\cal K})} + \| B\|_{V_{\beta-1,\delta-1}^{0,s}({\cal K})^3} \le
  c\, \| \zeta_l f\|_{V_{\beta,\delta}^{0,s}({\cal K})}\, .
\end{equation}
We introduce the set ${\cal K}_x=\{ \xi\in {\cal K}:\, |x|/32 < |\xi| <32|x|,\  |x-\xi|< r(x)/2\}$.
Note that there are the inequalities (\ref{3el4}) for $x\in \mbox{supp}\, \zeta_m$, $\xi\in \mbox{supp}\,
\zeta_l$, and $|x-\xi|< r(x)/2$.
Since
\[
\big| P(x,\xi)\, \nabla_x\big( \zeta_m(x)\, \chi^+(x,\xi)\big| + \big| \zeta_m(x)\, \chi^+(x,\xi)\,
  Q(x,\xi)\big| \le c\, r(x)^{-1}\, |x-\xi|^{-2}
\]
for $\xi\in \mbox{supp}\, \zeta_l$, we obtain
\begin{eqnarray*}
|A(x)|^s & \le & c\, r(x)^{-s} \Big(\int_{{\cal K}_x} |x-\xi|^{-2}\, \big| \zeta_l(\xi)\, f(\xi)\big|\, d\xi\Big)^s\\
& \le & c\, r(x)^{-s}\int_{{\cal K}_x} |x-\xi|^{-2}\, \big| \zeta_l(\xi)\, f(\xi)\big|^s\, d\xi\
  \Big( \int\limits_{|x-\xi|<r(x)/2} |x-\xi|^{-2}\, d\xi\Big)^{s-1}\\
& \le & c\, r(x)^{-1} \int_{{\cal K}_x} |x-\xi|^{-2}\, \big| \zeta_l(\xi)\, f(\xi)\big|^s\, d\xi.
\end{eqnarray*}
Consequently,
\begin{eqnarray*}
&& \| A\|^s_{V_{\beta,\delta}^{0,s}({\cal K})} \le c\, \int_{\cal K} |x|^{s\beta}\prod_{k=1}^n
  \Big( \frac{r_k(x)}{|x|}\Big)^{s\delta_k}\, r(x)^{-1}\Big( \int_{{\cal K}_x} |x-\xi|^{-2}\, \big|
  \zeta_l(\xi)\, f(\xi)\big|^s\, d\xi\Big)\, dx\, \\
&& \le c\, \int_{\cal K} |\xi|^{s\beta}\prod_{k=1}^n \Big( \frac{r_k(\xi)}{|\xi|}\Big)^{s\delta_k}\,
  r(\xi)^{-1}\, \big| \zeta_l(\xi)\, f(\xi)\big|^s\,\Big( \int\limits_{|x-\xi|<r(\xi)} |x-\xi|^{-2}\, dx\Big)\, d\xi
  \le c\, \| \zeta_l f\|^s_{V_{\beta,\delta}^{0,s}({\cal K})}\, .
\end{eqnarray*}
Analogously, we obtain
\[
|B(x)|^s \le c\, r(x)^{s-1} \int_{{\cal K}_x} |x-\xi|^{-2}\, \big| \zeta_l(\xi)\, f(\xi)\big|^s\, d\xi
\]
what implies the desired estimate for the norm of $B$ in $V_{\beta-1,\delta-1}^{0,s}({\cal K})^3$.
Estimate (\ref{2gl3}) is an immediate consequence of (\ref{3gl3}) and (\ref{4gl3}). \rule{1ex}{1ex}

\subsection{Existence und uniqueness of weak solutions}

Let $F \in V_{\beta,\delta}^{-1,s}({\cal K})^3$, $g\in W_{\beta,\delta}^{0,s}({\cal K})$ and
$h_k\in W_{\beta,\delta}^{1-1/s,s}(\Gamma_k)^{3-d_k}$, $k=1,\ldots,n$. We suppose that the following conditions
are satisfied.
\begin{itemize}
\item[(i)] The line $\mbox{Re}\, \lambda = 1-\beta-3/s$ does not contain eigenvalues of the pencil
  ${\mathfrak A}(\lambda)$.
\item[(ii)] $\max(0,1-\mu_k)<\delta_k+2/s < 1$ \ for $k=1,\ldots,n$.
\item[(iii)] The vector functions $h_j\in W_{\beta,\delta}^{1-1/s,s}(\Gamma_j)^{3-d_j}$ are such that
there exists a vector function $w\in W_{\beta,\delta}^{1,s}({\cal K})^3$ satisfying the condition
$S_j u=h_j$ on $\Gamma_j$, $j=1,\ldots,n$.
\end{itemize}
The last is a condition on the traces of the boundary data $h_j$ on the edges of ${\cal K}$. These traces
exist since $\delta_k+2/s<1$. Let $\Gamma_{j_+}$, $\Gamma_{j_-}$ be the sides of ${\cal K}$ adjacent
to the edge $M_j$. Then condition (iii) is equivalent to (\ref{compat1}).

\begin{Th} \label{gt1}
Let $F\in V_{\beta,\delta}^{-1,s}({\cal K})^3$, $g\in W_{\beta,\delta}^{0,s}({\cal K})$, and
$h_j \in W_{\beta,\delta}^{1-1/s,s}(\Gamma_j)^{3-d_j}$. Suppose that conditions {\em (i)--(iii)} are satisfied.
Then there exists a unique solution $(u,p)\in W_{\beta,\delta}^{1,s}({\cal K})^3 \times W_{\beta,\delta}^{0,s}
{\cal K})$ of problem {\em (\ref{g4}), (\ref{g5})}.
\end{Th}

{\it Proof}:
Without loss of generality, we may assume that $h_j=0$. Suppose that the functional $F$ is given in the form
(\ref{g3}), where $f^{(0)}=0$ and $f^{(k)}\in V_{\beta,\delta}^{0,s}({\cal K})^3$ for $k=1,2,3$.
We consider the operator
\[
{\cal X}\stackrel{def}{=} \Big(\prod_{k=1}^3 V_{\beta,\delta}^{0,s}({\cal K})^3\Big) \times
  V_{\beta,\delta}^{0,s}({\cal K}) \ni (f^{(1)},f^{(2)},f^{(3)},g)   \to {\cal O}(f^{(1)},f^{(2)},f^{(3)},g)=(u,p),
\]
where $u$ and $p$ are defined by (\ref{g6}), (\ref{g7}) (with $f^{(0)}=0$) and $G_{i,j}$ are the elements of
Green's matrix introduced in Section 3.4. Then by Lemma \ref{gl1},
\begin{equation} \label{1gt1}
\| \zeta_m {\cal O} \zeta_l(f^{(1)},f^{(2)},f^{(3)},g) \|_{\cal Y} \le c\, 2^{-|l-m|\varsigma}
  \, \|\zeta_l(f^{(1)},f^{(2)},f^{(3)},g)\|_{\cal X}
\end{equation}
for $|l-m|\ge 3$, where
\[
{\cal Y} \stackrel{def}{=} W_{\beta,\delta}^{1,s}({\cal K})^3\times  W_{\beta,\delta}^{0,s}({\cal K})
\]
and $c,\varsigma$ are positive constants independent of $f^{(k)},g,l,m$. In order to prove the same inequality for
$|l-m|\le 2$, we introduce the functions
\begin{eqnarray*}
&& u_i^\pm(x) = \sum_{j=1}^3 \sum_{k=1}^3 \int_{\cal K} \zeta_l(\xi)\, f^{(k)}_j(\xi)\, \chi^\pm(x,\xi)\,
  \partial_{\xi_k} G_{i,j}(x,\xi)\, d\xi  + \int_{\cal K} \zeta_l(\xi) g(\xi)\, \chi^\pm(x,\xi)\,
  G_{i,4}(x,\xi)\, d\xi,\\
&& p^\pm(x) = -\zeta_l(x)\, g(x) + \sum_{j=1}^3 \sum_{k=1}^3 \int_{\cal K} \zeta_l(\xi)\, f^{(k)}_j(\xi)\,
  \chi^\pm(x,\xi)\, \partial_{\xi_k} G_{4,j}(x,\xi)\, d\xi \\
&& \qquad\qquad  + \int_{\cal K} \zeta_l(\xi)\, g(\xi)\, \chi^\pm(x,\xi)G_{4,4}(x,\xi)\, d\xi,
\end{eqnarray*}
where $\chi^+$ and $\chi^-$ are defined by (\ref{e7}). Then
\[
b(u^+ +u^-,v) -\int_{\cal K} (p^+ +p^-)\nabla\cdot v\, dx =F(\zeta_l v) \quad \mbox{ for all }
  v \in V_{-\beta,-\delta}^{1,s'}({\cal K})^3,\ S_k v=0\ \mbox{ on }\Gamma_k,
\]
$-\nabla\cdot (u^+ +u^-)= \zeta_l g$ in ${\cal K}$, and $S_j (u^++u^-)=0$ on $\Gamma_j$.
Furthermore, by Lemma \ref{gl2}, we have $\zeta_m (u^-,p^-)\in {\cal Y}$ and
\begin{equation} \label{2gt1}
\| \zeta_m (u^-,p^-)\|_{\cal Y} \le c\, \|\zeta_l(f^{(1)},f^{(2)},f^{(3)},g)\|_{\cal X}
\end{equation}
if $|l-m|\le 3$, where $c$ is independent of $f,g,l,m$. From Lemmas \ref{el4} and \ref{gl3}
it follows that
\begin{equation} \label{3gt1}
\| \zeta_m u^+ \|_{V_{\beta-1,\delta-1}^{0,s}({\cal K})^3}
  + \| \zeta_m p^+ \|_{V_{\beta-1,\delta-1}^{-1,s}({\cal K})}
  \le c\, \|\zeta_l(f^{(1)},f^{(2)},f^{(3)},g)\|_{\cal X}
\end{equation}
The vector function $(u^+,p^+)$ is a solution of the problem
\begin{eqnarray*}
&& b(u^+,v) - \int_{\cal K} p^+\, \nabla\cdot v\, dx =  \tilde{F}(v) \quad \mbox{ for all }
  v \in V_{-\beta,-\delta}^{1,s'}({\cal K})^3,\ S_k v=0\ \mbox{ on }\Gamma_k,\ k=1,\ldots,n, \\
&& -\nabla\cdot u^+ = \zeta_l g + \nabla\cdot u^-\ \mbox{ in }{\cal K}, \quad S_j u^+ =-S_j u^-\
  \mbox{ on }\Gamma_j,\ j=1,\ldots,n,
\end{eqnarray*}
where
\[
\tilde{F}(v) = F(\zeta_l v) - b(u^-,v)+ \int_{\cal K} p^{-}\, \nabla\cdot v\, dx
\]
Obviously, $\zeta_m \tilde F \in V_{\beta,\delta}^{-1,s}({\cal K}$.
Let $\eta_m=\zeta_{m-1}+\zeta_m+\zeta_{m+1}$.
Then, by an estimate analogous to that in Corollary \ref{dc1},
\begin{eqnarray*}
&&\| \zeta_m u^+\|_{V_{\beta,\delta}^{1,s}({\cal K})^3}+ \|\zeta_m p^+\|_{V_{\beta,\delta}^{0,s}({\cal K})} \\
&& \le c\, \Big( \| \eta_m u^+\|_{V_{\beta-1,\delta-1}^{0,s}({\cal K})^3} + \|\eta_m p^+
  \|_{V_{\beta-1,\delta-1}^{-1,s}({\cal K})} + \| \eta_m \zeta_l(f^{(1)},f^{(2)},f^{(3)},g)\|_{\cal X} +
  \| \eta_m (u^-,p^-)\|_{\cal Y}\Big) \\
\end{eqnarray*}
for $|l-m|\le 2$. Due to (\ref{2gt1}) and (\ref{3gt1}), the right hand side of the last inequality can be estimated by
the norm of $\zeta_l(f^{(1)},f^{(2)},f^{(3)},g)$ in ${\cal X}$. Consequently,
\[
\|\zeta_m(u^++u^-,p^++p^-)\|_{\cal Y} \le c\, \| \zeta_l(f^{(1)},f^{(2)},f^{(3)},g)\|_{\cal X}\quad\mbox{for }|l-m|\le 2.
\]
Thus, estimate (\ref{1gt1}) is valid for arbitrary $l$ and $m$. Consequently, by Lemma \ref{fl1}, the operator
${\cal O}$ continuously maps ${\cal X}$ into ${\cal Y}$. This proves the existence of a solution of
problem (\ref{g4}), (\ref{g5}) in the case when $F$ has the form (\ref{g3}) with $f^{(0)}=0$,
$f^{(k)}\in V_{\beta,\delta}^{0,s}({\cal K})^3$ for $k=1,2,3$.

By what has been shown in the previous section, the mapping $f^{(0)} \to (v,q)$ defined by
\begin{equation} \label{g6a}
v_i(x) = \sum_{j=1}^3 \int_{\cal K} f^{(0)}_j(\xi)\, G_{i,j}(x,\xi) \, d\xi, \ i=1,2,3,\quad
   p(x) = -g(x) + \sum_{j=1}^3 \int_{\cal K}  f^{(0)}_j(\xi)\, G_{4,j}(x,\xi) \, d\xi
\end{equation}
is continuous from $V_{\beta+1,\delta+1}^{0,s}({\cal K})^3$ into the subspace
$W_{\beta+1,\delta+1}^{2,s}({\cal K})^3\times W_{\beta+1,\delta+1}^{1,s}({\cal K})$ of
${\cal Y}$. Thus, problem (\ref{g4}), (\ref{g5}) is solvable for arbitrary
$F\in V_{\beta,\delta}^{-1,s}({\cal K})$.

It remains to prove the uniqueness of the solution. Suppose $(u,p) \in {\cal Y}$ is a solution
of the homogeneous problem (\ref{g4}), (\ref{g5}). Then $(u,p)$ is also a solution of the homogeneous
problem (\ref{d1}), (\ref{d2}), and from Lemma \ref{dl4} it follows that $u \in W_{\beta+1,\delta+1}^{2,s}
({\cal K})^3$ and $p\in W_{\beta+1,\delta+1}^{1,s}({\cal K})$. Consequently, by Theorem \ref{ft2},
we have $u=0$ and $p=0$. The proof is complete. \rule{1ex}{1ex}

\begin{Rem} \label{gr1}
{\em In \cite{mr-03} the existence and uniqueness of weak solutions in $W_{\beta,0}^{1,2}({\cal K})^3
\times W_{\beta,0}^{0,2}({\cal K})$ was proved for arbitrary $F \in W_{\beta,0}^{-1,2}({\cal K})^3
=(W_{-\beta,0}^{1,2}({\cal K})^*)^3$, $g\in W_{\beta,0}^{0,2}({\cal K})$ and $h_j\in W_{\beta,\delta}^{1/2,2}
(\Gamma_j)$, $j=1,\ldots,n$. Note that every $F\in W_{\beta,0}^{-1,2}({\cal K})$ has the form
\[
F(v)=\int_{\cal K} f^{(0)}\cdot v\, dx + \sum_{k=1}^3 \int_{\cal K} f^{(k)}\, \partial_{x_k}v\cdot dx \
  \mbox{ for all }  v\in W_{-\beta,0}^{1,2}({\cal K})^3,
\]
where $f\in W_{\beta+1,0}^{0,2}({\cal K})$, $f^{(k)} \in W_{\beta,0}^{0,2}({\cal K})$, $k=1,2,3$.
It can be easily shown that the assertions of Lemmas \ref{gl1}--\ref{gl3} are also valid if $s=2$, $\delta=0$,
and $\beta$ satisfies (\ref{g8}). Consequently, the weak solution $(u,p)\in W_{\beta,0}^{1,2}({\cal K})^3
\times W_{\beta,0}^{0,2}({\cal K})$ has also the form (\ref{g6}), (\ref{g7}) if $h_j=0$.}
\end{Rem}

\subsection{Regularity assertions for weak solutions}

\begin{Le} \label{hl1}
Let $(u,p) \in W_{\beta',\delta'}^{1,\sigma}({\cal K})^3\times W_{\beta',\delta'}^{0,\sigma}({\cal K})$
be a solution of problem
\begin{eqnarray} \label{h1}
&& b(u,v) - \int_{\cal K} p\, \nabla\cdot v = F(v)\ \mbox{ for all }v\in V_{-\beta',-\delta'}^{1,\sigma'}
  ({\cal K})^3, \ S_j v=0 \mbox{ on }\Gamma_j,\ j=1,\ldots,n,\\ \label{h2}
&& -\nabla\cdot u = g \ \mbox{ in }{\cal K},\quad S_j u=h_j\ \mbox{ on }\Gamma_j,\ j=1,\ldots,n,
\end{eqnarray}
where $\sigma'=\sigma/(\sigma-1)$,
\[
F\in V_{\beta,\delta}^{-1,s}({\cal K})^3\cap V_{\beta',\delta'}^{-1,\sigma}({\cal K}), \quad
g\in W_{\beta,\delta}^{0,s}({\cal K})\cap W_{\beta',\delta'}^{0,\sigma}({\cal K}),\quad
h_j\in W_{\beta,\delta}^{1-1/s,s}(\Gamma_j)^{3-d_j} \cap W_{\beta',\delta'}^{1-1/\sigma,\sigma}(\Gamma_j)^{3-d_j}.
\]
If the closed strip between the lines $\mbox{\em Re}\,
\lambda =1-\beta-3/s$ and $\mbox{\em Re}\, \lambda =1-\beta'-3/\sigma$ is free of eigenvalues
of the pencil ${\mathfrak A}(\lambda)$ and $\delta,\delta'$ satisfy the inequalities
\begin{equation} \label{h3}
\max(0,1-\mu_k)<\delta_k+2/s<1, \quad \max(0,1-\mu_k)<\delta'_k+2/\sigma<1 \ \mbox{ for }k=1,\ldots,n
\end{equation}
(in the case $\sigma=2$ it is allowed that $\delta'_k=1$), then $(u,p)\in W_{\beta,\delta}^{1,s}({\cal K})^3\times W_{\beta,\delta}^{0,s}({\cal K})$.
\end{Le}

{\it Proof}: Under the assumptions on the lemma, the boundary data $h_j$ satisfy the
compatibility condition (\ref{compat1}). Therefore, there exists a vector function $v\in
W_{\beta,\delta}^{1,s}({\cal K})^3 \cap W_{\beta',\delta'}^{1,\sigma}({\cal K})^3$
satisfying $S_j v=h_j$ on $\Gamma_j$, $j=1,\ldots,n$. For this reason, we may restrict
ourselves to the case $h_j=0$. According to Theorem \ref{gt1} (see also Remark \ref{gr1}), there are unique solutions
of problem (\ref{g4}), (\ref{g5}) in $W_{\beta,\delta}^{1,s}({\cal K})^3\times
W_{\beta,\delta}^{0,s}({\cal K})$ and $W_{\beta',\delta'}^{1,s'}({\cal K})^3\times
W_{\beta',\delta'}^{0,s'}({\cal K})$. Both solutions are given by (\ref{g6}), (\ref{g7})
with the same Green matrix $G(x,\xi)$. This proves the lemma. \rule{1ex}{1ex} \\

The same result is true for weak solutions in $W_{\beta',0}^{1,2}({\cal K})^3\times
W_{\beta',0}^{0,2}({\cal K})$ (cf. Remark \ref{gr1}). Furthermore, the following generalization
of Lemma \ref{hl1} holds.

\begin{Th} \label{ht1}
Let $u,p,F,g,$ and $h_j$ be as in Lemma {\em \ref{hl1}}. We assume that there are no eigenvalues
of the pencil ${\mathfrak A}(\lambda)$ on the lines $\mbox{\em Re}\, \lambda =1-\beta-3/s$ and
$\mbox{\em Re}\, \lambda =1-\beta'-3/\sigma$ and that $\delta,\delta'$ satisfy the inequalities
{\em (\ref{h3})}. Then $(u,p)$ admits the decomposition
\begin{equation} \label{h4}
(u,p) =  \sum_{\nu=1}^N \sum_{j=1}^{I_\nu} \sum_{s=0}^{\kappa_{\nu,j}-1}
  c_{\nu,j,s} \sum_{\sigma=0}^s \frac{1}{\sigma!}\ (\log \rho)^\sigma\, \big( \rho^{\lambda_\nu}
  u^{(\nu,j,s-\sigma)}(\omega), \rho^{\lambda_\nu-1} p^{(\nu,j,s-\sigma)}(\omega)\big)+ (w,q)
\end{equation}
where $(w,q) \in W_{\beta,\delta}^{1,s}({\cal K})^3\times W_{\beta,\delta}^{0,s}({\cal K})$
is a weak solution of problem {\em (\ref{d1})--(\ref{d2})}, $\lambda_\nu$ are the eigenvalues
of the pencil ${\mathfrak A}$ between the lines $\mbox{\em Re}\, \lambda =1-\beta-3/s$ and
$\mbox{\em Re}\, \lambda =1-\beta'-3/\sigma$, and $\big( u^{(\nu,j,s)},p^{(\nu,j,s)}\big)$ are eigenvectors
and generalized eigenvectors corresponding to the eigenvalue $\lambda_\nu$.
\end{Th}

{\it Proof}: As in the proof of Lemma \ref{hl1}, we may restrict ourselves to the case $h_j=0$.
Let $\{ F_i\}\subset C_0^\infty(\bar{\cal K}\backslash \{ 0\})^3$, $\{ g_i\}\subset
C_0^\infty(\bar{\cal K}\backslash \{ 0\})$ be sequences converging to $F$ in
$V_{\beta,\delta}^{-1,s}({\cal K})^3\cap V_{\beta',\delta'}^{-1,\sigma}({\cal K})$ and
$g$ in $W_{\beta,\delta}^{0,s}({\cal K})\cap W_{\beta',\delta'}^{0,\sigma}({\cal K})$, respectively.
According to \cite[Th.3.2]{mr-03}, there exist unique solutions
\[
(u^{(i)},p^{(i)}) \in W_{\beta'-3/2+3/\sigma,0}^{1,2}({\cal K})^3\times
    W_{\beta'-3/2+3/\sigma,0}^{0,2}({\cal K}) \mbox{ and }
(w^{(i)},q^{(i)}) \in W_{\beta-3/2+3/s,0}^{1,2}({\cal K})^3\times
    W_{\beta-3/2+3/s,0}^{0,2}({\cal K})
\]
of the problem
\begin{eqnarray*}
&& b(u,v) - \int_{\cal K} p\, \nabla\cdot v = F_i(v)\ \mbox{ for all }v\in
  C_0^\infty(\bar{\cal K}\backslash\{ 0\})^3, \ S_j v=0 \mbox{ on }\Gamma_j,\ j=1,\ldots,n,\\
&& -\nabla\cdot u = g_i \ \mbox{ in }{\cal K},\quad S_j u=0\ \mbox{ on }\Gamma_j,\ j=1,\ldots,n.
\end{eqnarray*}
By what has been shown above, the vector functions $(u^{(i)},p^{(i)})$ and $(w^{(i)},q^{(i)})$ belong also
to $W_{\beta',\delta'}^{1,\sigma}({\cal K})^3\times W_{\beta',\delta'}^{0,\sigma}({\cal K})$ and
$W_{\beta,\delta}^{1,s}({\cal K})^3\times W_{\beta,\delta}^{0,s}({\cal K})$, respectively. Furthermore, from
Theorem \ref{gt1} it follows that the sequence $\{(u^{(i)},p^{(i)})\}$ converges to $(u,p)$, while
$\{ (w^{(i)},q^{(i)})\}$ converges to a vector function $(w,q) \in W_{\beta,\delta}^{1,s}({\cal K})^3\times
W_{\beta,\delta}^{0,s}({\cal K})$.
Let ${\cal X}$ denote the linear span of the vector functions
\[
\sum_{\sigma=0}^s \frac{1}{\sigma!}\ (\log \rho)^\sigma\, \big( \rho^{\lambda_\nu}
  u^{(\nu,j,s-\sigma)}(\omega), \rho^{\lambda_\nu-1} p^{(\nu,j,s-\sigma)}(\omega)\big)
\]
appearing in (\ref{h4}). By \cite[Th.4.4]{mr-03}, we have $(u^{(i)}-w^{(i)},p^{(i)}-q^{(i)})\in {\cal X}$
and, consequently, also $(u-w,p-q)\in {\cal X}$. This proves the theorem. \rule{1ex}{1ex}

\begin{Th} \label{ht2}
Let $(u,p) \in W_{\beta',\delta'}^{1,\sigma}({\cal K})^3\times W_{\beta',\delta'}^{0,\sigma}({\cal K})$
be a solution of problem {\em (\ref{h1}), (\ref{h2})}. We suppose that
$F\in V_{\beta',\delta'}^{-1,\sigma}({\cal K})^3$ and
\begin{equation} \label{1ht2}
F(v) = \int_{\cal K} f\cdot v\, dx + \sum_{j=1}^n \int_{\Gamma_j} \phi_j \cdot v\, dx
  \ \mbox{ for all } v\in C_0^\infty(\bar{\cal K}\backslash\{ 0\})^3,\, S_j v=0\
  \mbox{ on } \Gamma_j,\ j=1,\ldots,n,
\end{equation}
where $f\in W_{\beta,\delta}^{0,s}({\cal K})^3$, $\phi_j \in W_{\beta,\delta}^{1-1/s,s}(\Gamma_j)^{d_j}$
Furthermore, we assume that $g\in W_{\beta,\delta}^{0,s}({\cal K})\cap W_{\beta',\delta'}^{1,\sigma}({\cal K})$,
$h_j\in W_{\beta,\delta}^{1-1/s,s}(\Gamma_j)^{3-d_j} \cap W_{\beta',\delta'}^{2-1/\sigma,\sigma}(\Gamma_j)^{3-d_j}$,
there are no eigenvalues of the pencil ${\mathfrak A}(\lambda)$ in the closed strip between the lines
$\mbox{\em Re}\, \lambda =1-\beta'-3/\sigma$ and $\mbox{\em Re}\, \lambda =2-\beta-3/s$, $\delta$ and $\delta'$
satisfy the inequalities
\[
\max(0,1-\mu_k)>\delta'_k+2/\sigma<1, \quad \max(0,2-\mu_k)>\delta_k+2/s <2 \ \mbox{for } k=1,\ldots,n
\]
(in the case $\sigma=2$ it is allowed that $\delta'_k=1$), and $g,h_j,\phi_j$ satisfy the compatibility
condition {\em (iii)} of Section {\em 3.6}.
Then $u\in W_{\beta,\delta}^{2,s}({\cal K})^3$ and $p\in W_{\beta,\delta}^{1,s}({\cal K})$.
\end{Th}

{\it Proof}:
Suppose first that $\max(2-\mu_k,1)<\delta_k+2/s<2$ for $k=1,\ldots,n$. Then $W_{\beta,\delta}^{1-1/s,s}
(\Gamma_j)=V_{\beta,\delta}^{1-1/s,s}(\Gamma_j)$ and, therefore, the functional $F$ defined by
(\ref{1ht2}) belongs to $V_{\beta-1,\delta-1}^{-1,s}({\cal K})^3$. Using Theorem \ref{ht1},
we obtain $u\in W_{\beta-1,\delta-1}^{1,s}({\cal K})^3$ and $p\in W_{\beta-1,\delta-1}^{0,s}({\cal K})$,
and from the second part of Lemma \ref{dl4} we conclude that  $u\in W_{\beta,\delta}^{2,s}({\cal K})^3$,
$p\in W_{\beta,\delta}^{1,s}({\cal K})$.

If $\delta_k+2/s\le 1$ for at least one $k$, then in a first step we obtain
$u\in W_{\beta,\delta''}^{2,s}({\cal K})^3$ and $p\in W_{\beta,\delta''}^{1,s}({\cal K})$,
where $\delta''_k$ are arbitrary numbers satisfying $\max(2-\mu_k,1)<\delta''_k+2/s<2$ and
$\delta''_k\ge \delta_k$. Then Theorem \ref{ft3} implies the assertion of the theorem.
\rule{1ex}{1ex}

\begin{Le} \label{hl2}
Let $g\in W_{\beta,\delta}^{l-1,s}({\cal K})$, $h_j \in W_{\beta,\delta}^{l-1/s,s}(\Gamma_j)^{3-d_j}$,
$\phi_j \in W_{\beta,\delta}^{l-1-1/s,s}(\Gamma_j)^{d_j}$, $l\ge 3$, $-2/s < \delta_k\le 1-2/s$.
Suppose there exist $u\in W_{\beta-l+2,\varepsilon-2/s}^{2,s}({\cal K})^3$ and
$p \in W_{\beta-l+2,\varepsilon-2/s}^{1,s}({\cal K})$, $0< \varepsilon<1$, such that
\begin{equation} \label{1hl1}
\nabla \cdot u + g \in V_{\beta-l+2,\varepsilon-2/s}^{1,s}({\cal K}), \quad
  S_j u =h_j, \ \ N_j(u,p)=\phi_j \ \mbox{ on }\Gamma_j.
\end{equation}
Then there exist $v\in W_{\beta-l+2,\delta'}^{2,s}({\cal K})^3$, $q\in W_{\beta-l+2,\delta'}^{1,s}
({\cal K})$, where $\delta'=\delta$ if $l=3$, $\delta'_k = \varepsilon-2/s$ if $l\ge 4$, such that
\begin{equation} \label{2hl1}
\nabla \cdot v + (\rho\partial_\rho+1)g \in V_{\beta-l+2,\delta'}^{1,s}({\cal K}), \quad
  S_j v = \rho\partial_\rho h_j, \ \ N_j(v,q)=(\rho\partial_\rho+1)\phi_j \ \mbox{ on }\Gamma_j.
\end{equation}
\end{Le}

{\it Proof}: We prove the lemma for the Dirichlet problem. The proof for other boundary conditions
proceeds analogously. The existence of $u$ and $p$ satisfying (\ref{1hl1}) is equivalent to the trace
conditions in Lemma \ref{dl3}. We assume, without loss of generality, that $M_k$ coincides with the
$x_3$-axis. Then the trace conditions on $M_k$ have the form
\begin{equation} \label{3hl1}
h_{k_+}|_{M_k} = h_{k_-}|_{M_k},\quad n_{k_-}\cdot(\partial_r h_{k_+})|_{M_k}+ n_{k_+}\cdot
  (\partial_r h_{k_-})|_{M_k}= \big( g|_{M_k}+ \partial_{x_3}h_{3,k_+}|_{M_k}\big)\, \sin\theta_k.
\end{equation}
Here $\Gamma_{k_+}$ and $\Gamma_{k_-}$ are the sides adjacent to the edge $M_k$, $n_{k_+}$ and $n_{k_-}$
denote the exterior normals to these sides, $\theta_k$ is the inner angle at $M_k$, and
$h_{3,k_+}$ denotes the third component of the vector $h_{k_+}$.

Suppose first that $l\ge 4$ or $l=3$ and $\delta_k<1-2/s$. Then the traces of $\partial_{x_3}\partial_r
h_{k_\pm}$ and $\partial_{x_3}g$ on $M_k$ exist, and from (\ref{3hl1}) it follows that
$\partial_{x_3}h_{k_+}|_{M_k}= \partial_{x_3}h_{k_-}|_{M_k}$ and
\[
n_{k_-}\cdot (x_3\partial_{x_3}+1)\, (\partial_r h_{k_+})|_{M_k}+ n_{k_+}\cdot(x_3\partial_{x_3}+1)\,
  (\partial_r h_{k_-})|_{M_k} = (x_3\partial_{x_3}+1)\, \big( g|_{M_k}+ \partial_{x_3}h_{3,k_+}|_{M_k}\big)\,
  \sin\theta_k.
\]
Since $\rho\partial_\rho= r\partial_r+ x_3\partial_{x_3} = x_1\partial_{x_1} + x_2\partial_{x_2}
+x_3\partial_{x_3}$ and $x_1=x_2=0$ on $M_k$, from the last equalities it follows that
$(\rho \partial_\rho h_{k_+})|_{M_k}= (\rho \partial_\rho h_{k_-})|_{M_k}$ and
\[
n_{k_-}\cdot(\partial_r \rho \partial_\rho h_{k_+})|_{M_k}+ n_{k_+}\cdot
  (\partial_r \rho \partial_\rho h_{k_-})|_{M_k}= \big( (\rho \partial_\rho+1)g|_{M_k}+ \partial_{x_3}
  (\rho \partial_\rho h_{3,k_+})|_{M_k}\big)\, \sin\theta_k.
\]
This is the trace condition on $M_k$ for the existence of $v$ and $q$ satisfying (\ref{2hl1}). In the case
$l=3$, $\delta_k=1-2/s$ the validity of the trace condition can be proved analogously by means of
Lemma \ref{al5}. \rule{1ex}{1ex}

\begin{Th} \label{ht3}
Let $(u,p) \in W_{\beta',\delta'}^{1,\sigma}({\cal K})^3\times W_{\beta',\delta'}^{0,\sigma}({\cal K})$
be a solution of problem {\em (\ref{h1}), (\ref{h2})}. We suppose that
$F\in V_{\beta',\delta'}^{-1,\sigma}({\cal K})^3$ has the representation {\em (\ref{1ht2})},
where $f\in W_{\beta,\delta}^{l-2,s}({\cal K})^3$, $\phi_j \in W_{\beta,\delta}^{l-1-1/s,s}(\Gamma_j)^{d_j}$.
Furthermore, we assume that $g\in W_{\beta,\delta}^{l-1,s}({\cal K})\cap W_{\beta',\delta'}^{0,\sigma}({\cal K})$,
$h_j\in W_{\beta,\delta}^{l-1/s,s}(\Gamma_j)^{3-d_j} \cap W_{\beta',\delta'}^{1-1/\sigma,\sigma}(\Gamma_j)^{3-d_j}$,
there are no eigenvalues of the pencil ${\mathfrak A}(\lambda)$ in the closed strip between the lines
$\mbox{\em Re}\, \lambda =1-\beta'-3/\sigma$ and $\mbox{\em Re}\, \lambda =l-\beta-3/s$, $\delta$ and $\delta'$
satisfy the inequalities
\begin{equation} \label{h5}
\max(0,1-\mu_k)<\delta'_k+2/\sigma<1,\quad \max(0,l-\mu_k)<\delta_k+2/s<l, \ \mbox{ for }k=1,\ldots,n
\end{equation}
(in the case $\sigma=2$ it is allowed that $\delta'_k=1$),
and $g,h_j,\phi_j$ satisfy condition {\em (iii)} of Section {\em 3.6} with $\beta''=\beta-l+2$ and
$\delta''_k=\max(\delta_k-l+2,\frac 12 -\frac 2s)$ instead of $\beta$ and $\delta_k$, respectively.
Then $u\in W_{\beta,\delta}^{l,s}({\cal K})^3$ and $p\in W_{\beta,\delta}^{l-1,s}({\cal K})$.
\end{Th}

{\it Proof}:
By Theorem \ref{ht2}, the assertion of the theorem is true for $l=2$. We suppose that the assertion is true for
$l=m-1\ge 2$ and show that it is true for $l=m$.
Let first $\delta_k+2/s >1$ for $k=1,\ldots,n$. Then $W_{\beta,\delta}^{j,s}({\cal K}) \subset
W_{\beta-1,\delta-1}^{j-1,s}({\cal K})$, $W_{\beta,\delta}^{j+1-1/s,s}(\Gamma_k) \subset
W_{\beta,\delta}^{j-1/s,s}(\Gamma_k)$ for $j\ge 1$ and from the induction hypothesis it follows that
$u \in W_{\beta-1,\delta-1}^{l-1,s}({\cal K})^3$, $p \in W_{\beta-1,\delta-1}^{l-2,s}({\cal K})$.
Applying Lemma \ref{dl4}, we obtain $u\in W_{\beta,\delta}^{l,s}({\cal K})^3$ and
$p\in W_{\beta,\delta}^{l-1,s}({\cal K})$.

Suppose now that $\delta_k+2/s \le 1$ for all $k$. Then, in particular, $\mu_k>l-1$ for all $k$.
Since $W_{\beta,\delta}^{j,s}({\cal K}) \subset W_{\beta-1,\varepsilon-2/s}^{j-1,s}({\cal K})$ for $j\ge 1$,
$\varepsilon>0$, it follows from the induction hypothesis that
$u\in W_{\beta-1,\varepsilon-2/s}^{l-1,s}({\cal K})^3$ and $p\in W_{\beta-1,\varepsilon-2/s}^{l-2,s}
({\cal K})$. Using again Lemma \ref{dl4}, we conclude that $u\in W_{\beta,\varepsilon+1-2/s}^{l,s}
({\cal K})^3$ and $p\in W_{\beta,\varepsilon+1-2/s}^{l-1,s}({\cal K})$. Consequently,
$\rho\partial_\rho u \in W_{\beta-1,\varepsilon+1-2/s}^{l-1,s}({\cal K})^3$ and
$\rho\partial_\rho p\in W_{\beta-1,\varepsilon+1-2/s}^{l-2,s}({\cal K})$. Since the vector function
$(\rho\partial_\rho u, \rho\partial_\rho p + p)$ is a solution of the problem
\begin{eqnarray*}
&& -\Delta (\rho\partial_\rho u)+ \nabla(\rho\partial_\rho p +p) = (\rho\partial_\rho +2) f
  \in W_{\beta-1,\delta}^{l-3,s}({\cal K})^3, \quad
-\nabla\cdot (\rho\partial_\rho u) = (\rho\partial_\rho +1)g \in W_{\beta-1,\delta}^{l-2,s}({\cal K}), \\
&& S_j \rho\partial_\rho u =  \rho\partial_\rho h_j \in W_{\beta-1,\delta}^{l-1-1/s,s}(\Gamma_j), \quad
N_j(\rho\partial_\rho u,\rho\partial_\rho p+p)= (\rho\partial_\rho +1)\phi_j
  \in W_{\beta-1,\delta}^{l-2-1/s,s}(\Gamma_j),
\end{eqnarray*}
the induction hypothesis and Lemma \ref{hl2} imply $\rho\partial_\rho u \in
W_{\beta-1,\delta}^{l-1,s}({\cal K})^3$ and $\rho\partial_\rho p\in W_{\beta-1,\delta}^{l-2,s}({\cal K})$.
This together with the inclusion $(u,p) \in W_{\beta-1,\varepsilon-2/s}^{l-1,s}({\cal K})^3\times
W_{\beta-1,\varepsilon-2/s}^{l-2,s}({\cal K})$ and Lemma \ref{dl5} yields $u\in
W_{\beta,\delta}^{l,s}({\cal K})^3$ and $p\in W_{\beta,\delta}^{l-1,s}({\cal K})$.

Finally, we assume that $\delta_k+2/s\le 1$ for some but not all $k$. Then let $\psi_1,\ldots,\psi_n$
be smooth functions on $\bar{\Omega}$ such that $\psi_k\ge 0$, $\psi_k=1$ near $M_j\cap S^2$, and
$\sum \psi_k =1$. We extend $\psi_k$ to ${\cal K}$ by the equality $\psi_k(x)=\psi_k(x/|x|)$. Then
$\partial_x^\alpha \psi_k(x) \le c\, |x|^{-|\alpha|}$. From the results proved above it follows that
$\psi_k u \in W_{\beta,\delta}^{l,s}({\cal K})^3$ and $\psi_k p \in W_{\beta,\delta}^{l-1,s}({\cal K})^3$
for $k=1,\ldots,n$. This completes the proof. \rule{1ex}{1ex}

\begin{Co} \label{hc1}
Let $u,p,F,g,h_j$ be as in Theorem {\em \ref{ht3}}. We assume that there are no eigenvalues of the pencil
${\mathfrak A}(\lambda)$ on the lines $\mbox{\em Re}\, \lambda =1-\beta'-3/\sigma$ and
$\mbox{\em Re}\, \lambda =l-\beta-3/s$, and that $\delta,\delta'$ satisfy the inequalities {\em (\ref{h5})}.
Then $(u,p)$ admits the decomposition {\em (\ref{h4})}, where $w\in W_{\beta,\delta}^{l,s}({\cal K})^3$,
$q\in W_{\beta,\delta}^{l-1,s}({\cal K})$, and $\lambda_\nu$ are the eigenvalues of the pencil
${\mathfrak A}(\lambda)$ between the lines $\mbox{\em Re}\, \lambda =1-\beta'-3/\sigma$ and
$\mbox{\em Re}\, \lambda =l-\beta-3/s$, and $\big( u^{(\nu,j,s)},p^{(\nu,j,s)}\big)$ are eigenvectors
and generalized eigenvectors corresponding to the eigenvalue $\lambda_\nu$.
\end{Co}

{\it Proof}:
Under our assumptions on $\delta'$, the functional $F$ belongs to $V_{\beta-l+1,\delta''}^{1,s}({\cal K})^3$
with arbitrary $\delta''=(\delta_1'',\ldots,\delta''_n)$, $\delta''_k\ge \delta_k-l+1$, $\max(0,1-\mu_k)<
\delta''_k+2/s<1$. Consequently, by Theorem \ref{ht1}, $(u,p)$ has the representation (\ref{h4}),
where $(w,q)\in W_{\beta-l+1,\delta''}^{1,s}({\cal K})^3\times W_{\beta-l+1,\delta''}^{0,s}({\cal K})$
is a solution of problem (\ref{h1}), (\ref{h2}). Applying Theorem \ref{ht3}, we obtain $w\in
W_{\beta,\delta}^{l,s}({\cal K})^3$ and $q\in W_{\beta,\delta}^{l-1,s}({\cal K})$. \rule{1ex}{1ex}

\setcounter{equation}{0}
\setcounter{Le}{0}
\setcounter{Th}{0}
\setcounter{Co}{0}
\setcounter{Rem}{0}
\section{The problem in a bounded domain}

Let ${\cal G}$ be a bounded domain of polyhedral type in ${\Bbb R}^3$. This means that
\begin{itemize}
\item[(i)] the boundary $\partial{\cal G}$ consists of smooth (of class $C^\infty$)
  open two-dimensional manifolds $\Gamma_j$ (the faces of ${\cal G}$), $j=1,\ldots,n$,
  smooth curves $M_k$ (the edges), $k=1,\ldots,m$, and corners $x^{(1)},\ldots,x^{(d)}$,
\item[(ii)] for every $\xi\in M_k$ there exist a neighborhood ${\cal U}_\xi$ and
  a diffeomorphism (a $C^\infty$ mapping) $\kappa_\xi$ which maps
  ${\cal G}\cap {\cal U}_\xi$ onto ${\cal D}_\xi \cap B_1$, where ${\cal D}_\xi$ is a
  dihedron of the form (\ref{dih}) and $B_1$ is the unit ball,
\item[(iii)] for every corner $x^{(j)}$ there exist a neighborhood ${\cal U}_j$ and
  a diffeomorphism $\kappa_j$ mapping ${\cal G}\cap {\cal U}_j$ onto ${\cal K}_j \cap
  B_1$, where ${\cal K}_j$ is a cone with vertex at the origin.
\end{itemize}
We consider the problem
\begin{eqnarray} \label{i1}
&& -\Delta u  + \nabla p = f, \quad - \nabla \cdot u = g \ \mbox{ in }{\cal G}, \\ \label{i2}
&& S_j u=h_j, \quad N_j(u,p) = \phi_j \ \mbox{ on }\Gamma_j, \ j=1,\ldots,n
\end{eqnarray}
where $S_j$ and $N_j$ as well as the numbers $d_j\in \{0,1,2,3\}$ are defined as in Section 3.

\subsection{Sobolev spaces in \boldmath${\cal G}$\unboldmath}

We denote by $\rho_j(x)$ the distance of $x$ to the corner $x^{(j)}$, by $\rho(x)$ the distance
to the set $X=\{ x^{(1)},\ldots,x^{(d)}\}$, and by $r_k(x)$ the distance to the edge $M_k$.
Then $W_{\beta,\delta}^{l,s}({\cal G})$ is defined as the weighted Sobolev space with the norm
\[
\| u\|_{W_{\beta,\delta}^{l,s}({\cal G})} = \Big( \int_{\cal G}
  \sum_{|\alpha|\le l} \big| \partial_x^\alpha u\big|^s \ \prod_{j=1}^d
  \rho_j^{s(\beta_j-l+|\alpha|)}\ \prod_{k=1}^m \big( \frac{r_k}{\rho}
  \big)^{s\delta_k} \, dx \Big)^{1/s}.
\]
Here $1<s<\infty$, $\beta=(\beta_1,\ldots,\beta_d)\in {\Bbb R}^d$, $\delta=(\delta_1,\ldots,\delta_m)
\in {\Bbb R}^m$, $\delta_k>-2/s$ for $k=1,\ldots,m$, and $l$ is a nonnegative integer.
Note that the space $W_{0,0}^{1,2}({\cal G})$ (where both $\beta$ and $\delta$ are zero)
coincides with the nonweighted Sobolev space $W^{1,2}({\cal G})$.

For arbitrary $\beta\in {\Bbb R}^d$, $\delta\in {\Bbb R}^m$, $1<s<\infty$ and integer $l\ge 0$ let
$V_{\beta,\delta}^{l,s}({\cal G})$ be the weighted Sobolev space with the norm
\[
\| u\|_{V_{\beta,\delta}^{l,s}({\cal G})} = \Big( \int_{\cal G}
  \sum_{|\alpha|\le l} \big| \partial_x^\alpha u\big|^s \ \prod_{j=1}^d
  \rho_j^{s(\beta_j-l+|\alpha|)}\ \prod_{k=1}^m \big( \frac{r_k}{\rho}
  \big)^{s(\delta_k-l+|\alpha|)} \, dx \Big)^{1/s}.
\]
The dual space of $V_{\beta,\delta}^{l,s}({\cal G})$ is denoted by
$V_{-\beta,-\delta}^{-l,s'}({\cal G})$, where $s'=s/(s-1)$.

Finally, we denote the trace spaces on $\Gamma_j$ for $V_{\beta,\delta}^{l,s}({\cal G})$ and
$W_{\beta,\delta}^{l,s}({\cal G})$ by $V_{\beta,\delta}^{l-1/s,s}(\Gamma_j)$
and $W_{\beta,\delta}^{l-1/s,s}(\Gamma_j)$, respectively.

\subsection{Model problems and corresponding operator pencils}

We introduce the operator pencils generated by problem (\ref{i1}), (\ref{i2}) for the
edge points and vertices of the domain ${\cal G}$.

1) Let $\xi$ be a point on an edge $M_k$, and let $\Gamma_{k_+},\Gamma_{k_-}$ be the faces of
${\cal G}$ adjacent to $\xi$. Then by ${\cal D}_\xi$ we denote the dihedron which is
bounded by the half-planes $\Gamma_{k_\pm}^\circ$ tangent to $\Gamma_{k_\pm}$ at
$\xi$. The angle between the half-planes $\Gamma_{k_\pm}^\circ$ is denoted by $\theta(\xi)$.
We consider the model problem
\begin{eqnarray*}
&& -\Delta u + \nabla p =f,\quad -\nabla\cdot u=g \quad\mbox{in }{\cal D}_\xi , \\
&& S_{k_\pm} u= h_{k_\pm},\quad N_{k_\pm}(u,p)=\phi_{k_\pm}\ \mbox{ on } \Gamma_{k_\pm}^\circ .
\end{eqnarray*}
The operator pencil corresponding to this model problem (see Section 2.2) is denoted
by $A_\xi(\lambda)$. Furthermore, let $\lambda_1(\xi)$ be the eigenvalue with smallest
positive real part of this pencil, while $\lambda_2(\xi)$ is the eigenvalue with smallest real part greater
than 1. We define
\[
\mu(\xi) = \left\{ \begin{array}{ll} \mbox{Re}\, \lambda_1(\xi) & \mbox{if }d_{k_+} +d_{k_-} \mbox{ is odd }\
  \mbox{ or }  \ d_{k_+} + d_{k_-} \mbox{ is even and } \theta(\xi) \ge \pi/m_k, \\
  \mbox{Re}\, \lambda_2(\xi) & \mbox{if } d_{k_+} + d_{k_-} \mbox{ is even and } \alpha_k < \pi/m_k, \end{array}\right.
\]
where $m_k=1$ if $d_{k_+} = d_{k_-}$, $m_k=2$ if $d_{k_+} \not= d_{k_-}$. Finally, let
\begin{equation} \label{i6}
\mu_k = \inf_{\xi\in M_k} \mu(\xi).
\end{equation}

2) Let $x^{(j)}$ be a corner of ${\cal G}$ and let $I_j$ be the set of all indices
$k$ such that $x^{(j)}\in \ol{\Gamma}_k$. By our assumptions, there exist a
neighborhood ${\cal U}$ of $x^{(j)}$ and a diffeomorphism $\kappa$ mapping
${\cal G}\cap {\cal U}$ onto ${\cal K}_j\cap B_1$ and $\Gamma_k\cap {\cal U}$ onto
$\Gamma_k^\circ\cap B_1$ for $k\in I_j$, where ${\cal K}_j$ is a polyhedral cone
with vertex $0$ and $\Gamma_k^\circ$ are the faces of this cone.
Without loss of generality, we may assume that the Jacobian matrix $\kappa'(x)$ is
equal to the identity matrix $I$ at the point $x^{(j)}$.
We consider the model problem
\begin{eqnarray*}
&& -\Delta  u + \nabla p = f,\quad -\nabla\cdot u = g\quad \mbox{ in }{\cal K}_j,\\
&& S_k u=h_k,\quad N_k(u,p)=\phi_k \ \mbox{ on } \Gamma_k^\circ\mbox{ for }k \in I_j.
\end{eqnarray*}
The operator pencil generated by this model problem (see Section 3.2) is denoted by
${\mathfrak A}_j(\lambda)$.

\subsection{Existence of weak solutions}

We introduce the spaces
\[
V = \{ u\in W^{1,2}({\cal G})^3: \ S_j u =0\mbox{ on }\Gamma_j,\ j=1,\ldots,n\},\quad
V_0 = \{ u\in V:\ \nabla\cdot u=0\}.
\]
Furthermore, we denote by $L_V$ the set of all $u\in V$ such that $\varepsilon_{i,j}(u)=0$ for
$i,j=1,2,3$. It can be easily seen that $L_V$ is contained in the span of all all constant vectors
and of the vectors $(x_2,-x_1,0),\ (0,x_3,-x_2), \ (-x_3,0,x_1)$. In particular, we have
$L_V\subset V_0$.

Let the bilinear form $b$ be defined as
\[
b(u,v)= 2 \int_{\cal G} \sum_{i,j=1}^3 \varepsilon_{i,j}(u)\, \varepsilon_{i,j}(v)\, dx.
\]
We consider the problem
\begin{eqnarray} \label{i3}
&& b(u,v) - \int_{\cal G} p\, \nabla\cdot v \, dx = F(v)\ \mbox{ for all } v\in V \\ \label{i4}
&& -\nabla\cdot u =g\ \mbox{ in }{\cal G},\quad S_j u=h_j\ \mbox{ on }\Gamma_j,\ j=1,\ldots,n,
\end{eqnarray}
where $F$ is a given linear an continuous functional on $V$, $g\in L_2({\cal G})$,
$h_j\in W^{1/2,2}(\Gamma_j)^{3-d_j}$. We assume that the vector functions $h_j$ are such that
there exists a vector function $v\in W^{1,2}({\cal G})^3$ satisfying the boundary conditions
$S_j v =h_j$ on $\Gamma_j$ for $j=1,\ldots,n$. This means, the boundary data $h_j$ must satisfy
a certain trace condition on the edges of the domain (see Section 3.1).

\begin{Le} \label{il1}
Let $g \in L_2({\cal G})$, and let $h_j\in W^{1/2,2}(\Gamma_j)^{3-d_j}$ are such that there exists
a vector function $v\in W^{1,2}({\cal G})^3$, $S_j v =h_j$ on $\Gamma_j$ for $j=1,\ldots,n$. In the
case when $d_j \in \{ 0,2\}$ for all $j$, we assume additionally that
\begin{equation} \label{1il1}
\int_{\cal G} g\, dx + \sum_{j:\, d_j=0} \int_{\Gamma_j} h_j\cdot n\, dx +
  \sum_{j:\, d_j=2} \int_{\Gamma_j} h_j\, dx =0.
\end{equation}
Then there exists a vector function $u \in W^{1,2}({\cal G})^3$ such that $\nabla\cdot u = -g$
and $S_j u =h_j$ on $\Gamma_j$ for $j=1,\ldots,n$.
\end{Le}

{\it Proof:}
Let $v\in W^{1,2}({\cal G})^3$, $S_j v =h_j$ on $\Gamma_j$ for $j=1,\ldots,n$. We have to show
that there exists a vector function $w\in V$ such that $\nabla\cdot w = -g - \nabla\cdot v$.
Then $u=v+w$ is the desired vector function.

Let first $d_j \in \{ 0,2\}$ for all $j$. By \cite[Ch.1,Cor.2.4]{Girault}, there exists a
vector function $w\in \stackrel{\circ}{W}\!\!{}^{1,2}({\cal G})^3\subset V$ satisfying
$\nabla\cdot w = -g - \nabla\cdot v$ if
\[
\int_{\cal G} (g+\nabla\cdot v)\, dx =0.
\]
The last condition is equivalent to (\ref{1il1}).

We consider the case when $d_j\in \{1,3\}$ for at least one $j=j_0$. Let $\phi\in C_0^\infty(\Gamma_{j_0})$
be a function the integral of which over $\Gamma_{j_0}$ is equal to 1.
Then there exists a vector function $\psi\in W^{1,2}({\cal G})^3$ such that
\[
\psi=0 \ \mbox{ on }\Gamma_j\mbox{ for }j\not=0, \quad \psi_n=\phi,\ \psi_\tau=0\ \mbox{ on }\Gamma_{j_0}
\]
(see Lemma \ref{dl1}). Since $j_0 \in \{ 1,3\}$, the vector function $\psi$ belongs to $V$. We introduce the function
\[
g' = g+\nabla\cdot v - c\, \nabla\cdot \psi,\quad\mbox{where } c= \int_{\cal G} (g+\nabla\cdot v)\, dx.
\]
Since
\[
\int_{\cal G} g'\, dx= c\, \Big(1- \int_{\cal G} \nabla\cdot \psi\, dx\Big)
  = c\, \Big( 1 -\int_{\Gamma_{j_0}} \phi\, dx\Big) =0,
\]
there exists a vector function $w'\in V$ such that $-\nabla\cdot w'= g'$. Consequently,
$w=w'-c\psi$ satisfies the equation $\nabla\cdot w = -g-\nabla\cdot v$. The result follows. \rule{1ex}{1ex}\\

The necessity of condition (\ref{1il1}) in Lemma \ref{il1} is obvious. Moreover, since
$b(u,v)=0$ and $\nabla\cdot v=0$ for $v\in L_V$, for the solvability of problem (\ref{i3}), (\ref{i4})
it is necessary that
\begin{equation} \label{i5}
F(v) = 0 \quad\mbox{for all } v\in L_V .
\end{equation}

\begin{Th} \label{it1}
Let $g$ and $h_j$ be as in Lemma {\em \ref{il1}}, and let $F\in V^*$ be a functional satisfying the
condition {\em (\ref{i5})}. Then there exists a solution $(u,p)\in W^{1,2}({\cal G})^3 \times L_2({\cal G})$
of problem {\em (\ref{i3}), (\ref{i4})}. Here $p$ is uniquely determined if $d_j \in \{ 1,3\}$ for at least
one $j$ and unique up to constants if $d_j\in \{ 0,2\}$ for all $j$. The vector function $u$
is unique up to elements from $L_V$.
\end{Th}

{\it Proof}:
1) Let first $g=0$ and $h_j=0$ for $j=1,\ldots,n$. We denote by $L_V^\bot$
the orthogonal complement of $L_V$ in $V_0$. By Korn's inequality, we have
\begin{equation} \label{1it1}
b(u,\bar{u}) \ge c\, \| u\|^2_{W^{1,2}({\cal G})^3}\quad\mbox{for all } v\in L_V^\bot .
\end{equation}
Consequently, there exists a unique vector function $u \in L_V^\bot$ such that
$b(u,v)= F(v)$ for all $v\in L_V^\bot$.
Since both $b(u,v)$ and $F(v)$ vanish for $v\in L_V$, it follows that
\begin{equation} \label{2it1}
b(u,v) = F(v) \quad \mbox{for all }v\in V_0.
\end{equation}
Let $V_0^\bot$ denote the orthogonal complement of $V_0$ in $V$. By Lemma \ref{il1},
the operator $B=-\mbox{div}$ is an isomorphism from $V_0^\bot$ onto $L_2({\cal G})$ if
$d_j\in \{1,3\}$ for at least one $j$ and onto the space
\[
\stackrel{\circ}{L}\!\!{}_2({\cal G}) = \{ q\in L_2({\cal G}): \ \int_{\cal G} q(x)\, dx=1\}
\]
if $d_j\in \{ 0,2\}$ for all $j$. Suppose that $d_j\in \{ 0,2\}$ for all $j$. Then we consider
the mapping
\[
L_2({\cal G}) \ni q \to \ell(q) = F(B^{-1} \stackrel{\circ}{q}) - b(u,B^{-1}\stackrel{\circ}{q}), \quad
  \mbox{where } \stackrel{\circ}{q} = q - \frac{1}{|{\cal G}|} \int_{\cal G} q(x)\, dx
  \in \stackrel{\circ}{L}\!\!{}_2({\cal G}).
\]
Obviously, $\ell$ defines a linear and continuous functional on $L_2({\cal G})$. Consequently, there exists
a function $p\in L_2({\cal G})$ such that
\[
\int_{\cal G} p\, q\, dx = \ell(q)\quad\mbox{for all }q\in L_2({\cal G}).
\]
Consequently,
\begin{equation} \label{3it1}
- \int_{\cal G} p\, \nabla\cdot v\,  dx = \ell(-\nabla\cdot v) = F(v)-b(u,v)
  \quad\mbox{for all } v\in V_0^\bot.
\end{equation}
In the case when $d_j\in \{1,3\}$ for at least one $j$, the existence of $p\in L_2({\cal G})$ satisfying
(\ref{2it1}) follows analogously from the continuity of the mapping
\[
L_2({\cal G}) \ni q \to \ell(q) = F(B^{-1} q) - b(u,B^{-1}q) \in {\Bbb C}.
\]
Combining (\ref{2it1}) and (\ref{3it1}), we conclude that $u$ and $p$ satisfies (\ref{i3}). This proves
the existence of a solution.

We prove the uniqueness. Let $u\in V_0$ and $p\in L_2({\cal G})$ satisfy (\ref{i3}) with $F=0$. Then,
in particular, $b(u,\bar{u})=0$. Obviously, $b(u,\bar{u})=b(u-w,\bar{u}-\bar{w})$, where
$w$ is the orthogonal projection of $u\in V_0$ onto $L_V$. Using (\ref{1it1}), we obtain $u-w=0$,
i.e., $u\in L_V$. However, then $b(u,v)=0$ for all $v\in V$ and, therefore,
\[
\int_{\cal G} p\, \nabla\cdot v =0\ \mbox{ for all }v\in V.
\]
If $d_j\in \{ 1,3\}$ for at least one $j$, then $v$ can be chosen such that $\nabla v= \bar{p}$,
and we obtain $p=0$. If $d_j\in \{ 0,2\}$ for all $j$, then we obtain
\[
\int_{\cal G} p\, q\, dx = 0 \quad\mbox{for all } q\in L_2({\cal G}),\ \int_{\cal G} q\, dx =0.
\]
From this we conclude that $p$ is constant. The proof is complete. \rule{1ex}{1ex}

\subsection{Regularity assertions for weak solutions}

Our goal is to show that the solution $(u,p) \in W^{1,2}({\cal G})^3 \times L_2({\cal G})$ of problem
(\ref{i3}), (\ref{i4}) belongs to $W^{1,s}_{\beta,\delta}({\cal G})^3\times W_{\beta,\delta}^{0,s}({\cal G})$
under certain conditions on $F$, $g$, $h_j$, $\beta$ and $\delta$.
For this end, we consider the perturbed Stokes problem in the cone ${\cal K}_j$
\begin{eqnarray} \label{i7}
&& b_1(u,v) + \int_{{\cal K}_j} p\, L_1v\, dx = F(v)\  \mbox{ for all } v\in W^{1,2}({\cal K}_j)^3,\
  S_k v = 0  \mbox{ on }\Gamma_k^\circ,\ k\in I_j, \\ \label{i8}
&& L_1 u =g\ \mbox{ in }{\cal K}_j, \quad S_k u =h_k \ \mbox{ on }\Gamma_k^\circ,\ k\in I_j,
\end{eqnarray}
where
\begin{eqnarray*}
&& b_1(u,v)= 2\int_{{\cal K}_j}\sum_{k,l=1}^3\varepsilon_{k,l}(u)\, \varepsilon_{k,l}(v)\, dx
  + \sum_{\mu,\nu,k,l=1}^3 \int_{{\cal K}_j} b_{\mu,\nu,k,l}(x)
  \, \frac{\partial u_k}{\partial x_\mu}\, \frac{\partial u_l}{\partial x_\nu}\, dx, \\
&&  L_1 v = -\nabla \cdot v + \sum_{k,l=1}^3 c_{k,l}(x)\frac{\partial v_k}{\partial x_l} \, .
\end{eqnarray*}
We assume that
\begin{equation} \label{i9}
\sum_{\mu,\nu,k,l} |b_{\mu,\nu,k,l}(x)|+ \sum_{k,l} |c_{k,l}(x)| < \varepsilon
\end{equation}
with sufficiently small $\varepsilon$.

\begin{Le} \label{il2}
Let $(u,p)\in W_{\beta',0}^{1,2}({\cal K}_j)^3 \times W_{\beta',0}^{0,2}({\cal K}_j)$ be a solution
of problem {\em (\ref{i7}), (\ref{i8})}, where
\begin{equation} \label{1il2}
F\in W_{\beta',0}^{-1,s}({\cal K}_j)^3\cap V_{\beta,\delta}^{-1,s}({\cal K}_j)^3, \ \
  g\in W_{\beta',0}^{0,2}({\cal K}_j) \cap W_{\beta,\delta}^{0,s}({\cal K}_j), \ \
  h_k \in W_{\beta',0}^{1/2,2}(\Gamma_k^\circ) \cap W^{1-1/s,s}_{\beta,\delta}(\Gamma_k^\circ)
\end{equation}
for $k\in I_j$. Suppose that there are no eigenvalues of the pencil ${\mathfrak A}_j(\lambda)$ in the closed strip
between the lines $\mbox{\em Re}\lambda = -\beta'-1/2$ and $\mbox{\em Re}\lambda = 1-\beta-3/s$,
the components of $\delta$ satisfy the inequalities $\max(1-\mu_k,0)<\delta_k+2/s<1$, and that the number
$\varepsilon$ in {\em (\ref{i9})} is sufficiently small. Then
$u\in W_{\beta,\delta}^{1,s}({\cal K}_j)^3$ and $p \in W_{\beta,\delta}^{0,s}({\cal K}_j)$.
\end{Le}

{\it Proof:}
Let ${\cal W}_{s,\beta,\delta}$ be the space of all
\[
\{ h_k\}_{k\in I_j} \in \prod_{k\in I_j} W^{1-1/s,s}_{\beta,\delta}(\Gamma_k^\circ)^{3-d_k}
\]
such that there exists a vector function $u\in W_{\beta,\delta}^{1,s}({\cal K}_j)$ satisfying
$S_k u =h_k$ on $\Gamma_k^\circ$ for $k\in I_j$. This is a subspace of vector-functions on
$\Gamma_k^\circ$ satisfying certain compatibility conditions on the edges of the cone ${\cal K}_j$
(see Sections 2.1 and 3.1). We define $A$ as the operator
\begin{equation} \label{2il2}
W_{\beta',0}^{1,2}({\cal K}_j)^3 \times W_{\beta',0}^{0,2}({\cal K}_j) \ni (u,p) \to
  (F,g,h_j) \in W_{\beta',0}^{-1,s}({\cal K}_j)^3 \times W_{\beta',0}^{0,2}({\cal K}_j) \times
  {\cal W}_{2,\beta',0}
\end{equation}
where $F$, $g$ and $h_k$ are given by (\ref{i7}), (\ref{i8}). Furthermore, let $A_0$ be the operator
(\ref{2il2}), where
\[
F(v)=2\int_{{\cal K}_j} \sum_{i,j=1}^3 \varepsilon_{i,j}(u)\, \varepsilon_{i,j}(v)\, dx
  -\int_{{\cal K}_j} p\, \nabla\cdot v\, dx,\quad g=-\nabla\cdot u,\quad h_k=S_k u.
\]
By \cite[Th.4.2]{mr-03}, the operator $A_0$ is an isomorphism. Furthermore, it follows from
Theorem \ref{gt1} and Lemma \ref{hl1} that $A_0$ is an isomorphism
\begin{eqnarray}
&& \hspace{-2em}\Big( W_{\beta',0}^{1,2}({\cal K}_j)^3\times W_{\beta',0}^{0,2}({\cal K}_j)\Big)
  \cap \Big( W_{\beta,\delta}^{1,s}({\cal K}_j)^3\times W_{\beta,\delta}^{0,s}({\cal K}_j)\Big)
   \nonumber \\ \label{3il2}
&& \to \Big( W_{\beta',0}^{-1,s}({\cal K}_j)^3 \times W_{\beta',0}^{0,2}({\cal K}_j) \times
  {\cal W}_{2,\beta',0} \Big) \cap \Big( V_{\beta,\delta}^{-1,s}({\cal K}_j)^3
  \times W_{\beta,\delta}^{0,s}({\cal K}_j) \times {\cal W}_{s,\beta,\delta}\Big)
\end{eqnarray}
Due to (\ref{i9}), the operator norm (\ref{3il2}) of $A-A_0$ is less than $c\varepsilon$. Hence for
sufficiently small $\varepsilon$, the operator  $A-A_0$ is also an isomorphism (\ref{3il2}).
The result follows. \rule{1ex}{1ex}

\begin{Th} \label{it2}
Let $(u,p)\in W^{1,2}({\cal G})^3\times L_2({\cal G})$ be a solution of problem {\em (\ref{i3}), (\ref{i4})},
where
\[
F\in V^* \cap V_{\beta,\delta}^{-1,s}({\cal K}_j)^3, \ \
  g\in L_2({\cal G}) \cap W_{\beta,\delta}^{0,s}({\cal G}), \ \
  h_k \in W^{1/2,2}(\Gamma_k) \cap W^{1-1/s,s}_{\beta,\delta}(\Gamma_k).
\]
Suppose that there are no eigenvalues of the pencils ${\mathfrak A}_j(\lambda)$, $j=1,\ldots,d$ in the closed
strip between the lines $\mbox{\em Re}\lambda = -1/2$ and $\mbox{\em Re}\lambda = 1-\beta-3/s$ and that
the components of $\delta$ satisfy the inequalities $\max(1-\mu_k,0)<\delta_k+2/s<1$. Then
$u\in W^{1,s}_{\beta,\delta}({\cal G})^3$, $p\in W_{\beta,\delta}^{0,s}({\cal G})$.
\end{Th}

{\it Proof:}
It suffices to prove the theorem for vector functions $(u,p)$ with small supports. For solutions with
arbitrary support the assertion then can be easily proved by means of a partition of unity on ${\cal G}$.
Let the support of $(u,p)$ be contained in a sufficiently small neighborhood ${\cal U}$ of the vertex
$x^{(j)}$, and let $\kappa$ be a diffeomorphism mapping ${\cal G}\cap {\cal U}$ onto ${\cal K}_j \cap
B$, where ${\cal K}_j$ is a cone with vertex at the origin and $B$ is a ball centered about the origin.
We assume that $\kappa'(x^{(j)})=I$. Then the vector function $(w(x),q(x))=\big(u\big(\kappa^{-1}(x)\big),
p\big(\kappa^{-1}(x)\big)\big)$ is a solution of a perturbed Stokes problem (\ref{i7}), (\ref{i8}),
where the coefficients $b_{\mu,\nu,k,l}$ and $c_{k,l}$ are zero at the origin and bounded by small
constants on the support of $(w,q)$. Applying Lemma \ref{il2}, we obtain
$(w,q)\in W_{\beta_j,\delta}^{1,s}({\cal K}_j)^3\times W_{\beta_j,\delta}^{0,s}({\cal K}_j)$
and, therefore $(u,p) \in W^{1,s}_{\beta,\delta}({\cal G})^3\times W_{\beta,\delta}^{0,s}({\cal G})$.
For vector functions $(u,p)$ with support in a neighborhood of an edge point, the assertion
of the theorem can be proved analogously. \rule{1ex}{1ex} \\

Analogously, the following theorem can be proved (cf. Theorem \ref{ht3}).

\begin{Th} \label{it3}
Let $(u,p)\in W_{\beta',0}^{1,2}({\cal G})^3 \times W_{\beta',0}^{0,2}({\cal G})$ be a solution
of problem {\em (\ref{i7}), (\ref{i8})}, where $g\in W_{\beta,\delta}^{l-1,s}({\cal G})$,
$h_j\in W_{\beta,\delta}^{l-1/s,s}(\Gamma_j)^{3-d_j}$, and
$F\in V^*$ has the representation
\[
F(v) = \int_{\cal G} f\cdot v\, dx + \sum_{j=1}^n \int_{\Gamma_j} \phi_j\cdot v\, dx
  \quad\mbox{for all }v\in V
\]
with $f\in W_{\beta,\delta}^{l-2,s}(\Gamma_j)$, $\phi_j \in W_{\beta,\delta}^{l-1-1/s}(\Gamma_j)^{d_j}$.
We suppose that there are no eigenvalues of the pencils ${\mathfrak A}_j(\lambda)$, $j=1,\ldots,d$, in
the closed strip between the lines $\mbox{\em Re}\, \lambda= -1/2$ and $\mbox{\em Re}\, \lambda
=l-\beta-3/s$ and that $\max(l-\mu_k,0)<\delta_k+2/s<l$ for $k=1,\ldots,m$.
Furthermore, we assume that $g$, $h_j$ and $\phi_j$ satisfy compatibility conditions on the edges $M_k$ which
guarantee that there exist $w\in W_{\beta-l+2,\delta'}^{2,s}({\cal G})^3$ and $q\in
W_{\beta-l+2,\delta'}^{1,s}({\cal G})$, $\delta'_k=\max(\delta_k-l+2,\frac 12 -\frac 2s)$, such that
\[
S_jw=h_j, \ \ N_j(w,q)=\phi_j \ \mbox{ on }\Gamma_j,\ j=1,\ldots,n,\quad
  \nabla\cdot w +g \in V_{\beta-l+2,\delta'}^{1,s}({\cal G}).
\]
Then $u\in W_{\beta,\delta}^{l,s}({\cal G})^3$ and $p\in W_{\beta,\delta}^{l-1,s}({\cal G})$.
\end{Th}

\subsection{Examples}

Here we establish some regularity assertions for weak solutions of special boundary value problems
for the Stokes system in the class of the nonweighted spaces $W^{l,s}({\cal G})$.
Let ${\cal G}$ be a polyhedron with sides $\Gamma_j$, $j=1,...,n$, and edges $M_k$, $k=1,\ldots,m$.
We denote the angle at the edge $M_k$ by $\theta_k$.
For the sake of simplicity, we restrict ourselves to homogeneous boundary conditions
\begin{equation} \label{h6}
  S_j u=0,\quad N_j(u,p)=0\quad\mbox{on }\Gamma_j,\ j=1,\ldots,n.
\end{equation}
Analogous results are valid for inhomogeneous boundary conditions provided the boundary data
satisfy certain compatibility conditions on the edges. \\

{\em The Dirichlet problem for the Stokes system}.
Let $f\in W^{-1,2}({\cal G})^3$ and $g\in L_2({\cal G})$ satisfy the compatibility conditions of Theorem \ref{it1}.
Then there exists a solution $(u,p) \in W^{1,2}({\cal G})^3\times L_2({\cal G})$ of the Dirichlet problem
\[
-\Delta u+\nabla p = f,\ \ -\nabla u=g\ \mbox{in }{\cal G},\quad u=0\ \mbox{on }\Gamma_j,\ j=1,\ldots,n.
\]
Here $u$ is unique and $p$ is unique up to a constant (see also \cite[Th.5.1]{Girault}).
It is known that there are no eigenvalues of the pencils ${\mathfrak A}_j(\lambda)$
in the strip $-1 \le \mbox{Re}\, \lambda \le 0$ (see \cite[Th.5.5.6]{kmr2}). In the case, when ${\cal G}$
is convex, then even the strip $-2 < \mbox{Re}\, \lambda < 1$ does not contain eigenvalues of the pencils
${\mathfrak A}_j(\lambda)$ (see \cite[Th.5.5.5]{kmr2}). Moreover, it can be easily verified that
$\mu_k>1/2$, $\mu_k> 2/3$ if $\theta_k < 3\, \mbox{arccos}\frac 14 \approx
1.2587\pi$, $\mu_k>1$ if $\theta_k <\pi$, and $\mu_k> 4/3$ if $\theta_k<\frac 34 \pi$.
Using these results together with Theorems \ref{it2} and \ref{it3}, we obtain the
following assertions.
\begin{itemize}
\item If $f\in (W^{1,s'}({\cal G})^*)^3$ and $g\in L_s({\cal G})$, $2< s \le 3$, $s'=s/(s-1)$, then
  $(u,p) \in W^{1,s}({\cal G})^3\times L_s({\cal G})$.
  If the polyhedron ${\cal G}$ is convex, then this assertion is true for all $s>2$.
\item If $f\in W^{-1,2}({\cal G})^3 \cap L_s({\cal G})^3$ and $g\in L_2({\cal G}) \cap W^{1,s}({\cal G})$,
  $1<s\le 4/3$, then $(u,p)\in W^{2,s}({\cal G})^3\times W^{1,s}({\cal G})$. If $\theta_k< 3\,
  \mbox{arccos}\frac 14 \approx 1.2587\pi$ for $k=1,\ldots,m$, then this result is true for $1<s\le 3/2$.
  If ${\cal G}$ is convex, then this result is valid for $1<s\le 2$ provided $g$ satisfies (\ref{condg})
  if $s=2$. If, moreover, the angles at the edges are less than $\frac 34 \pi$, then the result holds even
  for $1<s<3$ provided $g$ satisfies (\ref{condg}) if  $s=2$ and $g=0$ on $M_k$, $k=1,\ldots,m$, if $s>2$.
\end{itemize}
Here we used also the facts that $W^{1,s'}({\cal G})=V_{0,0}^{1,s'}({\cal G})$ for $s'<2$ and
$W^{1,s}({\cal G})=W_{0,0}^{1,s}({\cal G})$ for $s<3$.
In the case $s=2$ the $W^{2,s}$-regularity result for convex polyhedrons was also proved by
Dauge \cite{Dauge-89}, for convex two-dimensional polygonal domains we refer to Kellogg and Osborn \cite{Kellogg}.\\

{\em The Neumann problem for the Stokes system}.
We consider the weak solution $u\in W^{1,2}({\cal G})^3\times L_2({\cal G})$ of the Neumann problem
\[
-\Delta u+\nabla p = f,\ \ -\nabla u=g\ \mbox{in }{\cal G},\quad \frac{\partial u}{\partial n}=0\
   \mbox{on }\Gamma_j,\ j=1,\ldots,n.
\]
For this problem it is known that the strip $-1 \le \mbox{Re}\, \lambda \le 0$ contains only the eigenvalues
$\lambda=0$ and $\lambda=1$ of the operator pencils ${\mathfrak A}_j(\lambda)$ (see \cite[Th.6.3.2]{kmr2})
if ${\cal G}$ is a Lipschitz polyhedron. The numbers $\mu_k$ are the same as for the Dirichlet problem.
Therefore, the following assertions are valid.
\begin{itemize}
\item If $f\in (W^{1,s'}({\cal G})^*)^3$ and $g\in L_s({\cal G})$, $2< s <3$,
  then $(u,p)\in W^{1,s}({\cal G})^3\times L_s({\cal G})$.
\item If $f\in (W^{1,2}({\cal G})^*)^3\cap L_s({\cal G})^3$ and $g\in L_2({\cal G}) \cap W^{1,s}({\cal G})$,
  $1<s\le 4/3$, then $(u,p) \in W^{2,s}({\cal G})^3\times W^{1,s}({\cal G})$.
  If the angles $\theta_k$ are less than  $3\, \mbox{arccos}\frac 14$, then this result is true for $1<s< 3/2$.
\end{itemize}

{\em The mixed problem with Dirichlet and Neumann boundary conditions}.
We assume that on each side $\Gamma_j$ either the Dirichlet condition $u=0$ or the Neumann condition
$\frac{\partial u}{\partial n}=0$ is given. If on the adjoining sides of the edge $M_k$ the same boundary
conditions are given, then $\mu_k > 1/2$. If on one of the adjoining sides the Dirichlet condition and on
the other side the Neumann condition is given, then $\mu_k >1/4$. This implies the following result.
\begin{itemize}
\item If $f\in (W^{1,2}({\cal G})^*)^3\cap L_s({\cal G})^3$ and $g\in L_2({\cal G}) \cap W^{1,s}({\cal G})$,
  $1<s\le 8/7$, then  the weak solution $(u,p)$ belongs to $W^{2,s}({\cal G})^3\times W^{1,s}({\cal G})$.
\end{itemize}

{\em The mixed problem with boundary conditions} (i)--(iii).
Let $(u,p) \in W^{1,2}({\cal G})^3\times L_2({\cal G})$ be a weak solution of problem
(\ref{i1}), (\ref{h6}), where $d_k\le 2$ for all $k$ (i.e., the Neumann condition does not appear in the
boundary conditions). We assume that the Dirichlet condition is given on at least one of the
adjoining sides of every edge. Then, by \cite[Th.6.1.5]{kmr2}, the strip $-1\le \mbox{Re}\, \lambda\le 0$
is free of eigenvalues of the pencils ${\mathfrak A}_j(\lambda)$. Furthermore, we have $\mu_k> 1/2$
if the Dirichlet condition is given on both adjoining sides of the edge $M_k$. For the other indices $k$,
we have $\mu_k > 1/4$ and $\mu_k >1/3$ if $\theta_k < \frac 32 \pi$.
\begin{itemize}
\item If $f\in (W^{1,s'}({\cal G})^*)^3$ and $g\in L_s({\cal G})$, $2< s \le 8/3$,
  then $(u,p)\in W^{1,s}({\cal G})^3\times L_s({\cal G})$. Suppose that $\theta_k<\frac 32 \pi$ if
  the boundary condition (ii) or (iii) is given on one of the adjoining sides of the edge $M_k$.
  Then this result is even true for $2<s \le 3$.
\item If $f\in (W^{1,2}({\cal G})^*)^3\cap L_s({\cal G})^3$ and $g\in L_2({\cal G}) \cap W^{1,s}({\cal G})$,
  $1<s\le 8/7$, then $(u,p) \in W^{2,s}({\cal G})^3\times W^{1,s}({\cal G})$. Suppose that $\theta_k
  < 3\, \mbox{arccos}\frac 14$ if the Dirichlet condition is given on both adjoining sides of $M_k$,
  $\theta_k < \frac 32 \, \mbox{arccos}\frac 14$ if the boundary condition (i) is given on one of the
  adjoining sides of $M_k$, and $\theta_k < \frac 34 \pi$ if the boundary condition (ii) is given
  on one of the adjoining sides of $M_k$. Then the last result is true for $1< s\le 3/2$.
\end{itemize}
Note that in the last case, we have $\mu_k >2/3$ for $k=1,\ldots,m$.

Finally, we consider problem (\ref{i1}), (\ref{h6}) when the Dirichlet condition is given on the sides
$\Gamma_1,\ldots,\Gamma_{n-1}$, while the boundary condition (ii) is given on $\Gamma_n$.
Let $I$ be the set of all $k$ such that $M_k \subset \bar{\Gamma}_n$ and $I'=\{1,\ldots,n\}\backslash I.$
We suppose that the polyhedron ${\cal G}$ is convex and $\theta_k < \pi/2$
for $k\in I$. Then $\mu_k>1$ for all $k$, and the strip $-1/2\le \mbox{Re}\, \lambda <1$ is free of eigenvalues
of the pencils ${\mathfrak A}_j(\lambda)$ (see \cite[Th.6.2.7]{kmr2}). If $\theta_k< \frac 38 \pi$ for $k\in I$
and $\theta_k< \frac 34 \pi$ for $k\in I'$, then even $\mu_k>4/3$. This implies the following result.
\begin{itemize}
\item Let $f\in (W^{1,2}({\cal G})^*)^3\cap L_s({\cal G})^3$ and $g\in L_2({\cal G})\cap W^{1,s}({\cal G})$,
  $s>1$. In the case $s>2$, we suppose that $g|_{M_k}=0$ for all $k$, while condition
  (\ref{condg}) is assumed to be valid for $s=2$.
  Then the weak solution $(u,p)\in W^{1,2}({\cal G})^3 \times L_2({\cal G})$ of problem (\ref{i1}),
  (\ref{h6}) belongs to $W^{2,s}({\cal G})^3\times W^{1,s}({\cal G})$ for $1<s\le 2$. If
  $\theta_k< \frac 38 \pi$ for $k\in I$ and $\theta_k< \frac 34 \pi$ for $k\in I'$, then the result
  holds even for $1<s<3$.
\end{itemize}

\end{document}